\documentclass[journal,a4paper,oneside,onecolumn]{IEEEtran} 
\usepackage{amsfonts,amssymb}
\usepackage[utf8]{inputenc} 
\usepackage[T1]{fontenc}
\usepackage{url}
\usepackage{ifthen}
\usepackage{cite}
\usepackage[cmex10]{amsmath}
\usepackage{mathabx}
\newtheorem{thm}{Theorem}
\newtheorem{lem}{Lemma}
\newtheorem{df}{Definition}
\newtheorem{rem}{Remark}
\newtheorem{example}{Example}

\usepackage{curve2e}
\usepackage{color}

\newcommand{\fS}{\mathfrak{S}}
\newcommand{\fC}{\mathfrak{C}}

\newcommand{\NN}{\mathbb{N}}

\newcommand{\B}{\mathcal{B}}
\newcommand{\C}{\mathcal{C}}
\newcommand{\E}{\mathcal{E}}
\newcommand{\F}{\mathcal{F}}
\newcommand{\G}{\mathcal{G}}
\newcommand{\I}{\mathcal{I}}
\newcommand{\J}{\mathcal{J}}
\newcommand{\K}{\mathcal{K}}
\newcommand{\M}{\mathcal{M}}
\newcommand{\Q}{\mathcal{Q}}
\newcommand{\cS}{\mathcal{S}}
\newcommand{\T}{\mathcal{T}}
\newcommand{\U}{\mathcal{U}}
\newcommand{\V}{\mathcal{V}}
\newcommand{\W}{\mathcal{W}}
\newcommand{\X}{\mathcal{X}}
\newcommand{\Y}{\mathcal{Y}}
\newcommand{\Z}{\mathcal{Z}}

\newcommand{\bb}{\boldsymbol{b}}
\newcommand{\cc}{\boldsymbol{c}}
\newcommand{\mm}{\boldsymbol{m}}
\newcommand{\bp}{\boldsymbol{p}}
\newcommand{\uu}{\boldsymbol{u}}
\newcommand{\vv}{\boldsymbol{v}}
\newcommand{\ww}{\boldsymbol{w}}
\newcommand{\xx}{\boldsymbol{x}}
\newcommand{\yy}{\boldsymbol{y}}
\newcommand{\zz}{\boldsymbol{z}}
\newcommand{\CC}{\boldsymbol{C}}
\newcommand{\UU}{\boldsymbol{U}}
\newcommand{\VV}{\boldsymbol{V}}
\newcommand{\WW}{\boldsymbol{W}}
\newcommand{\XX}{\boldsymbol{X}}
\newcommand{\YY}{\boldsymbol{Y}}
\newcommand{\ZZ}{\boldsymbol{Z}}
\newcommand{\zero}{\boldsymbol{0}}
\newcommand{\one}{\boldsymbol{1}}
\newcommand{\aalpha}{\boldsymbol{\alpha}}
\newcommand{\bbeta}{\boldsymbol{\beta}}
\newcommand{\bcF}{\boldsymbol{\F}}

\newcommand{\hU}{\widehat{U}}
\newcommand{\hZ}{\widehat{Z}}
\newcommand{\hW}{\widehat{W}}
\newcommand{\hX}{\widehat{X}}
\newcommand{\hu}{\widehat{u}}
\newcommand{\hWW}{\widehat{\WW}}
\newcommand{\hww}{\widehat{\ww}}

\newcommand{\tW}{\widetilde{W}}

\newcommand{\chu}{\widecheck{u}}
\newcommand{\chU}{\widecheck{U}}
\newcommand{\chW}{\widecheck{W}}
\newcommand{\chww}{\widecheck{\ww}}

\newcommand{\limn}{\lim_{n\to\infty}}
\newcommand{\limsupn}{\limsup_{n\to\infty}}
\newcommand{\liminfn}{\liminf_{n\to\infty}}
\newcommand{\pliminfn}{\operatornamewithlimits{\mathrm{p-liminf}}_{n\to\infty}}
\newcommand{\plimsupn}{\operatornamewithlimits{\mathrm{p-limsup}}_{n\to\infty}}
\newcommand{\Prod}{\operatornamewithlimits{\text{\LARGE $\times$}}}
\newcommand{\oH}{\overline{H}}
\newcommand{\uH}{\underline{H}}
\newcommand{\uI}{\underline{I}}
\newcommand{\oI}{\overline{I}}
\newcommand{\oT}{\overline{\T}}
\newcommand{\uT}{\underline{\T}}
\newcommand{\od}{\overline{d}}
\newcommand{\oO}{\overline{O}}
\newcommand{\oQ}{\overline{Q}}
\newcommand{\ugamma}{\underline{\gamma}}
\newcommand{\ogamma}{\overline{\gamma}}

\newcommand{\lrB}[1]{\left[{#1}\right]}
\newcommand{\lrb}[1]{\left\{{#1}\right\}}
\newcommand{\lrsb}[1]{\left({#1}\right)}
\newcommand{\lrbar}[1]{\left|{#1}\right|}

\newcommand{\markov}{\leftrightarrow}
\newcommand{\e}{\varepsilon}
\newcommand{\vphi}{\varphi}

\newcommand{\Prob}{\mathrm{Porb}}
\newcommand{\Error}{\mathrm{Error}}
\newcommand{\im}{\mathrm{Im}}

\newcommand{\ROP}{\mathcal{R}_{\mathrm{OP}}}

\newcommand{\ROPJB}{\mathcal{R}^{\mathrm{JB}}_{\mathrm{OP}}}
\newcommand{\RITJB}{\mathcal{R}^{\mathrm{JB}}_{\mathrm{IT}}}
\newcommand{\RCRNGJB}{\mathcal{R}^{\mathrm{JB}}_{\mathrm{CRNG}}}
\newcommand{\RITDSC}{\mathcal{R}^{\mathrm{DSC}}_{\mathrm{IT}}}
\newcommand{\RCRNGDSC}{\mathcal{R}^{\mathrm{DSC}}_{\mathrm{CRNG}}}
\newcommand{\RCRNGMDC}{\mathcal{R}^{\mathrm{MDC}}_{\mathrm{CRNG}}}

\newcommand{\RBTDSC}{\mathcal{R}^{\mathrm{DSC}}_{\mathrm{BT}}}
\newcommand{\RWKADSC}{\mathcal{R}^{\mathrm{DSC}}_{\mathrm{WKA}}}
\newcommand{\RECMDC}{\mathcal{R}^{\mathrm{MDC}}_{\mathrm{EC}}}
\newcommand{\RZBMDC}{\mathcal{R}^{\mathrm{MDC}}_{\mathrm{ZB}}}
\newcommand{\RHBDSI}{\mathcal{R}^{\mathrm{DSI}}_{\mathrm{HB}}}
\newcommand{\RCRNGDSCp}{\mathcal{R}'^{\mathrm{DSC}}_{\mathrm{CRNG}}}
\newcommand{\RCRNGMDCp}{\mathcal{R}'^{\mathrm{MDC}}_{\mathrm{CRNG}}}
\newcommand{\RCRNGDSIp}{\mathcal{R}'^{\mathrm{DSI}}_{\mathrm{CRNG}}}

\newcommand{\helper}{\mathrm{helper}}

\newcommand{\ic}{i^{\complement}}
\newcommand{\jc}{j^{\complement}}
\newcommand{\Ipc}{\I'^{\complement}}
\newcommand{\Ipcj}{\I_j'^{\complement}}

\allowdisplaybreaks

\title{
 Distributed Source Coding,\\
 Multiple Description Coding,
 and\\
 Source Coding with Side Information at Decoders\\
 Using Constrained-Random Number Generators
}
\author{
 \IEEEauthorblockN{Jun~Muramatsu~\IEEEmembership{Senior Member,~IEEE}}
 \\
 \IEEEauthorblockA{
	NTT Communication Science Laboratories, NTT Corporation\\
	2-4 Hikaridai, Seika-cho, Soraku-gun, Kyoto 619-0237, Japan.\\
	E-mail: jun.muramatsu@ieee.org.
 }
}
\begin{document}
\maketitle

\begin{abstract}
This paper investigates a unification of distributed source coding,
multiple description coding,
and source coding with side information at decoders.
The equivalence between
the multiple-decoder extension of 
distributed source coding
with decoder side information
and the multiple-source extension
of multiple description coding
with decoder side information
is clarified.
Their multi-letter rate-distortion region
for arbitrary general correlated sources
is characterized in terms of entropy functions.
We construct a code based on constrained-random number generators
and show its achievability.
{\bf Note: } 
A part of this paper
(Sections \ref{sec:jb} and \ref{sec:converse-jb}, and some appendices)
inherits the contents from {\tt arXiv:2401.13232[cs.IT]}.
\end{abstract}

\section{Introduction}

This paper investigates
a unification of distributed source coding, multiple description coding,
and source coding with side information at decoders.

\begin{figure}
\begin{center}
 \unitlength 0.55mm
 \begin{picture}(155,85)(-8,10)
	\put(0,57){\makebox(0,0){$X^n$}}
	
	\put(27,71){\makebox(0,0){Encoder}}
	\put(20,47){\framebox(14,20){$\Phi^{(n)}$}}
	\put(5,57){\vector(1,0){15}}
	\put(34,57){\vector(1,0){10}}
	\put(52,57){\makebox(0,0){$M^{(n)}$}}
	
	\put(60,57){\line(12,26){12}}
	\put(60,57){\line(12,-32){12}}
	
	\put(114,92){\makebox(0,0){Decoders}}
	\put(72,83){\vector(1,0){10}}
	\put(90,83){\makebox(0,0){$M^{(n)}_{\I_1}$}}
	\put(97,83){\vector(1,0){10}}
	\put(90,73){\makebox(0,0){$Y^n_1$}}
	\put(97,73){\vector(1,0){10}}
	\put(107,68){\framebox(14,20){$\Psi^{(n)}_1$}}
	\put(60,57){\vector(1,0){22}}
	\put(90,57){\makebox(0,0){$M^{(n)}_{\I_2}$}}
	\put(97,57){\vector(1,0){10}}
	\put(90,47){\makebox(0,0){$Y^n_2$}}
	\put(97,47){\vector(1,0){10}}
	\put(107,42){\framebox(14,20){$\Psi^{(n)}_2$}}
	\put(90,37){\makebox(0,0){$\vdots$}}
	\put(114,37){\makebox(0,0){$\vdots$}}
	\put(72,25){\vector(1,0){10}}
	\put(90,25){\makebox(0,0){$M^{(n)}_{\I_{|\J|}}$}}
	\put(97,25){\vector(1,0){10}}
	\put(90,15){\makebox(0,0){$Y^n_{|\J|}$}}
	\put(97,15){\vector(1,0){10}}
	\put(107,10){\framebox(14,20){$\Psi^{(n)}_{|\J|}$}}
	
	\put(121,78){\vector(1,0){10}}
	\put(132,77){\makebox(0,0)[l]{$Z^n_1$}}
	
	\put(121,52){\vector(1,0){10}}
	\put(132,51){\makebox(0,0)[l]{$Z^n_2$}}
	
	\put(136,37){\makebox(0,0){$\vdots$}}
	
	\put(120,20){\vector(1,0){10}}
	\put(131,19){\makebox(0,0)[l]{$Z^n_{|\J|}$}}
 \end{picture}
\end{center}
\caption{
 Source Coding with Side Information at Decoders
}
\label{fig:dsi}
\end{figure}
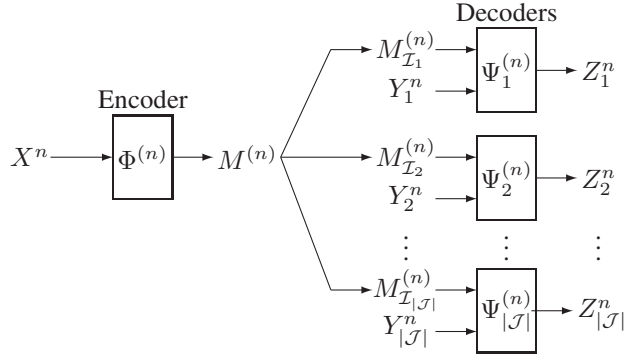

In distributed source coding (Fig.~\ref{fig:dsc}),
each encoder encodes one of the correlated sources in a distributed manner, 
and the decoder reproduces the sources within the allowed distortion limit.
In multiple description coding (Fig.~\ref{fig:mdc}),
each encoder encodes the same source into a different codeword
and each decoder reproduces a source within the allowed distortion limit.
In source coding with (non-causal) side information
at (many) decoders (Fig.~\ref{fig:dsi}),
decoders have access to each side information
in addition to the same codeword of an encoder
and reproduce each source within the allowed distortion limit.
It is expected that the decoding error probability asymptotically
approaches zero as the block length goes to infinity.
We consider general correlated sources,
for which we do not assume conditions
such as consistency, stationarity, or ergodicity.

\begin{figure}
\begin{center}
 \unitlength 0.55mm
 \begin{picture}(120,70)(0,0)
	\put(10,46){\makebox(0,0){$X^n_1$}}
	\put(10,32){\makebox(0,0){$\vdots$}}
	\put(10,14){\makebox(0,0){$X_{|\I|}^n$}}
	\put(16,46){\vector(1,0){10}}
	\put(16,14){\vector(1,0){10}}
	\put(33,65){\makebox(0,0){Encoders}}
	\put(26,39){\framebox(14,14){$\Phi^{(n)}_1$}}
	\put(33,32){\makebox(0,0){$\vdots$}}
	\put(26,7){\framebox(14,14){$\Phi^{(n)}_{|\I|}$}}
	\put(40,46){\vector(1,0){10}}
	\put(40,14){\vector(1,0){10}}
	\put(58,46){\makebox(0,0){$M^{(n)}_1$}}
	\put(58,32){\makebox(0,0){$\vdots$}}
	\put(58,14){\makebox(0,0){$M^{(n)}_{|\I|}$}}
	\put(66,46){\vector(1,0){10}}
	\put(66,14){\vector(1,0){10}}
	\put(83,65){\makebox(0,0){Decoder}}
	\put(76,0){\framebox(14,60){$\Psi^{(n)}$}}
	\put(90,51){\vector(1,0){10}}
	\put(90,9){\vector(1,0){10}}
	\put(100,51){\makebox(0,0)[l]{$Z^n_1$}}
	\put(104,32){\makebox(0,0){$\vdots$}}
	\put(100,7){\makebox(0,0)[l]{$Z^n_{|\K|}$}}
 \end{picture}
\end{center}
\caption{
 Distributed Source Coding
}
\label{fig:dsc}
\end{figure}
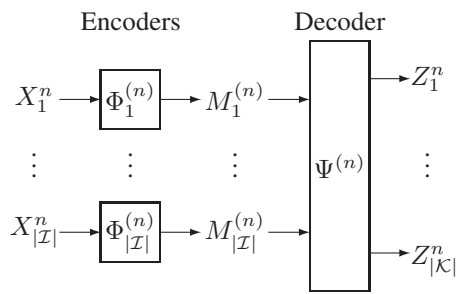

\begin{figure}
\begin{center}
 \unitlength 0.55mm
 \begin{picture}(155,85)(-8,10)
	\put(0,52){\makebox(0,0){$X^n$}}
	
	\put(5,52){\line(5,26){5}}
	\put(5,52){\line(5,-32){5}}
	
	\put(10,78){\vector(1,0){10}}
	\put(5,52){\vector(1,0){15}}
	\put(10,20){\vector(1,0){10}}
	
	\put(27,92){\makebox(0,0){Encoders}}
	\put(20,68){\framebox(14,20){$\Phi^{(n)}_1$}}
	\put(34,78){\vector(1,0){10}}
	\put(52,78){\makebox(0,0){$M^{(n)}_1$}}
	\put(20,42){\framebox(14,20){$\Phi^{(n)}_2$}}
	\put(34,52){\vector(1,0){10}}
	\put(52,52){\makebox(0,0){$M^{(n)}_2$}}
	\put(50,37){\makebox(0,0){$\vdots$}}
	\put(27,37){\makebox(0,0){$\vdots$}}
	\put(20,10){\framebox(14,20){$\Phi^{(n)}_{|\I|}$}}
	\put(34,20){\vector(1,0){10}}
	\put(52,20){\makebox(0,0){$M^{(n)}_{|\I|}$}}
	
	\put(60,78){\line(1,0){14}}
	\put(60,78){\line(14,-58){14}}
	\put(60,52){\line(14,26){14}}
	\put(60,52){\line(1,0){14}}
	\put(60,20){\line(1,0){14}}
	\put(60,20){\line(14,58){14}}
	
	\put(114,92){\makebox(0,0){Decoders}}
	\put(74,78){\vector(1,0){10}}
	\put(91,78){\makebox(0,0){$M^{(n)}_{\I_1}$}}
	\put(97,78){\vector(1,0){10}}
	\put(107,68){\framebox(14,20){$\Psi^{(n)}_1$}}
	\put(74,52){\vector(1,0){10}}
	\put(91,52){\makebox(0,0){$M^{(n)}_{\I_2}$}}
	\put(97,52){\vector(1,0){10}}
	\put(107,42){\framebox(14,20){$\Psi^{(n)}_2$}}
	\put(91,37){\makebox(0,0){$\vdots$}}
	\put(114,37){\makebox(0,0){$\vdots$}}
	\put(74,20){\vector(1,0){10}}
	\put(91,20){\makebox(0,0){$M^{(n)}_{\I_{|\J|}}$}}
	\put(97,20){\vector(1,0){10}}
	\put(107,10){\framebox(14,20){$\Psi^{(n)}_{|\J|}$}}
	
	\put(121,78){\vector(1,0){10}}
	\put(132,77){\makebox(0,0)[l]{$Z^n_1$}}
	
	\put(121,52){\vector(1,0){10}}
	\put(132,51){\makebox(0,0)[l]{$Z^n_2$}}
	
	\put(136,37){\makebox(0,0){$\vdots$}}
	
	\put(120,20){\vector(1,0){10}}
	\put(131,19){\makebox(0,0)[l]{$Z^n_{|\J|}$}}
 \end{picture}
\end{center}
\caption{
 Multiple Description Coding
}
\label{fig:mdc}
\end{figure}
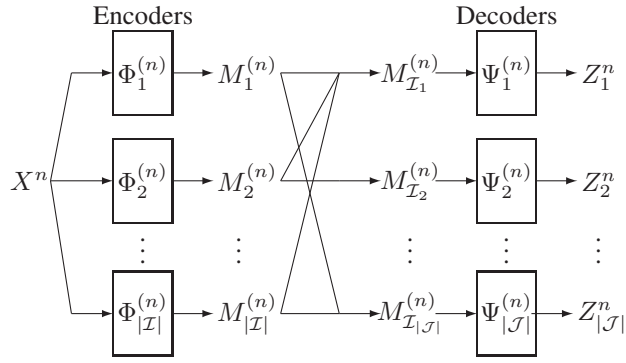

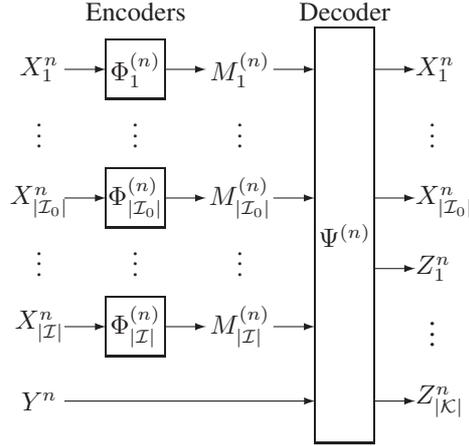
\begin{figure}
\begin{center}
 \unitlength 0.55mm
 \begin{picture}(120,115)(0,0)
	\put(10,95){\makebox(0,0){$X^n_1$}}
	\put(10,81){\makebox(0,0){$\vdots$}}
	\put(10,63){\makebox(0,0){$X^n_{|\I_0|}$}}
	\put(10,50){\makebox(0,0){$\vdots$}}
	\put(10,33){\makebox(0,0){$X_{|\I|}^n$}}
	\put(10,15){\makebox(0,0){$Y^n$}}
	\put(16,95){\vector(1,0){10}}
	\put(16,64){\vector(1,0){10}}
	\put(16,33){\vector(1,0){10}}
	\put(16,15){\vector(1,0){60}}
	\put(33,109){\makebox(0,0){Encoders}}
	\put(26,88){\framebox(14,14){$\Phi^{(n)}_1$}}
	\put(33,81){\makebox(0,0){$\vdots$}}
	\put(26,57){\framebox(14,14){$\Phi^{(n)}_{|\I_0|}$}}
	\put(33,50){\makebox(0,0){$\vdots$}}
	\put(26,26){\framebox(14,14){$\Phi^{(n)}_{|\I|}$}}
	\put(40,95){\vector(1,0){10}}
	\put(40,64){\vector(1,0){10}}
	\put(40,33){\vector(1,0){10}}
	\put(58,95){\makebox(0,0){$M^{(n)}_1$}}
	\put(58,81){\makebox(0,0){$\vdots$}}
	\put(58,64){\makebox(0,0){$M^{(n)}_{|\I_0|}$}}
	\put(58,50){\makebox(0,0){$\vdots$}}
	\put(58,33){\makebox(0,0){$M^{(n)}_{|\I|}$}}
	\put(66,95){\vector(1,0){10}}
	\put(66,64){\vector(1,0){10}}
	\put(66,33){\vector(1,0){10}}
	\put(83,109){\makebox(0,0){Decoder}}
	\put(76,5){\framebox(14,100){$\Psi^{(n)}$}}
	\put(90,95){\vector(1,0){10}}
	\put(90,64){\vector(1,0){10}}
	\put(100,95){\makebox(0,0)[l]{$X^n_1$}}
	\put(104,81){\makebox(0,0){$\vdots$}}
	\put(100,63){\makebox(0,0)[l]{$X^n_{|\I_0|}$}}
	\put(90,47){\vector(1,0){10}}
	\put(104,33){\makebox(0,0){$\vdots$}}
	\put(90,15){\vector(1,0){10}}
	\put(100,47){\makebox(0,0)[l]{$Z^n_1$}}
	\put(104,33){\makebox(0,0){$\vdots$}}
	\put(100,15){\makebox(0,0)[l]{$Z^n_{|\K|}$}}
 \end{picture}
\end{center}
\caption{
 Distributed Source Coding Formulated by Jana and Blahut
}
\label{fig:jana-blahut}
\end{figure}

The distributed lossless source coding
was introduced by Slepian and Wolf \cite{SW73}
and the distributed lossy source coding
by Berger \cite{B78} and Tung \cite{T78}.
Jana and Blahut \cite{JB08}
formulated the general case (Fig. \ref{fig:jana-blahut})
which includes distributed lossless source coding \cite{SW73},
lossy source coding with side information
at the decoder \cite{WZ76},
lossless source coding with coded side information \cite{AK75,KM79,GP79,HK80,W75},
and distributed lossy source coding \cite{B78,T78}.
With regard to distributed lossless source coding,
the single-letter region for two stationary memoryless correlated sources
was derived in \cite{SW73}
and the multi-letter region for two general correlated sources
was derived in \cite{MK95}.
The multi-letter region for three or more stationary ergodic sources
was identified in \cite{C75}.
With regard to lossy source coding with side information
at the decoder,
the single-letter region for two stationary memoryless correlated sources
was derived in \cite{WZ76},
and the multi-letter region for two general correlated sources
in \cite{IM02}.
With regard to lossless source coding with coded side information,
the single-letter region for stationary memoryless correlated sources
was derived in \cite{W75,AK75}
and the multi-letter region for two general correlated sources
was derived in \cite{MK95},
where it was assumed that there is one target source and
one side-information source called a helper.
The single-letter region for stationary memoryless correlated sources
was derived in \cite{GP79}
for an arbitrary number of target sources
and one helper.
Although the case of one target source
and an arbitrary number of helpers was introduced in \cite{KM79},
the single-letter region for stationary memoryless correlated sources
is still unknown.
Distributed lossy source coding
for two stationary memoryless correlated sources
was introduced in \cite{B78,T78},
where inner and outer regions were derived.
The multi-letter region for general correlated sources
was derived in \cite{YQ06a,YQ06b}.
However, the single-letter region
for general stationary memoryless correlated sources
remains unknown.

The multiple description coding (Fig.~\ref{fig:mdc})
was introduced by Gersho and Witsenhausen
(see \cite{EC82}, \cite[pp. 335-336]{EK11}),
where there is no side information at decoders.
For the case of two descriptions,
the best known single-letter inner region
for a stationary memoryless source
was introduced in \cite{VKG03,ZB87},
where the equivalence of the regions
given in \cite{VKG03} and \cite{ZB87}
was shown in \cite{WCZCP11}.
The case of three of more descriptions
was studied in \cite{SP18,VAR16,VKG03}.
Multiple-description coding
includes successive-refinement coding \cite{EC91}
as a special case.
It should be noted that
the single-letter regions for a general stationary memoryless source
and the multi-letter region for a general source remain unknown.

The source coding with side information at decoders (Fig.~\ref{fig:dsi}),
which is an extension of
lossy source coding with side information at the decoder \cite{WZ76},
was introduced by Heegard and Berger \cite{HB85}.
Inner regions for a stationary memoryless source
were derived in \cite{HB85} for two decoders
and in \cite{TCG11} for three or more decoders.
The multi-letter region for general correlated sources
was derived in \cite{MU12}.
However, the single-letter region
for general stationary memoryless sources remain unknown.

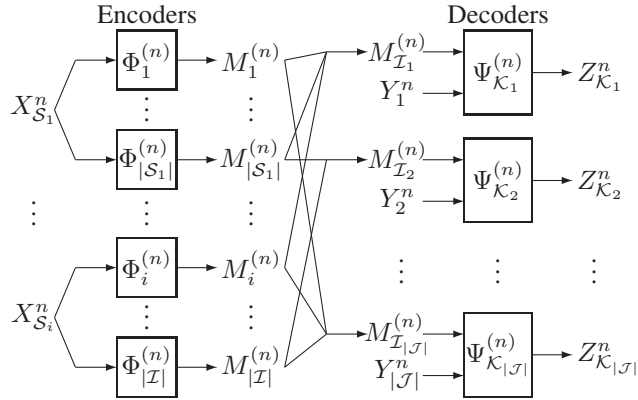
\begin{figure}
\begin{center}
 \unitlength 0.55mm
 \begin{picture}(155,95)(-8,0)
	\put(27,92){\makebox(0,0){Encoders}}
	
	\put(0,69){\makebox(0,0){$X^n_{\cS_1}$}}
	\put(5,69){\line(5,12){5}}
	\put(5,69){\line(5,-12){5}}
	\put(10,81){\vector(1,0){10}}
	\put(10,57){\vector(1,0){10}}
	
	\put(20,74){\framebox(14,14){$\Phi^{(n)}_1$}}
	\put(34,81){\vector(1,0){10}}
	\put(52,81){\makebox(0,0){$M^{(n)}_1$}}
	\put(27,71){\makebox(0,0){$\vdots$}}
	\put(52,71){\makebox(0,0){$\vdots$}}
	\put(20,50){\framebox(14,14){$\Phi^{(n)}_{|\cS_1|}$}}
	\put(34,57){\vector(1,0){10}}
	\put(52,57){\makebox(0,0){$M^{(n)}_{|\cS_1|}$}}
	
	\put(0,46){\makebox(0,0){$\vdots$}}
	\put(27,46){\makebox(0,0){$\vdots$}}
	\put(52,46){\makebox(0,0){$\vdots$}}
	
	\put(0,19){\makebox(0,0){$X^n_{\cS_i}$}}
	\put(5,19){\line(5,12){5}}
	\put(5,19){\line(5,-12){5}}
	\put(10,31){\vector(1,0){10}}
	\put(10,7){\vector(1,0){10}}
	
	\put(20,24){\framebox(14,14){$\Phi^{(n)}_i$}}
	\put(34,31){\vector(1,0){10}}
	\put(52,31){\makebox(0,0){$M^{(n)}_i$}}
	\put(27,21){\makebox(0,0){$\vdots$}}
	\put(52,21){\makebox(0,0){$\vdots$}}
	\put(20,0){\framebox(14,14){$\Phi^{(n)}_{|\I|}$}}
	\put(34,7){\vector(1,0){10}}
	\put(52,7){\makebox(0,0){$M^{(n)}_{|\I|}$}}
	
	\put(60,81){\line(10,2){10}}
	\put(60,81){\line(10,-66){10}}
	\put(60,57){\line(10,26){10}}
	\put(60,57){\line(10,-0){10}}
	
	\put(60,31){\line(10,52){10}}
	\put(60,31){\line(10,-16){10}}
	\put(60,7){\line(10,50){10}}
	\put(60,7){\line(10,8){10}}
	
	\put(111,92){\makebox(0,0){Decoders}}
	\put(70,83){\vector(1,0){10}}
	\put(87,83){\makebox(0,0){$M^{(n)}_{\I_1}$}}
	\put(93,83){\vector(1,0){10}}
	\put(87,73){\makebox(0,0){$Y^n_1$}}
	\put(93,73){\vector(1,0){10}}
	\put(103,68){\framebox(16,20){$\Psi^{(n)}_{\K_1}$}}
	\put(70,57){\vector(1,0){10}}
	\put(87,57){\makebox(0,0){$M^{(n)}_{\I_2}$}}
	\put(93,57){\vector(1,0){10}}
	\put(87,47){\makebox(0,0){$Y^n_2$}}
	\put(93,47){\vector(1,0){10}}
	\put(103,42){\framebox(16,20){$\Psi^{(n)}_{\K_2}$}}
	
	\put(119,78){\vector(1,0){10}}
	\put(130,77){\makebox(0,0)[l]{$Z^n_{\K_1}$}}
	
	\put(119,52){\vector(1,0){10}}
	\put(130,51){\makebox(0,0)[l]{$Z^n_{\K_2}$}}
	
	\put(88,32){\makebox(0,0){$\vdots$}}
	\put(111,32){\makebox(0,0){$\vdots$}}
	\put(134,32){\makebox(0,0){$\vdots$}}
	
	\put(70,15){\vector(1,0){10}}
	\put(87,15){\makebox(0,0){$M^{(n)}_{\I_{|\J|}}$}}
	\put(93,15){\vector(1,0){10}}
	\put(87,5){\makebox(0,0){$Y^n_{|\J|}$}}
	\put(93,5){\vector(1,0){10}}
	\put(103,0){\framebox(16,20){$\Psi^{(n)}_{\K_{|\J|}}$}}
	\put(119,10){\vector(1,0){10}}
	\put(130,9){\makebox(0,0)[l]{$Z^n_{\K_{|\J|}}$}}
 \end{picture}
\end{center}
\caption{
 Unified Extension of Distributed Source Coding,
 Multiple Description Coding,
 and Source Coding with Side Information at Decoders
}
\label{fig:dsc-mdc}
\end{figure}

In this paper, we consider the unified extension of
distributed source coding, multiple description coding,
and source coding with side information at decoders
as illustrated in Fig.~\ref{fig:dsc-mdc}.
The contributions of this paper are listed below:
\begin{itemize}
 \item
 The multi-letter rate-distortion region
 of the multiple-decoder extension of the distributed source coding
 with decoder side information
 for arbitrary general correlated sources
 is characterized in terms of entropy functions.
 It should be noted that the codewords are generated independently.
 \item
 The multi-letter rate-distortion region
 of the multiple-source extension of the multiple description coding
 with decoder side information
 for arbitrary general correlated sources
 is characterized in terms of entropy functions.
 It should be noted that
 cooperation is allowed among encoders
 that have access to the same source.
 \item
 It is shown that the two rate-distortion regions are equivalent.
 From this fact, we have two characterizations of
 the unified rate-distortion region.
 \item
 It is shown that the multi-letter rate-distortion region
 is achievable with a code based on constrained-random number
 generators \cite{CRNG,CRNGVLOSSY}.
 When random variables are assumed to be stationary memoryless,
 the best known single-letter inner regions are achievable
 by using this type of code.
\end{itemize}

It should be noted that
the multi-letter region is not computable because it contains limits.
The computable single-letter region
for stationary memoryless correlated sources
is still unknown.
The main argument of this paper
is the optimality of the code
based on constrained-random number generators
and the derivation of the multi-letter region
based on sup/inf entropy rates.

This paper is organized as follows.
Basic definitions and notations are introduced
in Section \ref{sec:definition}.
The rate-distortion regions are defined in Section \ref{sec:dsc-mdc}.
Section \ref{sec:revisited} makes
a comparison with previous studies on distributed source coding,
multiple-description coding,
and source coding with side information at decoders.
Section \ref{sec:jb} discusses
the formulation by Jana and Blahut
and makes a comparison with previous results.
Proofs of some relations between regions
are given in Section \ref{sec:converse}
and \ref{sec:RCRNGDSCsubsetRCRNGMDC}.
Code construction is introduced in Section \ref{sec:construction}
and the proof of its achievability is given in Section \ref{sec:proof-crng}.

\section{Definitions and Notations}
\label{sec:definition}

Sets are written in calligraphic style (e.g. $\U$)
and a member of a set is written in corresponding roman style (e.g. $u$).
If $\U$ is a set and $\V_u$ is also a set for each $u\in\U$,
we use the notation $\V_{\U}\equiv\Prod_{u\in\U}\V_u$.
We use the notation $v_{\U}\equiv\{v_u\}_{u\in\U}$
to represent the set of elements
(e.g. sequences, random variables, functions) $v_u$ with index $u\in\U$.
We use the notation $|\U|$ to represent the cardinality of $\U$.
We use the notation $2^{\U}\setminus\{\emptyset\}$
to represent the family of all non-empty subsets of $\U$.

A random variable and its realization
are denoted in roman (e.g. $U$ and $u$),
where the range of the random variable is written
in corresponding calligraphic style (e.g. $\U$).
For a given $n\in\NN\equiv\{1,2,\ldots\}$,
which denotes block length,
an $n$-dimensional vector random variable
is denoted by superscript $n$ (e.g. $U^n$),
where the range of the random variable is written in corresponding
calligraphic style (e.g. $\U^n$).
A $n$-dimensional vector (the realization of a vector random variable)
is denoted by superscript $n$ or in boldface (e.g. $u^n$ or $\uu$).

Unless otherwise stated,
we assume that the sequence of correlated random variables
$(\WW_{\I},\XX_{\I},\YY_{\J},\ZZ_{\K})\equiv
\{(W^n_{\I},X^n_{\I},Y_{\J}^n,$ $Z^n_{\K})\}_{n=1}^{\infty}$
is a joint general source,
for which we do not assume conditions
such as consistency, stationarity, or ergodicity.
We assume that the alphabets $\X^n_i$ and $\Y^n_j$ of $X_i^n$ and $Y^n_j$
are the Cartesian products of set $\X_i$ and $\Y_j$, respectively.
On the other hand,
the alphabets $\W^n_i$ and $\Z^n_k$ of $W^n_i$ and $Z^n_k$
are not restricted to the Cartesian product
of sets $\W_i$ and $\Z_k$, respectively.
We use the notations $\W^n_i$ and $\Z^n_k$ to make it
easier to understand the correspondence
with the stationary memoryless case.
We also assume that $\W^n_i$ is a finite set
but $\X_i^n$, $\Y^n_j$, and $\Z^n_k$
are allowed to be infinite sets under appropriate conditions,
where the summations are replaced with integrals.
We use the information-spectrum methods \cite{HAN,HV93}
summarized in Appendix \ref{sec:ispec}.
In Section \ref{sec:revisited},
we consider the case where $(\WW_{\I},\XX_{\I},\YY_{\J},\ZZ_{\K})$
is stationary memoryless.
For given random variables $U$, $U'$, and $V$,
$H(U)$ denotes the entropy,
$H(U|V)$ denotes the conditional entropy,
$I(U;U')$ denotes the mutual information,
and $I(U;U'|V)$ denotes the conditional mutual information.

Finally, let $\chi$ be the support function defined as
\begin{equation*}
 \chi(\mathrm{statement})
 \equiv
 \begin{cases}
	1
	&\text{if the $\mathrm{statement}$ is true}
	\\
	0
	&\text{if the $\mathrm{statement}$ is false}.
 \end{cases}
\end{equation*}

\section{Unified Extension of Distributed Source Coding, Multiple Description Coding, and Source Coding with Side Information at Decoders}
\label{sec:dsc-mdc}

In this section,
we define the unification (Fig.~\ref{fig:dsc-mdc}) of
distributed source coding, multiple description coding,
and source coding with side information at decoders.

Let $\I$ be the index set of encoders/codewords.
Let $\XX_{\I}\equiv\{\XX_i\}_{i\in\I}$
be a set of correlated general sources,
where $\XX_i\equiv\{X^n_i\}_{n=1}^{\infty}$
is the source observed by the $i$-th encoder.
We assume that the $i$-th encoder observes source $X^n_i$
and transmits the codeword $M^{(n)}_i\in\M^{(n)}_i$.

Let $\J$ be the index set of decoders.
For each $j\in\J$,
let $\I_j$ be a subset of $\I$ representing the index set of codewords
transmitted to the $j$-th decoder.
For each $j\in\J$,
let $\YY_j\equiv\{Y^n_j\}_{n=1}^{\infty}$,
be (non-causal) side information available only at the $j$-th decoder.
Let $\ZZ_{\K}\equiv\{\ZZ_k\}_{k\in\K}$ be a set of reproductions,
where $\ZZ_k\equiv\{Z^n_k\}_{n=1}^{\infty}$.
For each $j\in\J$,
let $\K_j$ be the index set of reproductions of the $j$-th decoder,
where we assume that $\{\K_j\}_{j\in\J}$ forms a partition of $\K$,
that is, $\K=\bigcup_{j\in\J}\K_j$
and $\K_j\cap\K_{j'}=\emptyset$ if $j\neq j'$.
We assume that the $j$-th decoder reproduces $Z^n_{\K_j}$
after observing the set of codewords
$M^{(n)}_{\I_j}\equiv\{M^{(n)}_i\}_{i\in\I_j}$
and the uncoded side information $Y^n_j$.
It should be noted that
side information $Y^n_j$ can be included in the target sources
by assuming that $Y^n_j$ is encoded and decoded with infinite rate.

For each $k\in\K$ and $n\in\NN$,
let $d^{(n)}_k:\X^n_{\I}\times\Y^n_{\J}\times\Z^n_k\to[0,\infty)$
be a distortion measure.
For given triplet of general sources $(\XX_{\I},\YY_{\J},\ZZ_{\K})$,
let $\od_k(\XX_{\I},\YY_{\J},\ZZ_k)$ be defined as
\begin{equation*}
 \od_k(\XX_{\I},\YY_{\J},\ZZ_k)
 \equiv
 \plimsupn d^{(n)}_k(X^n_{\I},Y^n_{\J},Z^n_k)
\end{equation*}
for each $k\in\K$.
Let $R_{\I}\equiv\{R_i\}_{i\in\I}$ and $D_{\K}\equiv\{D_k\}_{k\in\K}$
be sets of positive numbers.

\subsection{Operational Definition}
Here, we introduce the operational definition of the rate-distortion region.
\begin{df}
\label{df:ROP-edsc}
A rate-distortion pair $(R_{\I},D_{\K})$ is {\em achievable}
for a given set of distortion measures
$\{d^{(n)}_k\}_{k\in\K,n\in\NN}$
iff there is a sequence of codes
$\{(\{\vphi^{(n)}_i\}_{i\in\I},\{\psi^{(n)}_k\}_{k\in\K})\}_{n=1}^{\infty}$
consisting of
encoding functions $\vphi^{(n)}_i:\X_i^n\to\M_i^{(n)}$
and reproducing functions
$\psi^{(n)}_k:\M^{(n)}_{\I_j}\times\Y^n_j\to\Z^{(n)}_k$
that satisfy
\begin{gather}
 \limsupn \frac {\log|\M^{(n)}_i|}n \leq R_i
 \ \text{for all}\ i\in\I
 \label{eq:rate}
 \\
 \limn
 \Prob\lrsb{
	d^{(n)}_k(X_{\I}^n,Y^n_{\J},Z_k^n)
	> D_k+\delta
 }=0
 \ \text{for all $k\in\K$ and $\delta>0$},
 \label{eq:lossy}
\end{gather}
where $\M^{(n)}_i$ is a finite set for all $i\in\I$
and $Z^n_k\equiv\psi_k^{(n)}(\{\vphi^{(n)}_i(X_i^n)\}_{i\in\I_j},Y^n_j)$
is the $k$-th reproduction for each $j\in\J$ and $k\in\K_j$.
It should be noted that the $j$-th decoder
has a set of decoding functions $\{\psi_k^{(n)}\}_{k\in\K_j}$.
The {\em rate-distortion region} $\ROP$
under the maximum-distortion criterion is defined as
the closure of the set of all achievable rate-distortion pairs.
\end{df}

\subsection{Information-Theoretical Definition
 Based on Distributed Source Coding}
Here, we introduce the information-theoretical definition
of the rate-distortion region
based on the multiple-decoder extension of distributed source coding.
Let $(\WW_{\I},\ZZ_{\K})$ be a set of general sources,
where $\WW_i\equiv \{W^n_i\}_{n=1}^{\infty}$ for each $i\in\I$
and $\ZZ_k\equiv \{Z^n_k\}_{n=1}^{\infty}$ for each $k\in\K$.

\begin{df}
\label{df:RIT-DSC}
Let $\RITDSC(\WW_{\I},\ZZ_{\K})$
be defined as the set of all $(R_{\I},D_{\K})$ satisfying
\begin{align}
 \sum_{i\in\I'_j}R_i
 &\geq
 \oH(\WW_{\I'_j}|\WW_{\Ipcj},\YY_j)
 -
 \sum_{i\in\I'_j}
 \uH(\WW_i|\XX_i)
 \label{eq:RIT-DSC-RI'}
 \\
 D_k
 &\geq
 \od_k(\XX_{\I},\YY_{\J},\ZZ_k)
 \label{eq:RIT-DSC-D}
\end{align}
for all $j\in\J$, $\I'_j\in2^{\I_j}\setminus\{\emptyset\}$, and $k\in\K$.
Region $\RITDSC$ is defined by the union of $\RITDSC(\WW_{\I},\ZZ_{\K})$
over all general sources $(\WW_{\I},\ZZ_{\K})$
that satisfy the following conditions:
\begin{gather}
 (W^n_{\I\setminus\{i\}},X^n_{\I\setminus\{i\}},Y^n_{\J})
 \markov
 X^n_i
 \markov
 W^n_i
 \label{eq:RIT-DSC-markov-encoder}
 \\
 (W^n_{\I\setminus\I_j},X^n_{\I},Y^n_{\J\setminus\{j\}},Z^n_{\K\setminus\K_j})
 \markov
 (W^n_{\I_j},Y^n_j)
 \markov
 Z^n_{\K_j}
 \label{eq:RIT-DSC-markov-decoder}
\end{gather}
for all $i\in\I$, $j\in\J$, and $n\in\NN$.
Optionally, $Z^n_{\K_j}$ is allowed to be restricted to
being a deterministic function of $(W^n_{\I_j},Y^n_j)$.
\end{df}

It should be noted that
condition (\ref{eq:RIT-DSC-markov-encoder}) represents
that all codewords are generated independently
even when some encoders have access to the identical/synchronized source.

Here, we define the achievable rate-distortion region
by using the code with constrained-random number generators.
\begin{df}
\label{df:RCRNG-DSC}
Let $\RCRNGDSC(\WW_{\I},\ZZ_{\K})$
be defined as the set of all $(R_{\I},D_{\K})$ where 
there are real-valued variables
$\{r_i\}_{i\in\I}$ that satisfy
\begin{align}
 0
 \leq
 r_i
 &\leq
 \uH(\WW_i|\XX_i)
 \label{eq:RCRNG-DSC-ri}
 \\
 \sum_{i\in\I'_j}[r_i+R_i]
 &\geq
 \oH(\WW_{\I'_j}|\WW_{\Ipcj},\YY_j)
 \label{eq:RCRNG-DSC-sum[ri+Ri]}
\end{align}
for all $i\in\I$, $j\in\J$, and $\I'_j\in2^{\I_j}\setminus\{\emptyset\}$,
and (\ref{eq:RIT-DSC-D}) for all $k\in\K$.
Then region $\RCRNGDSC$ is defined by the union of
$\RCRNGDSC(\WW_{\I},\ZZ_{\K})$
over all general sources $(\WW_{\I},\ZZ_{\K})$
satisfying (\ref{eq:RIT-DSC-markov-encoder})
and (\ref{eq:RIT-DSC-markov-decoder})
for all $i\in\I$, $j\in\J$, and $n\in\NN$.
Optionally, $Z^n_{\K_j}$ is allowed to
be restricted to being a deterministic function of $(W^n_{\I_j},Y^n_j)$.
\end{df}

\begin{rem}
We introduce auxiliary variables $\{r_i\}_{i\in\I}$
in the definition of $\RCRNGDSC(\WW_{\I},\ZZ_{\K})$
because relations
(\ref{eq:RCRNG-DSC-ri}) and (\ref{eq:RCRNG-DSC-sum[ri+Ri]})
are related directly to the proof of
both the converse and achievability.
It should be noted that both the righthand side of (\ref{eq:RCRNG-DSC-ri})
and (\ref{eq:RCRNG-DSC-sum[ri+Ri]})
are monomial entropy functions,
that is, they have explicit operational interpretations
as explained in Remark \ref{rem:interpretation}
in Section \ref{sec:construction}.
It should be noted that we can obtain $\{r_i\}_{i\in\I}$
that satisfy (\ref{eq:RCRNG-DSC-ri}) and (\ref{eq:RCRNG-DSC-sum[ri+Ri]})
by using the linear programming
when $\{R_i\}_{i\in\I}$ and the right hand sides of 
(\ref{eq:RCRNG-DSC-ri}) and (\ref{eq:RCRNG-DSC-sum[ri+Ri]})
are given as concrete real numbers.
\end{rem}

\subsection{Information-Theoretical Definition
 Based on Multiple Description Coding}
Here, we introduce the information-theoretical definition
of the rate-distortion region based on 
the multiple source extension of multiple description coding.
Multiple description coding assumes that some encoders share an identical/synchronized source.

Let us assume that $\fS$ forms a partition of $\I$,
that is, $\bigcup_{\cS\in\fS}\cS=\I$ and $\cS\cap\cS'=\emptyset$
if $\cS\neq\cS'$.
Let us also assume that
the $i$-th encoder has access to source $\XX_i$,
which can also be identified by $\cS\in\fS$ satisfying $i\in\cS$.
We denote $\XX_{\cS}$ to satisfy $\XX_{\cS}=\XX_i$ for all $i\in\cS\in\fS$.

Here, we define the achievable rate-distortion region
by using the code with constrained-random number generators.
Let $(\WW_{\I},\ZZ_{\K})$ be a set of general sources,
where $\WW_i\equiv \{W^n_i\}_{n=1}^{\infty}$ for each $i\in\I$
and $\ZZ_k\equiv \{Z'^n_k\}_{n=1}^{\infty}$ for each $k\in\K$.
\begin{df}
\label{df:RCRNG-MDC}
Let $\RCRNGMDC(\WW_{\I},\ZZ_{\K})$
be defined as the set of all $(R_{\I},D_{\K})$ where 
there are real-valued variables $\{r_i\}_{i\in\I}$ satisfying
\begin{align}
 0
 \leq
 \sum_{i\in\cS'}r_i
 &\leq
 \uH(\WW_{\cS'}|\XX_{\cS})
 \label{eq:RCRNG-MDC-sum[ri]}
 \\
 \sum_{i\in\I'_j}[r_i+R_i]
 &\geq
 \oH(\WW_{\I'_j}|\WW_{\Ipcj},\YY_j)
 \label{eq:RCRNG-MDC-sum[ri+Ri]}
\end{align}
for all $\cS\in\fS$,
$\cS'\in2^{\cS}\setminus\{\emptyset\}$,
$j\in\J$, and $\I'_j\in2^{\I_j}\setminus\{\emptyset\}$,
and (\ref{eq:RIT-DSC-D}) for all $k\in\K$.
Then region $\RCRNGMDC$ is defined by the union of
$\RCRNGMDC(\WW_{\I},\ZZ_{\K})$
over all general sources $(\WW_{\I},\ZZ_{\K})$
that satisfy the following conditions:
\begin{gather}
 (W^n_{\I\setminus\cS},X^n_{\I\setminus\cS},Y^n_{\J})
 \markov
 X^n_{\cS}
 \markov
 W^n_{\cS}
 \label{eq:RCRNG-MDC-markov-encoder}
 \\
 (W^n_{\I\setminus\I_j},X^n_{\I},Y^n_{\J\setminus\{j\}},Z^n_{\K\setminus\K_j})
 \markov
 (W^n_{\I_j},Y^n_j)
 \markov
 Z^n_{\K_j}
 \label{eq:RCRNG-MDC-markov-decoder}
\end{gather}
for all $\cS\in\fS$, $i\in\I$, $j\in\J$, and $n\in\NN$.
Optionally, $Z^n_{\K_j}$ is allowed to
be restricted to being the deterministic function of $(W^n_{\I_j},Y^n_j)$.
\end{df}

It should be noted that
the condition (\ref{eq:RCRNG-MDC-markov-encoder}) represents that
encoders $\{\vphi^{(n)}_i\}_{i\in\cS}$
observing the same source $\XX_{\cS}$ can cooperate with each other
in the sense that they can generate correlated sources
$W^n_{\cS}\equiv\{W^n_i\}_{i\in\cS}$.

\begin{rem}
We can obtain $\{r_i\}_{i\in\I}$
that satisfy (\ref{eq:RCRNG-MDC-sum[ri]})
and (\ref{eq:RCRNG-MDC-sum[ri+Ri]})
by using the linear programming
when $\{R_i\}_{i\in\I}$ and the right hand sides of 
(\ref{eq:RCRNG-MDC-sum[ri]})
and (\ref{eq:RCRNG-MDC-sum[ri+Ri]})
are given as concrete real numbers.
\end{rem}

\subsection{Main Theorem}

\begin{thm}
\label{thm:ROP=RIT=RCRNG=RCRNG'}
For a set of general correlated sources $(\XX_{\I},\YY_{\J})$,
we have
\begin{equation*}
 \ROP=\RITDSC=\RCRNGDSC=\RCRNGMDC.
\end{equation*}
\end{thm}
\begin{IEEEproof}
The theorem comes from the following facts:
\begin{itemize}
 \item 
 the relation $\RITDSC=\RCRNGDSC$,
 which is shown
 by using the Fourier-Motzkin method \cite[Appendix E]{EK11}
 to obtain the fact that
 (\ref{eq:RIT-DSC-RI'}) is equivalent to the existence
 of $\{r_i\}_{i\in\I}$ satisfying
 (\ref{eq:RCRNG-DSC-ri})	and (\ref{eq:RCRNG-DSC-sum[ri+Ri]});
 \item the converse $\ROP\subset\RCRNGDSC$,
 which is shown in Section \ref{sec:converse-dsc};
 \item the relation $\RCRNGDSC\subset\RCRNGMDC$,
 which is shown in Section \ref{sec:RCRNGDSCsubsetRCRNGMDC};
 \item the achievability $\RCRNGMDC\subset\ROP$,
 which is confirmed in
 Sections \ref{sec:construction} and \ref{sec:proof-crng}
 by constructing a code.
\end{itemize}
\end{IEEEproof}

\section{Distributed Source Coding,
 Multiple Description Coding,
 and Source Coding with Side Information at Decoders
 Revisited}
\label{sec:revisited}

In this section, we revisit distributed source coding,
multiple-description coding,
and source coding with side information at decoders
for correlated stationary memoryless sources.
For each $k\in\K$,
let $d_k:\X_{\I}\times\Y_{\J}\times\Z_k\to[0,\infty)$
be the bounded single letter distortion measure
and
\begin{equation*}
 d_k^{(n)}(\xx_{\I},\yy_{\J},\zz_k)\equiv
 \frac1n\sum_{l=1}^nd_k(x_{\I,l},y_{\J,l},z_{k,l})
\end{equation*}
for each $n\in\NN$,
$\xx_{\I}\equiv(x_{\I,1},\ldots,x_{\I,n})$,
$\yy_{\J}\equiv(y_{\J,1},\ldots,y_{\J,n})$,
and $\zz_k\equiv(z_{k,1},\ldots,z_{k,n})$.
Given the above we can replace
(\ref{eq:RIT-DSC-D}) by
\begin{equation*}
 D_k\geq E_{X_{\I}Y_{\J},Z_k}[d_k(X_{\I},Y_{\J},Z_k)]
\end{equation*}
from the law of large numbers.
In the following, we focus on the conditions for the rate vector
$R_{\I}\equiv\{R_i\}_{i\in\I}$
and the Markov conditions.
For a given $i\in\{1,2\}$, define $\ic,\jc\in\{1,2\}$
to satisfy $(i,\ic),(j,\jc)\in\{(1,2),(2,1)\}$.

\subsection{Distributed Source Coding}
\label{sec:dsc-example}

In this subsection, we revisit distributed source coding
introduced by Berger \cite{B78} and Tung \cite{T78}.
In distributed source coding,
it is assumed that sources are encoded independently,
the decoder receives all codewords,
and there is no side information at the decoder,
that is,
$\fS=\{\{i\}:i\in\I\}$, $|\J|=1$, $\K=\I$, and $\YY_{\J}$ is a constant.
In the following we assume that the distortion function $d_i$
does not depend on $\XX_{\I\setminus\{i\}}$.

The following examples are particular cases
of distributed source coding.
Let us assume that
$(\WW_{\I},\XX_{\I},\ZZ_{\K})$ is stationary memoryless
with generic random variable $(W_{\I},X_{\I},Z_{\K})$.

\begin{example}
\label{example:dsc2}
Here, let us consider the case of two sources,
that is, $\I=\K=\{1,2\}$.
From (\ref{eq:RIT-DSC-markov-encoder})--(\ref{eq:RCRNG-DSC-sum[ri+Ri]}),
we have relations
\begin{align}
 0
 \leq
 r_i
 &\leq
 H(W_i|X_i)
 \label{eq:RCRNG-DSC2-ri}
 \\
 r_i+R_i
 &\geq
 H(W_i|W_{\ic})
 \label{eq:RCRNG-DSC2-[ri+Ri]}
 \\
 r_1+R_1+r_2+R_2
 &\geq
 H(W_1,W_2)
 \label{eq:RCRNG-DSC2-sum[ri+Ri]}
\end{align}
and
\begin{gather*}
 (W_{\ic},X_{\ic})\markov X_i\markov W_i
 \\
 (X_1,X_2)\markov (W_1,W_2)\markov Z_i
\end{gather*}
for all $i\in\{1,2\}$.
By using the Fourier-Motzkin method \cite[Appendix E]{EK11} to eliminate
$\{r_1,r_2\}$ and redundant inequalities,
we have the equivalent conditions of
(\ref{eq:RCRNG-DSC2-ri})--(\ref{eq:RCRNG-DSC2-sum[ri+Ri]})
as
\begin{align*}
 R_i
 &\geq
 H(W_i|W_{\ic})
 -
 H(W_i|X_i)
 \\
 R_1+R_2
 &\geq
 H(W_1,W_2)
 -H(W_1|X_1)
 -H(W_2|X_2).
\end{align*}
By introducing a time-sharing random variable $T$,
which is shared by encoders and a decoder,
we have the inner region $\RCRNGDSCp$ derived as the
union of the region
\begin{equation*}
 \RCRNGDSCp(W_{\I},Z_{\K},T)
 \equiv
 \lrb{
	(R_1,R_2,D_1,D_2):
	\begin{aligned}
	 R_i
	 &\geq
	 H(W_i|W_{\ic},T) - H(W_i|X,T)
	 \\
	 R_1+R_2
	 &\geq
	 H(W_1,W_2,T) - H(W_1|X,T) - H(W_2|X,T)
	 \\
	 D_i
	 &\geq
	 E_{X_iZ_i}[d_i(X_i,Z_i)]
	 \\
	 &\text{for all}\ i\in\{1,2\}
	\end{aligned}
 }
\end{equation*}
over all $(W_{\I},Z_{\K},T)$ satisfying that
\begin{gather}
 \text{$T$ is independent of $(X_1,X_2)$}
 \label{eq:crng-dsc2-T}
 \\
 (W_{\ic},X_{\ic})\markov (X_i,T) \markov W_i
 \label{eq:crng-dsc2-markov-Wi}
 \\
 (X_1,X_2)\markov (W_1,W_2,T)\markov Z_i.
 \label{eq:crng-dsc2-markov-Zi}
\end{gather}
Here, let us define the Berger-Tung single-letter inner region 
$\RBTDSC$ \cite{B78,T78} as the union of the region
\begin{equation*}
 \RBTDSC(W_{\I},Z_{\K},T)
 \equiv
 \lrb{
	(R_1,R_2,D_1,D_2):
	\begin{aligned}
	 R_i
	 &\geq
	 I(X_i;W_i|W_{\ic},T)
	 \\
	 R_1+R_2
	 &\geq
	 I(X_1,X_2;W_1,W_2|T)
	 \\
	 D_i
	 &\geq
	 E_{X_iZ_i}[d_i(X_i,Z_i)]
	 \\
	 &\text{for all}\ i\in\{1,2\}
	\end{aligned}
 }
\end{equation*}
over all $(W_{\I},Z_{\K},T)$ satisfying
(\ref{eq:crng-dsc2-T})--(\ref{eq:crng-dsc2-markov-Zi}).
Then the region $\RCRNGDSCp$
is equal to the Berger-Tung single-letter inner region
by letting $Z_i\equiv W_i$ for each $i\in\I$
because
\begin{align}
 H(W_i|W_{\ic},T)
 -
 H(W_i|X_i,T)
 &=
 H(W_i|W_{\ic},T)
 -
 H(W_i|W_{\ic},X_i,T)
 \notag
 \\
 &=
 I(X_i;W_i|W_{\ic},T)
 \\
 H(W_1,W_2|T)
 -H(W_1|X_1,T)
 -H(W_2|X_2,T)
 &=
 H(W_1,W_2|T)
 -H(W_1|X_1,X_2,T)
 -H(W_2|W_1,X_1,X_2,T)
 \notag
 \\
 &=
 H(W_1,W_2|T)
 -H(W_1,W_2|X_1,X_2,T)
 \notag
 \\
 &=
 I(X_1,X_2;W_1,W_2|T),
\end{align}
where the first equalities comes from the fact that
(\ref{eq:crng-dsc2-markov-Wi}) implies
\begin{align*}
 H(W_i|X_i,T)
 &=
 H(W_i|W_{\ic},X_i,T)
 \\
 H(W_1|X_1,T)
 &=
 H(W_1|X_1,X_2,T)
 \\
 H(W_2|X_2,T)
 &=
 H(W_2|W_1,X_1,X_2,T).
\end{align*}
It should be noted here that
the Berger-Tung single-letter inner region 
is sub-optimal for particular cases \cite{SP21,WKA11}.
We could conclude that
the sub-optimality is caused by
restricting $(\WW_{\I},\XX_{\I},\ZZ_{\K})$
to being stationary memoryless,
where it has been reported that
the Berger-Tung single-letter inner region
can be improved by considering multi-letter extensions \cite{SP21}.
\end{example}

\begin{example}
Here, let us consider the case of three sources and two reproductions,
that is, $\I=\{0,1,2\}$ and $\K=\{1,2\}$.
From (\ref{eq:RIT-DSC-markov-encoder})--(\ref{eq:RCRNG-DSC-sum[ri+Ri]}),
we have the relations
\begin{align}
 0
 \leq
 r_0
 &\leq
 H(W_0|X_0)
 \label{eq:RCRNG-DSC3-r0}
 \\
 0
 \leq
 r_i
 &\leq
 H(W_i|X_i)
 \label{eq:RCRNG-DSC3-ri}
 \\
 r_0+R_0
 &\geq
 H(W_0|W_1,W_2)
 \label{eq:RCRNG-DSC3-[r0+R0]}
 \\
 r_i+R_i
 &\geq
 H(W_i|W_0,W_{\ic})
 \label{eq:RCRNG-DSC3-[ri+Ri]}
 \\
 r_0+R_0+r_i+R_i
 &\geq
 H(W_0,W_i|W_{\ic})
 \label{eq:RCRNG-DSC3-[r0+R0+Ri+Ri]}
 \\
 r_1+R_1+r_2+R_2
 &\geq
 H(W_1,W_2|W_0)
 \label{eq:RCRNG-DSC3-[r1+R1+R2+R2]}
 \\
 r_0+R_0+r_1+R_1+r_2+R_2
 &\geq
 H(W_0,W_1,W_2)
 \label{eq:RCRNG-DSC3-sum[ri+Ri]}
\end{align}
and
\begin{gather}
 (X_1,X_2,W_1,W_2)\markov X_0\markov W_0
 \label{eq:RCRNG-DSC3-markov-W0}
 \\
 (X_0,X_{\ic},W_0,W_{\ic})\markov X_i\markov W_i
 \label{eq:RCRNG-DSC3-markov-Wi}
 \\
 (X_0,X_1,X_2)\markov (W_0,W_1,W_2)\markov (Z_0,Z_1,Z_2)
 \label{eq:RCRNG-DSC3-markov-decoder}
\end{gather}
for all $i\in\{1,2\}$.
By using the Fourier-Motzkin method \cite[Appendix E]{EK11}
to eliminate $\{r_0,r_1,r_2\}$ and redundant inequalities,
we have the equivalent conditions of
(\ref{eq:RCRNG-DSC3-r0})-(\ref{eq:RCRNG-DSC3-sum[ri+Ri]})
as
\begin{align}
 R_0
 &\geq
 H(W_0|W_1,W_2)
 -
 H(W_0|X_0)
 \label{eq:crng-dsc3-R0}
 \\
 R_i
 &\geq
 H(W_i|W_0,W_{\ic})
 -
 H(W_i|X_i)
 \label{eq:crng-dsc3-Ri}
 \\
 R_0+R_i
 &\geq
 H(W_0,W_i|W_{\ic})
 -H(W_0|X_0)
 -H(W_i|X_i)
 \label{eq:crng-dsc3-R1+R2}
 \\
 R_1+R_2
 &\geq
 H(W_1,W_2|W_0)
 -H(W_1|X_1)
 -H(W_2|X_2)
 \\
 R_0+R_1+R_2
 &\geq
 H(W_0,W_1,W_2)
 -H(W_0|X_0)
 -H(W_1|X_1)
 -H(W_2|X_2).
 \label{eq:crng-dsc3-R0+R1+R2}
\end{align}
Here, let us assume that
\begin{equation}
 X_{\ic}\markov X_i\markov X_0
 \ \text{for all}\ i\in\{1,2\}.
 \label{eq:common}
\end{equation}
From Lemma \ref{lem:common} in Appendix \ref{sec:common},
condition (\ref{eq:common}) imply that $X_0$
can be shared by the both encoders
because it can be generated
using only one of $X_1$ and $X_2$.
Furthermore, let us assume that
\begin{align}
 R'_{0i}
 &\geq
 0
 \label{eq:crng-dsc3-R'0i}
 \\
 R_0
 &=
 R'_{01}+R'_{02}
 \\
 R'_i
 &=
 R_i+R'_{0i}
 \label{eq:crng-dsc3-R'i}
\end{align}
for all $i\in\{1,2\}$,
where the $i$-th encoder generates $(W_0,W_i)$ from $(X_0,X_i)$,
obtains codeword of $(W_0,W_i)$,
and sends the codeword of $W_i$ and a part of the codeword of $W_0$.
These relations correspond to  rate-splitting for the case of two sources
introduced in Example \ref{example:dsc2}.
By using the Fourier-Motzkin method \cite[Appendix E]{EK11}
to eliminate $\{R_0,R_1,R_2,R'_{01},R'_{02}\}$ and redundant inequalities,
we have the conditions for $(R'_1,R'_2)$ equivalent
to (\ref{eq:crng-dsc3-R0})--(\ref{eq:crng-dsc3-R0+R1+R2})
and (\ref{eq:crng-dsc3-R'0i})--(\ref{eq:crng-dsc3-R'i})
as
\begin{align*}
 R'_i
 &\geq
 H(W_i|W_0,W_{\ic}) - H(W_i|X)
 \\
 R'_1+R'_2
 &\geq
 H(W_0,W_1,W_2) - H(W_0|X) - H(W_1|X) - H(W_2|X)
\end{align*}
for all $i\in\{1,2\}$.
Then we have the inner region $\RCRNGDSCp$
derived as the union of the region
\begin{equation*}
 \RCRNGDSCp(X_0,W_{\I},Z_{\K})
 \equiv
 \lrb{
	(R_1,R_2,D_1,D_2):
	\begin{aligned}
	 R_i
	 &\geq
	 H(W_i|W_0,W_{\ic}) - H(W_i|X)
	 \\
	 R_1+R_2
	 &\geq
	 H(W_0,W_1,W_2) - H(W_0|X) - H(W_1|X) - H(W_2|X)
	 \\
	 D_i
	 &\geq
	 E_{X_iZ_i}[d_i(X_i,Z_i)]
	 \\
	 &\text{for all}\ i\in\{1,2\}
	\end{aligned}
 }
\end{equation*}
over all $(X_0,W_{\I},Z_{\K})$ satisfying
(\ref{eq:RCRNG-DSC3-markov-W0})--(\ref{eq:RCRNG-DSC3-markov-decoder})
and (\ref{eq:common}).
Here, let us define 
the Wagner-Kelly-Altu\u{g} single-letter inner region
$\RWKADSC$ \cite{WKA11}
as the union of the region
\begin{equation*}
 \RWKADSC(X_0,W_{\I},Z_{\K})
 \equiv
 \lrb{
	(R_1,R_2,D_1,D_2):
	\begin{aligned}
	 R_i
	 &\geq
	 I(X_i;W_i|W_0,W_{\ic})
	 \\
	 R_1+R_2
	 &\geq
	 I(X_1,X_2;W_0,W_1,W_2)
	 \\
	 D_i
	 &\geq
	 E_{X_iZ_i}[d_i(X_i,Z_i)]
	 \\
	 &\text{for all}\ i\in\{1,2\}
	\end{aligned}
 }
\end{equation*}
over all $(X_0,W_{\I},Z_{\K})$ satisfying
(\ref{eq:RCRNG-DSC3-markov-W0})--(\ref{eq:RCRNG-DSC3-markov-decoder})
and $X_0$ is a common component of $X_1$ and $X_2$
for which 
there is a pair of functions $(\xi_1,\xi_2)$ such that
\begin{equation}
 X_0=\xi_i(X_i)\ \text{for all}\ i\in\{1,2\}.
 \label{eq:crng-dsc3-X0}
\end{equation}
We have the fact that
the common component $X_0$
satisfies condition (\ref{eq:common}) and
\begin{align}
 H(W_i|W_0,W_{\ic}) - H(W_i|X)
 &= 
 H(W_i|W_0,W_{\ic}) - H(W_i|X,W_0,W_{\ic})
 \notag
 \\
 &=
 I(X_i;W_i|W_0,W_{\ic})
 \\
 H(W_0,W_1,W_2) - H(W_0|X) - H(W_1|X) - H(W_2|X)
 &=
 H(W_0,W_1,W_2) - H(W_0|X) - H(W_1|W_0,X)
 \notag
 \\*
 &\quad
 - H(W_2|W_0,W_1,X)
 \notag
 \\
 &=
 H(W_0,W_1,W_2) - H(W_0,W_1,W_2|W_0,W_1,X)
 \notag
 \\
 &=
 I(X_0,X_1,X_2;W_0,W_1,W_2)
 \notag
 \\
 &=
 I(X_1,X_2;W_0,W_1,W_2),
 \label{eq:crng-dsc3-proof-I(X0,X1,X2;W0,W1,W2)}
\end{align}
where the first equalities comes from 
the fact that (\ref{eq:RCRNG-DSC3-markov-Wi}) implies
\begin{align*}
 H(W_1|X)
 &=
 H(W_1|W_0,X)
 \\
 H(W_2|X)
 &=
 H(W_2|W_0,W_1,X)
\end{align*}
and the last inequality
of (\ref{eq:crng-dsc3-proof-I(X0,X1,X2;W0,W1,W2)})
comes from (\ref{eq:crng-dsc3-X0}).
Then we have the fact that
$\RCRNGDSCp$
derived as above,
which is the case of 
two sources introduced in Example \ref{example:dsc2},
includes the Wagner-Kelly-Altu\u{g} single-letter inner region $\RWKADSC$,
where the region is achievable with the code
using constrained-random number generators.
It is a future challenge to clarify whether
$\RCRNGDSCp$ is strictly larger than $\RWKADSC$ or not.
\end{example}

\subsection{Multiple-Description Coding}
\label{sec:mdc-example}

In this subsection, we revisit
multiple-description coding
introduced by Gersho and Witsenhausen
(see \cite{EC82}, \cite[pp. 335-336]{EK11}).
In multiple description coding,
it is assumed that
all encoders have access to the same source,
each decoder reproduces a different source,
and there is no side information at the decoder;
that is,
$\fS=\{\I\}$,
$\XX_i=\XX_{\I}$ for all $i\in\I$,
$\K_j=\{j\}$,
and $\YY_j$ is constant for all $j\in\J$.
Successive-refinement coding \cite{EC91} is a special case
of the multiple-description coding,
where $\I=\J=\{1,\ldots,|\I|\}$ and $\I_j=\{1,\ldots,j\}$
for all $j\in\J$.

The following examples are particular cases
of multiple-description coding.
Let us assume that $(\WW_{\I},\XX_{\I},\ZZ_{\K})$ is stationary memoryless
with generic random variable $(W_{\I},X,Z_{\K})$,
where $X_i\equiv X$ for all $i\in\I$
and $\J=\K$ from the assumption.

\begin{example}
\label{example:mdc2}
The original multiple-description coding
is the case of two codewords and three reproductions,
where $\I\equiv\{1,2\}$, $\J\equiv\{1,2,12\}$,
$\I_j\equiv\{j\}$ for each $j\in\{1,2\}$,
and $\I_{12}\equiv\{1,2\}$.
From (\ref{eq:RCRNG-MDC-sum[ri]})--(\ref{eq:RCRNG-MDC-markov-decoder}),
we have the relations
\begin{align}
 0
 \leq
 r_i
 &\leq
 H(W_i|X)
 \label{eq:RCRNG-MDC2-ri}
 \\
 r_1+r_2
 &\leq
 H(W_1,W_2|X)
 \label{eq:RCRNG-MDC2-sum[ri]}
 \\
 r_i+R_i
 &\geq
 H(W_i)
 \label{eq:RCRNG-MDC2-sum[ri+Ri]}
 \\
 r_1+R_1+r_2+R_2
 &\geq
 H(W_1,W_2)
 \label{eq:RCRNG-MDC2-[r1+R1+r2+R2]}
\end{align}
and
\begin{gather*}
 (Z_1,Z_2,Z_{12})\markov X\markov (W_1,W_2)
 \\
 (W_{\ic},X,Z_{12},Z_{\ic})\markov W_i\markov Z_i
 \\
 (X,Z_1,Z_2)\markov (W_1,W_2)\markov Z_{12}
\end{gather*}
for all $i\in\{1,2\}$.
By using the Fourier-Motzkin method \cite[Appendix E]{EK11}
to eliminate $\{r_1,r_2\}$ and redundant inequalities, we have
equivalent conditions to
(\ref{eq:RCRNG-MDC2-ri})--(\ref{eq:RCRNG-MDC2-[r1+R1+r2+R2]})
as
\begin{align*}
 R_i
 &\geq
 H(W_i)
 -H(W_i|X)
 \\
 R_1+R_2
 &\geq
 H(W_1) + H(W_2)
 -H(W_1,W_2|X)
\end{align*}
for all $i\in\{1,2\}$.
By introducing a time-sharing random variable $T$,
which is shared by encoders and decoders,
we have the inner region $\RCRNGMDCp$
derived as the union of the region
\begin{equation*}
 \RCRNGMDCp(W_{\I},Z_{\K})
 \equiv
 \lrb{
	(R_1,R_2,D_1,D_2,D_{12}):
	\begin{aligned}
	 R_i
	 &\geq
	 H(W_i|T)
	 -H(W_i|X,T)
	 \\
	 R_1+R_2
	 &\geq
	 H(W_1|T) + H(W_2|T)
	 -H(W_1,W_2|X,T)
	 \\
	 D_i
	 &\geq
	 E_{X_1Z_i}[d_i(X_1,Z_i)]
	 \\
	 D_{12}
	 &\geq
	 E_{XZ_{12}}[d_{12}(X,Z_{12})]
	 \\
	 &\text{for all}\ i\in\{1,2\}
	\end{aligned}
 }
\end{equation*}
over all $(W_{\I},Z_{\K},T)$ satisfying
\begin{gather}
 \text{$T$ is independent of $X$}
 \label{eq:crng-mdc2-T}
 \\
 (Z_1,Z_2,Z_{12})\markov (X,T)\markov (W_1,W_2)
 \\
 (W_{\ic},X,Z_{12},Z_{\ic})\markov (W_i,T)\markov Z_i
 \label{eq:crng-mdc2-markov-Zi}
 \\
 (X,Z_1,Z_2)\markov (W_1,W_2,T)\markov Z_{12}.
 \label{eq:crng-mdc2-markov-Z12}
\end{gather}
Here, let us define the El Gamal-Cover single-letter inner region
$\RECMDC$ \cite{EC82}
as the union of the region
\begin{equation*}
 \RECMDC(W_{\I},Z_{\K},T)
 \equiv
 \lrb{
	(R_1,R_2,D_1,D_2):
	\begin{aligned}
	 R_i
	 &\geq
	 I(X;Z_i|T)
	 \\
	 R_1+R_2
	 &\geq
	 I(X;Z_1,Z_2,Z_{12}|T)
	 +
	 I(Z_1;Z_2|T)
	 \\
	 D_i
	 &\geq
	 E_{XZ_i}[d_i(X,Z_i)]
	 \\
	 D_{12}
	 &\geq
	 E_{XZ_{12}}[d_i(X,Z_{12})]
	 \\
	 &\text{for all}\ i\in\{1,2\}
	\end{aligned}
 }
\end{equation*}
over all $(W_{\I},Z_{\K},T)$ satisfying
(\ref{eq:crng-mdc2-T})--(\ref{eq:crng-mdc2-markov-Z12}).
Then we have $\RCRNGMDCp\subset\RECMDC$ from the fact that
\begin{align}
 H(W_i|T)-H(W_i|X,T)
 &=
 I(X;W_i|T)
 \notag
 \notag
 \\
 &=
 H(X|T)
 -H(X|W_i,T)
 \notag
 \\
 &=
 H(X|T)
 -H(X|W_i,Z_i,T)
 \notag
 \\
 &=
 I(X;W_i,Z_i|T)
 \notag
 \\
 &\geq
 I(X;Z_i|T)
 \label{eq:crng-mdc2-proof-EC-Ri}
\end{align}
and
\begin{align}
 H(W_1|T)+H(W_2|T)
 &=
 H(W_1,Z_1|T)
 - H(Z_1|W_1,T)
 + H(W_2,Z_2|T)
 - H(Z_2|W_2,T)
 \notag
 \\
 &=
 H(W_1,Z_1|T)
 - H(Z_1|W_1,W_2,T)
 + H(W_2,Z_2|T)
 - H(Z_2|W_1,W_2,Z_1,T)
 \notag
 \\
 &=
 H(W_1,Z_1|T)
 + H(W_2,Z_2|T)
 - H(Z_1,Z_2|W_1,W_2,T)
 \notag
 \\
 &=
 H(W_1,Z_1|T)
 + H(W_2,Z_2|T)
 - H(W_1,W_2,Z_1,Z_2|T)
 + H(W_1,W_2|T)
 \notag
 \\
 &=
 I(W_1,Z_1;W_2,Z_2|T)
 + H(W_1,W_2|T)
 \notag
 \\
 &\geq
 I(Z_1;Z_2|T)
 + H(W_1,W_2|T)
 \label{eq:crng-mdc2-proof-EC-R1+R2-1}
 \\
 H(W_1,W_2|T)
 - H(W_1,W_2|X,T)
 &=
 I(X;W_1,W_2|T)
 \notag
 \\
 &=
 H(X|T)
 -H(X|W_1,W_2,T)
 \notag
 \\
 &=
 H(X|T)
 -H(X,Z_1,Z_2,Z_{12}|W_1,W_2,T)
 +H(Z_1,Z_2,Z_{12}|W_1,W_2,X,T)
 \notag
 \\
 &=
 H(X|T)
 -H(X|W_1,W_2,Z_1,Z_2,Z_{12},T)
 -H(Z_1,Z_2,Z_{12}|W_1,W_2,T)
 \notag
 \\*
 &\quad
 +H(Z_1|W_1,W_2,X,T)
 +H(Z_2|W_1,W_2,X,Z_1,T)
 +H(Z_{12}|W_1,W_2,X,Z_1,Z_2,T)
 \notag
 \\
 &=
 H(X|T)
 -H(X|W_1,W_2,Z_1,Z_2,Z_{12},T)
 -H(Z_1,Z_2,Z_{12}|W_1,W_2,T)
 \notag
 \\*
 &\quad
 +H(Z_1|W_1,W_2,T)
 +H(Z_2|W_1,W_2,Z_1,T)
 +H(Z_{12}|W_1,W_2,Z_1,Z_2,T)
 \notag
 \\
 &=
 H(X|T)
 -H(X|W_1,W_2,Z_1,Z_2,Z_{12},T)
 \notag
 \\
 &=
 I(X;W_1,W_2,Z_1,Z_2,Z_{12}|T)
 \notag
 \\
 &\geq
 I(X;Z_1,Z_2,Z_{12}|T)
 \label{eq:crng-mdc2-proof-EC-R1+R2-2}
\end{align}
implies
\begin{align}
 H(W_1|T)+H(W_2|T)
 - H(W_1,W_2|X,T)
 &=
 I(Z_1;Z_2|T)
 + H(W_1,W_2|T)
 - H(W_1,W_2|X,T)
 \notag
 \\
 &=
 I(Z_1;Z_2|T)
 +I(X;Z_1,Z_2,Z_{12}|T),
\end{align}
where the third equality of (\ref{eq:crng-mdc2-proof-EC-Ri}),
the second equality of (\ref{eq:crng-mdc2-proof-EC-R1+R2-1}),
and the fifth equality of (\ref{eq:crng-mdc2-proof-EC-R1+R2-2})
come from the fact that
(\ref{eq:crng-mdc2-markov-Zi}) and (\ref{eq:crng-mdc2-markov-Z12})
implies
\begin{align}
 H(X|W_i,T)
 &=
 H(X|W_i,Z_i,T)
 \notag
 \\
 H(Z_1|W_1,T)
 &=
 H(Z_1|W_1,W_2,T)
 \notag
 \\
 H(Z_2|W_2,T)
 &=
 H(Z_2|W_1,W_2,Z_1,T)
 \notag
 \\
 H(Z_1|W_1,W_2,X,T)
 &=
 H(Z_1|W_1,T)
 \notag
 \\
 &=
 H(Z_1|W_1,W_2,T)
 \\
 H(Z_2|W_1,W_2,X,Z_1,T)
 &=
 H(Z_2|W_2,T)
 \notag
 \\
 &=
 H(Z_2|W_1,W_2,Z_1,T)
 \\
 H(Z_{12}|W_1,W_2,X,Z_1,Z_2,T)
 &=
 H(Z_{12}|W_1,W_2,T)
 \notag
 \\
 &=
 H(Z_{12}|W_1,W_2,Z_1,Z_2,T).
\end{align}
Since the El Gamal-Cover single-letter inner region
is sub-optimal for a particular case \cite[Sec.\ 13.6]{EK11},
we have the fact that the region $\RCRNGMDCp$
is also sub-optimal.
\end{example}

\begin{example}
Here, let us consider the case of three codewords,
where
$\I\equiv\{0,1,2\}$, $\J\equiv\{1,2,12\}$,
$\I_j\equiv\{0,j\}$ for each $j\in\{1,2\}$,
and $\I_{12}\equiv\{0,1,2\}$.
From (\ref{eq:RCRNG-MDC-sum[ri]})--(\ref{eq:RCRNG-MDC-markov-decoder}),
we have the relations
\begin{align}
 0
 \leq
 r_0
 &\leq
 H(W_0|X)
 \label{eq:RCRNG-MDC3-r0}
 \\
 0
 \leq
 r_i
 &\leq
 H(W_i|X)
 \label{eq:RCRNG-MDC3-ri}
 \\
 r_0+r_i
 &\leq
 H(W_0,W_i|X)
 \label{eq:RCRNG-MDC3-[r0+ri]}
 \\
 r_1+r_2
 &\leq
 H(W_1,W_2|X)
 \label{eq:RCRNG-MDC3-[r1+r2]}
 \\
 r_0+r_1+r_2
 &\leq
 H(W_0,W_1,W_2|X)
 \label{eq:RCRNG-MDC3-sum[ri]}
 \\
 r_0+R_0
 &\geq
 H(W_0|W_i)
 \label{eq:RCRNG-MDC3-[r0+R0]i}
 \\
 r_i+R_i
 &\geq
 H(W_i|W_0)
 \label{eq:RCRNG-MDC3-[ri+Ri]i}
 \\
 r_0+R_0+r_i+R_i
 &\geq
 H(W_0,W_i)
 \label{eq:RCRNG-MDC3-[r0+R0+ri+Ri]i}
 \\
 r_0+R_0
 &\geq
 H(W_0|W_1,W_2)
 \label{eq:RCRNG-MDC3-[r0+R0]all}
 \\
 r_i+R_i
 &\geq
 H(W_i|W_0,W_{\ic})
 \label{eq:RCRNG-MDC3-[ri+Ri]all}
 \\
 r_0+R_0+r_i+R_i
 &\geq
 H(W_0,W_i|W_{\ic})
 \label{eq:RCRNG-MDC3-[r0+R0+ri+Ri]all}
 \\
 r_1+R_1+r_2+R_2
 &\geq
 H(W_1,W_2|W_0)
 \label{eq:RCRNG-MDC3-[r1+R1+r2+R2]all}
 \\
 r_0+R_0+r_1+R_1+r_2+R_2
 &\geq
 H(W_0,W_1,W_2)
 \label{eq:RCRNG-MDC3-sum[ri+Ri]all}
\end{align}
and
\begin{gather}
 (Z_0,Z_1,Z_{12})\markov X\markov (W_0,W_1,W_2)
 \label{eq:RCRNG-MDC3-markov-W}
 \\
 (W_{\ic},X,Z_{\ic},Z_{12})\markov (W_0,W_i)\markov Z_i
 \label{eq:RCRNG-MDC3-markov-Zi}
 \\
 (X,Z_1,Z_2)\markov (W_0,W_1,W_2)\markov Z_{12}
 \label{eq:RCRNG-MDC3-markov-Z12}
\end{gather}
for all $i\in\{1,2\}$.
By using the Fourier-Motzkin method \cite[Appendix E]{EK11}
to eliminate $\{r_0,r_1,r_2\}$ and redundant conditions,
we have conditions equivalent to
(\ref{eq:RCRNG-MDC3-r0})--(\ref{eq:RCRNG-MDC3-sum[ri+Ri]all})
as
\begin{align}
 R_0
 &\geq
 \max\{0,H(W_0|W_i) - H(W_0|X)\}
 \notag
 \\
 R_i
 &\geq
 \max\{0,H(W_i|W_0) - H(W_i|X)\}
 \label{eq:RCRNG-MDC3-Ri}
 \\
 R_0 + R_i
 &\geq
 H(W_0,W_i) - H(W_0,W_i|X)
 \notag
 \\
 R_1 + R_2
 &\geq
 H(W_1|W_0) + H(W_2|W_0)
 - H(W_0,W_1,W_2|X)
 \label{eq:RCRNG-MDC3-[R1+R2]}
 \\
 R_0 + R_1 + R_2
 &\geq
 H(W_0,W_i) + H(W_{\ic}|W_0)
 - H(W_0,W_1,W_2|X)
 \label{eq:RCRNG-MDC3-[R0+R1+R2]}
 \\
 2R_0 + R_1 + R_2
 &\geq
 H(W_0,W_1) + H(W_0,W_2)
 - H(W_0|X) - H(W_0,W_1,W_2|X)
 \notag
\end{align}
for all $i\in\{1,2\}$, 
where it is sufficient to add the condition
(\ref{eq:RCRNG-MDC3-[R0+R1+R2]}) when $i$ is either $1$ or $2$.
Here, let us assume that $R_0\equiv0$,
which corresponds to the case when
$\I=\{1,2\}$, $\J=\{1,2,12\}$, $\I_j=\{j\}$ for each $j\in\{1,2\}$
and $\I_{12}=\{1,2\}$.
Then we have the inner region $\RCRNGMDCp$,
which is the case of two codewords introduced
in Example \ref{example:mdc2},
derived as the union of the region
\begin{equation*}
 \RCRNGMDCp(W_0,W_{\I},Z_{\K})
 \equiv
 \lrb{
	(R_1,R_2,D_1,D_2,D_{12}):
	\begin{aligned}
	 R_i
	 &\geq
	 H(W_0,W_i)-H(W_0,W_i|X)
	 \\
	 R_1+R_2
	 &\geq
	 H(W_0,W_1)+H(W_0,W_2)
	 \notag
	 \\
	 &\quad
	 -H(W_0|X)-H(W_0,W_1,W_2|X)
	 \\
	 D_i
	 &\geq
	 E_{XZ_i}[d_i(X,Z_i)]
	 \\
	 D_{12}
	 &\geq
	 E_{XZ_{12}}[d_{12}(X,Z_{12})]
	 \\
	 &\text{for all}\ i\in\{1,2\}
	\end{aligned}
 }
\end{equation*}
over all $(W_0,W_{\I},Z_{\K})$ satisfying
(\ref{eq:RCRNG-MDC3-markov-W})--(\ref{eq:RCRNG-MDC3-markov-Z12})
and
\begin{equation}
 0
 \geq
 \max\{0,H(W_0|W_i)-H(W_0|X)\},
 \label{eq:RCRNG-MDC3-R0=0}
\end{equation}
where (\ref{eq:RCRNG-MDC3-Ri})--(\ref{eq:RCRNG-MDC3-[R0+R1+R2]})
are redundant when $R_0=0$ because
\begin{align}
 &
 H(W_0,W_i)-H(W_0,W_i|X)
 \notag
 \\*
 &\quad
 =
 \max\{0,H(W_i|W_0)+H(W_0)-H(W_0|W_i,X)-H(W_i|X)\}
 \notag
 \\
 &\quad
 \geq
 \max\{0,H(W_i|W_0)-H(W_i|X)\}
 \\
 &
 H(W_0,W_1) + H(W_0,W_2)
 - H(W_0|X) - H(W_0,W_1,W_2|X)
 \notag
 \\*
 &\quad
 =
 H(W_0,W_i) + H(W_{\ic}|W_0) + H(W_0) - H(W_0|X) - H(W_0,W_1,W_2|X)
 \notag
 \\
 &\quad
 \geq
 H(W_0,W_i) + H(W_{\ic}|W_0) - H(W_0,W_1,W_2|X)
 \notag
 \\
 &\quad
 \geq
 H(W_1|W_0) + H(W_2|W_0) - H(W_0,W_1,W_2|X).
\end{align}
Here, let us define the Zhang-Berger inner region $\RZBMDC$
\cite[Eq.~(13.9)]{EK11},\cite{VKG03,WCZCP11,ZB87}
for the case of two codewords
and three reproductions
introduced in Example \ref{example:mdc2}.
It is defined as the union of the region
\begin{equation*}
 \RZBMDC(W'_0,W'_{\I},Z'_{\K})
 \equiv
 \lrb{
	(R_1,R_2,D_1,D_2,D_{12}):
	\begin{aligned}
	 R_i
	 &\geq
	 I(X;W'_0,W'_i)
	 \\
	 R_1 + R_2
	 &\geq
	 I(X;W'_1,W'_2|W'_0) + 2 I(X;W'_0)
	 + I(W'_1;W'_2|W'_0)
	 \\
	 D_i
	 &\geq
	 E_{XZ'_i}[d_i(X,Z'_i)]
	 \\
	 D_{12}
	 &\geq
	 E_{XZ'_{12}}[d_i(X,Z'_{12})]
	 \\
	 &\text{for all}\ i\in\{1,2\}
	\end{aligned}
 }
\end{equation*}
over all $(W'_0,W'_{\I},Z'_{\K})$ satisfying
\begin{gather*}
 (Z'_1,Z'_2,Z'_{12})
 \markov
 X
 \markov
 (W'_0,W'_1,W'_2)
 \\
 (W'_{\ic},X,Z'_{\ic},Z'_{12})
 \markov
 (W'_0,W'_i)
 \markov
 Z'_i
 \\
 (X,Z'_1,Z'_2)
 \markov
 (W'_0,W'_1,W'_2)
 \markov
 Z'_{12}.
\end{gather*}
By letting
\begin{align*}
 W'_i
 &\equiv
 W_i
\end{align*}
for each $i\in\{0,1,2\}$
and
\begin{equation*}
 Z'_j
 \equiv
 Z_j
\end{equation*}
for each $j\in\{1,2,12\}$,
we have the relation $\RCRNGMDCp\subset\RZBMDC$
from the fact that
\begin{align}
 H(W_0,W_i) - H(W_0,W_i|X)
 &=
 I(X;W_0,W_i)
 \notag
 \\
 &=
 I(X;W'_0,W'_i)
 \\
 H(W_0,W_1) + H(W_0,W_2)
 - H(W_0|X) - H(W_0,W_1,W_2|X)
 &=
 I(W_1;W_2|W_0)
 + H(W_0) + H(W_0,W_1,W_2)
 \notag
 \\*
 &\quad
 - H(W_0|X) - H(W_0,W_1,W_2|X)
 \notag
 \\
 &=
 I(W_1;W_2|W_0)
 + I(X;W_0)
 + I(X;W_0,W_1,W_2)
 \notag
 \\
 &=
 I(X;W_1,W_2|W_0)
 + 2 I(X;W_0)
 + I(W_1;W_2|W_0)
 \notag
 \\
 &=
 I(X;W'_1,W'_2|W'_0) + 2 I(X;W'_0)
 + I(W'_1;W'_2|W_0).
\end{align}
Conversely, by letting
\begin{align*}
 W_0
 &\equiv
 W'_0
 \\
 W_i
 &\equiv
 (W'_0,W'_i)
\end{align*}
for each $i\in\{1,2\}$
and
\begin{equation*}
 Z_j
 \equiv
 Z'_j
\end{equation*}
for each $j\in\{1,2,12\}$,
we have the relation $\RCRNGMDCp\supset\RZBMDC$
from the fact that
\begin{align}
 I(X;W'_0,W'_i)
 &=
 H(W'_0,W'_i) - H(W'_0,W'_i|X)
 \notag
 \\
 &=
 H(W_0,W_i) - H(W_0,W_i|X)
 \\
 I(X;W'_1,W'_2|W'_0) + 2 I(X;W'_0)
 + I(W'_1;W'_2|W'_0)
 &=
 H(W'_0,W'_1) + H(W'_0,W'_2)
 - H(W'_0|X) - H(W'_0,W'_1,W'_2|X)
 \notag
 \\
 &=
 H(W_0,W_1) + H(W_0,W_2)
 - H(W_0|X) - H(W_0,W_1,W_2|X)
\end{align}
and the relation
\begin{align}
 H(W_0|W_i)-H(W_0|X)
 &=
 H(W'_0|W'_0,W'_i)
 -H(W'_0|X)
 \notag
 \\
 &=
 -H(W'_0|X)
 \notag
 \\
 &\leq
 0
\end{align}
implies (\ref{eq:RCRNG-MDC3-R0=0}).
From the above observations, we can conclude that
$\RCRNGMDCp$ derived as above,
which is the case of two codewords and three reproductions
introduced in Example \ref{example:mdc2},
is equal to the Zhang-Berger inner region $\RZBMDC$,
where the region is achievable with the code
using constrained-random number generators.
\end{example}

\subsection{Source Coding with Side Information at Decoders}

In this subsection, we revisit
source coding with side information at decoders
introduced by Heegard and Berger \cite{HB85}.
The source coding with side information at decoders
is the case where
$\fS=\{\I\}$, $|\I|=1$, and $\I_j=\I$ 
and $\K_j\equiv\{j\}$ for each $j\in\J$.

Here, let us consider the case of two decoders,
where $\J\equiv\{1,2\}$.
We omit the dependence of $\XX$, 
$\WW$, and $R$ on $i\in\I$
because $|\I|=1$.
In the following we assume that the distortion function $d_j$
does not depend on $\YY_{\J\setminus\{j\}}$.
From (\ref{eq:RIT-DSC-RI'}) and (\ref{eq:RIT-DSC-D}),
we have
\begin{align*}
 R
 &\geq
 \oH(\WW|\YY_j)-\uH(\WW|\XX)
 \\
 D_j
 &\geq
 \od_j(\XX,\YY_j,\ZZ_j)
\end{align*}
for all $j\in\J$.
This region is equal to the region introduced in \cite{MU12}
specified by
\begin{align*}
 R
 &\geq
 \oI(\XX;\WW)-\uI(\WW;\YY_j)
 \\
 D_j
 &\geq
 \od_j(\XX,\YY_j,\ZZ_j)
\end{align*}
for all $j\in\J$,
where it is shown in \cite{MU12}
that the region specified by above inequalities
is equal to $\ROP$.

In the following examples,
let us assume that
$(\WW_{\I},\XX_{\I},\YY_{\J},\ZZ_{\J})$ is stationary memoryless
with generic random variable $(W_{\I},X_{\I},Y_{\J},Z_{\J})$,
where $X_i\equiv X$ for all $i\in\I$ and $\K=\J$ from the assumption.
\begin{example}
\label{example:dsi2}.
Here, let us assume that
$\fS=\{\I\}$, $|\I|=1$ and $\I_j=\I$
and $\K_j=\{j\}$ for all $j\in\J\equiv\{1,2\}$.
From (\ref{eq:RIT-DSC-markov-encoder})--(\ref{eq:RCRNG-DSC-sum[ri+Ri]}),
we have relations
\begin{align}
 0
 \leq
 r
 &\leq
 H(W|X)
 \label{eq:RIT-SCI1-r}
 \\
 r+R
 &\geq
 H(W|Y_j)
 \label{eq:RIT-SCI1-[r+R]}
\end{align}
and
\begin{gather}
 Y^n_{\J}
 \markov
 X^n
 \markov
 W^n
 \label{eq:RIT-SCI1-markov-encoder}
 \\
 (X^n,Y^n_{\jc},Z^n_{\jc})
 \markov
 (W^n,Y^n_j)
 \markov
 Z^n_j
 \notag
\end{gather}
for all $j\in\{1,2\}$.
By using the Fourier-Motzkin method \cite[Appendix E]{EK11}
to eliminate $r$,
we have conditions equivalent to
(\ref{eq:RIT-SCI1-r}) and (\ref{eq:RIT-SCI1-[r+R]})
as
\begin{align}
 R 
 &\geq
 H(W|Y_j)- H(W|X)
 \notag
 \\
 &=
 H(W|Y_j) - H(W|X,Y_j)
 \notag
 \\
 &=
 I(W;X|Y_j)
\end{align}
for all $j\in\J$,
where the first equality comes from the fact that
(\ref{eq:RIT-SCI1-markov-encoder})
implies $H(W|X)=H(W|X,Y_j)$.
This inequality characterizes the region given in \cite{MU12}.
\end{example}

\begin{example}
Here, let us assume that
$\fS=\{\I\}$, $\I=\{0,1,2\}$,
$X_i=X$ for all $i\in\I$,
and $\I_j=\{0,j\}$ and $\K_j=\{j\}$ for all $j\in\J\equiv\{1,2\}$.
This is the case of multiple description
with side information at two decoders,
where there are three encoders
that have access to the same source $X$.
From (\ref{eq:RCRNG-MDC-sum[ri]})--(\ref{eq:RCRNG-MDC-markov-decoder}),
we have relations
\begin{align}
 0
 \leq
 r_0
 &\leq
 H(W_0|X)
 \label{eq:RCRNG-DSI3-r0}
 \\
 0
 \leq
 r_j
 &\leq
 H(W_j|X)
 \label{eq:RCRNG-DSI3-ri}
 \\
 r_0+r_j
 &\leq
 H(W_0,W_j|X)
 \label{eq:RCRNG-DSI3-[r0+ri]}
 \\
 r_1+r_2
 &\leq
 H(W_1,W_2|X)
 \label{eq:RCRNG-DSI3-[r1+r2]}
 \\
 r_0+r_1+r_2
 &\leq
 H(W_0,W_1,W_2|X)
 \label{eq:RCRNG-DSI3-sum[ri]}
 \\
 r_0+R_0
 &\geq
 H(W_0|W_j,Y_j)
 \label{eq:RCRNG-DSI3-[r0+R0]i}
 \\
 r_j+R_j
 &\geq
 H(W_j|W_0,Y_j)
 \label{eq:RCRNG-DSI3-[ri+Ri]i}
 \\
 r_0+R_0+r_j+R_j
 &\geq
 H(W_0,W_j|Y_j)
 \label{eq:RCRNG-DSI3-[r0+R0+ri+Ri]i}
\end{align}
and
\begin{gather}
 (Y_{\J},Z_{\J})\markov X\markov (W_0,W_1,W_2)
 \label{eq:RCRNG-DSI3-markov-W}
 \\
 (W_{\jc},X,Y_{\jc},Z_{\jc})\markov (W_0,W_j,Y_j)\markov Z_j
 \label{eq:RCRNG-DSI3-markov-Zi}
\end{gather}
for all $j\in\{1,2\}$.
By using the Fourier-Motzkin method \cite[Appendix E]{EK11}
to eliminate $\{r_0,r_1,r_2\}$ and redundant inequalities,
we have conditions equivalent to 
(\ref{eq:RCRNG-DSI3-r0})--(\ref{eq:RCRNG-DSI3-[r0+R0+ri+Ri]i})
as
\begin{align}
 R_0 
 &\geq
 H(W_0|W_j,Y_j) - H(W_0|X)
 \label{eq:crng-dsi3-R0}
 \\
 R_j
 &\geq
 H(W_j|W_0,Y_j) - H(W_j|X)
 \\
 R_0 + R_j 
 &\geq
 H(W_0,W_j|Y_j) - H(W_0,W_j|X)
 \\
 R_0 + R_j
 &\geq
 H(W_0|W_{\jc},Y_{\jc}) + H(W_j|W_0,Y_j) - H(W_0,W_j|X)
 \\
 R_1 + R_2
 &\geq
 H(W_1|W_0,Y_1) + H(W_2|W_0,Y_2) - H(W_1,W_2|X)
 \\
 R_0 + R_1 + R_2
 &\geq
 H(W_0,W_j|Y_j)
 + H(W_{\jc}|W_0,Y_{\jc}) - H(W_0,W_1,W_2|X)
 \\
 2 R_0 + R_1 + R_2
 &\geq
 H(W_0,W_1|Y_1) + H(W_0,W_2|Y_2)
 - H(W_0|X) - H(W_0,W_1,W_2|X)
 \label{eq:crng-dsi3-2R0+R1+R2}
\end{align}
for all $j\in\{1,2\}$.
Here, let us assume that
\begin{equation}
 R'
 =
 R_0+R_1+R_2,
 \label{eq:crng-dsi3-R'}
\end{equation}
where the set of encoders generates $(W_0,W_1,W_2)$ from $X$
and sends codeword of $(W_0,W_j)$ to the $j$-th decoder
that reproduces $(W_0,W_j)$.
This relation correspond to the case of 
one encoder introduced in Example \ref{example:dsi2}.
By using the Fourier-Motzkin method \cite[Appendix E]{EK11}
to eliminate $\{R_0,R_1,R_2\}$ and redundant inequalities,
we have the conditions for $R'$ equivalent
to (\ref{eq:crng-dsi3-R0})--(\ref{eq:crng-dsi3-R'})
as
\begin{equation*}
 R'
 \geq
 H(W_0,W_j|Y_j) + H(W_{\jc}|W_0,Y_{\jc})
 - H(W_0,W_1,W_2|X).
\end{equation*}
Then the region $\RCRNGDSIp$
of the source coding with side information at two decoders
is derived as the union of the region
\begin{equation*}
 \RCRNGDSIp(W_0,W_{\J},Z_{\J})
 \equiv
 \lrb{
	(R,D_1,D_2):
	\begin{aligned}
	 R
	 &\geq
	 H(W_0,W_j|Y_j) + H(W_{\jc}|W_0,Y_{\jc})
	 - H(W_0,W_1,W_2|X)
	 \\
	 D_j
	 &\geq
	 E_{XY_jZ_j}[d_j(X,Y_j,Z_j)]
	 \\
	 &\text{for all}\ j\in\{1,2\}
	\end{aligned}
 }
\end{equation*}
over all $(W_0,W_{\J},Z_{\J})$ satisfying
(\ref{eq:RCRNG-DSI3-markov-W}) and (\ref{eq:RCRNG-DSI3-markov-Zi}).
Here, let us define the Heegard-Berger single-letter inner region
$\RHBDSI$ \cite{HB85}
defined as the union of the region
\begin{equation*}
 \RHBDSI(W'_0,W'_{\J},Z'_{\J})
 \equiv
 \lrb{
	(R,D_1,D_2):
	\begin{aligned}
	 R
	 &\geq
	 I(X;W'_0|Y_j)
	 + I(X;W'_j|W'_0,Y_j)
	 + I(X;W'_{\jc}|W'_0,Y_{\jc})
	 \\
	 D_j
	 &\geq
	 E_{XY_jZ'_j}[d_j(X,Y_j,Z'_j)]
	 \\
	 &\text{for all}\ j\in\{1,2\}
	\end{aligned}
 }
\end{equation*}
over all $(W'_0,W'_{\J},Z'_{\J})$ satisfying
\begin{gather}
 (Y_{\J},Z'_{\J})\markov X\markov (W'_0,W'_1,W'_2)
 \label{eq:RCRNG-DSI3-markov-W'}
 \\
 (W'_{\jc},X,Y_{\jc},Z'_{\jc})\markov (W'_0,W'_j,Y_j)\markov Z'_j
 \label{eq:RCRNG-DSI3-markov-Z'i}
\end{gather}
for all $j\in\{1,2\}$.
By letting
\begin{equation*}
 W'_i
 \equiv
 W_i
\end{equation*}
for each $i\in\{0,1,2\}$
and
\begin{equation*}
 Z'_j
 \equiv
 Z_j
\end{equation*}
for each $j\in\{1,2\}$,
we have
$\RCRNGDSIp\subset\RHBDSI$
from the fact that
\begin{align}
 &
 H(W_0,W_j|Y_j) + H(W_{\jc}|W_0,Y_{\jc})
 - H(W_0,W_1,W_2|X)
 \notag
 \\*
 &=
 H(W_0|Y_j) + H(W_j|W_0,Y_j) + H(W_{\jc}|W_0,Y_{\jc})
 - H(W_0|X)
 - H(W_j|W_0,X)
 - H(W_{\jc}|W_0,W_j,X)
 \notag
 \\
 &=
 H(W_0|Y_j) + H(W_j|W_0,Y_j) + H(W_{\jc}|W_0,Y_{\jc})
 - H(W_0|X,Y_j)
 - H(W_j|W_0,X,Y_j)
 - H(W_{\jc}|W_0,W_j,X,Y_{\jc})
 \notag
 \\
 &=
 I(X;W_0|Y_j)
 + I(X;W_j|W_0,Y_j)
 + I(W_j,X;W_{\jc}|W_0,Y_{\jc})
 \notag
 \\
 &\geq
 I(X;W_0|Y_j)
 + I(X;W_j|W_0,Y_j)
 + I(X;W_{\jc}|W_0,Y_{\jc})
 \notag
 \\
 &=
 I(X;W'_0|Y_j)
 + I(X;W'_j|W'_0,Y_j)
 + I(X;W'_{\jc}|W'_0,Y_{\jc}),
\end{align}
where the second equality comes from
the fact that (\ref{eq:RCRNG-DSI3-markov-W}) implies
\begin{align}
 H(W_0|X)
 &=
 H(W_0|X,Y_j)
 \label{eq:H(W0|X)}
 \\
 H(W_j|W_0,X)
 &=
 H(W_j|W_0,X,Y_j)
 \label{eq:H(Wi|W0,X)}
 \\
 H(W_{\jc}|W_0,W_j,X)
 &=
 H(W_{\jc}|W_0,W_j,X,Y_{\jc}).
 \label{eq:H(Wic|W0,Wi,X)}
\end{align}
To show $\RCRNGDSIp\supset\RHBDSI$,
let us assume that a quintuple $(W'_0,W'_{\J},X,Y_{\J},Z'_{\J})$
satisfies
(\ref{eq:RCRNG-DSI3-markov-W'}) and (\ref{eq:RCRNG-DSI3-markov-Z'i}).
Then the joint distribution 
$\mu_{W'_0W'_{\J}XY_{\J}Z'_{\J}}$ is given as
\begin{equation*}
 \mu_{W'_0W'_{\I}XY_{\J}Z'_{\J}}
 (w'_0,w'_{\I},x,y_{\J},z'_{\J})
 =
 \lrB{
	\prod_{j\in\J}\mu_{Z'_j|W'_0,W'_j,Y_j}(z'_j|w'_0,w'_j,y_j)
 }
 \mu_{W'_0W'_1W'_2|X}(w'_0,w'_1,w'_2|x)	
 \mu_{XY_{\J}}(x,y_{\J}).
\end{equation*}
Let $(W_0,W_{\J},X,Y_{\J},Z_{\J})$ be a quintuple of random variables
which corresponds to the joint distribution
$\mu_{W_0W_{\J}XY_{\J},Z_{\J}}$
defined as
\begin{align*}
 &
 \mu_{W_0W_{\J}XY_{\J}Z_{\J}}
 (w_0,w_{\J},x,y_{\J},z_{\J})
 \notag
 \\*
 &\equiv
 \lrB{
	\prod_{j\in\J}\mu_{Z'_j|W'_0W'_jY_j}
	(z_j|w_0,w_j,y_j)
 }
 \lrB{
	\prod_{j\in\J}
	\mu_{W'_j|W'_0X}(w_j|w_0,x)
 }
 \mu_{W'_0|X}(w_0|x)
 \mu_{XY_{\J}}(x,y_{\J}),
\end{align*}
where $\mu_{W'_0|X}(w_0|x)$ and $\mu_{W'_i|W'_0X}(w_i|w_0,x)$
are defined as
\begin{align*}
 \mu_{W'_0|X}(w_0|x)
 &\equiv
 \sum_{w_1,w_2}\mu_{W'_0W'_1W_2|X}(w_0,w_1,w_2|x)	
 \\
 \mu_{W'_j|W'_0X}(w_j|w_0,x)
 &\equiv
 \frac{
	\sum_{w_{\jc}}\mu_{W'_0W'_1W_2|X}(w_0,w_1,w_2|x)	
 }{
	\mu_{W'_0|X}(w_0|x)
 }.
\end{align*}
It should be noted that
$\mu_{W'_j|W'_0X}(w_j|w_0,x)$
is defined when $\mu_{W'_0|X}(w_0|x)>0$.
Then we have the fact that $(W_0,W_{\J},Z_{\J})$ 
satisfies
(\ref{eq:RCRNG-DSI3-markov-W}), (\ref{eq:RCRNG-DSI3-markov-Zi}),
and
\begin{equation}
 W_1\markov (W_0,X)\markov W_2.
 \label{eq:RCRNG-DSI3-ci}
\end{equation}
Furthermore, we have the fact that
the joint distribution of 
$(W_0,W_j,X,Y_j,Z_j)$ is equal to
that of $(W'_0,W'_j,X,Y_j,Z'_j)$ for all $j\in\{1,2\}$
because
\begin{align}
 &
 \mu_{W_0W_jXY_jZ_j}
 (w_0,w_j,x,y_j,z_j)
 \notag
 \\*
 &=
 \sum_{w_{\jc},y_{\jc},z_{\jc}}
 \lrB{
	\prod_{j\in\J}\mu_{Z'_j|W'_0W'_jY_j}
	(z_j|w_0,w_j,y_j)
 }
 \lrB{
	\prod_{j\in\J}\mu_{W'_i|W'_0X}
	(w_j|w_0,x)
 }
 \mu_{W'_0|X}(w_0|x)
 \mu_{XY_{\J}}(x,y_{\J})
 \notag
 \\
 &=
 \mu_{Z'_j|W'_0W'_jY_j}(z_j|w_0,w_j,y_j)
 \mu_{W'_j|W'_0X}(w_j|w_0,x)
 \mu_{W'_0|X}(w_0|x)
 \mu_{XY_j}(x,y_j)
 \notag
 \\
 &=
 \mu_{Z'_j|W'_0W'_jY_j}(z_j|w_0,w_j,y_j)
 \mu_{W'_0W'_j|X}(w_0,w_j|x)
 \mu_{XY_j}(x,y_j)
 \notag
 \\
 &=
 \sum_{w_{\jc},y_{\jc},z_{\jc}}
 \lrB{
	\prod_{j\in\J}\mu_{Z'_j|W'_0W'_jY_j}
	(z_j|w_0,w_j,y_j)
 }
 \mu_{W'_0W'_1W'_2|X}(w_0,w_1,w_2|x)
 \mu_{XY_{\J}}(x,y_{\J})
 \notag
 \\
 &=
 \mu_{W'_0W'_jXY_jZ'_j}
 (w_0,w_j,x,y_j,z_j).
\end{align}
Then we have $\RCRNGDSIp\supset\RHBDSI$ from the relations
\begin{equation*}
 E_{XY_jZ'_j}[d_j(X,Y_j,Z'_j)]
 =	E_{XY_jZ_j}[d_j(X,Y_j,Z_j)]
\end{equation*}
and
\begin{align}
 &
 I(X;W'_0|Y_j)
 + I(X;W'_j|W'_0,Y_j)
 + I(X;W'_{\jc}|W'_0,Y_{\jc})
 \notag
 \\*
 &=
 I(X;W_0|Y_j)
 +	I(X;W_j|W_0,Y_j)
 + I(X;W_{\jc}|W_0,Y_{\jc})
 \notag
 \\
 &=
 H(W_0|Y_j)
 -	H(W_0|X,Y_j)
 +	H(W_j|W_0,Y_j)
 -	H(W_j|W_0,X,Y_j)
 +	H(W_{\jc}|W_0,Y_{\jc})
 -	H(W_{\jc}|W_0,X,Y_{\jc})
 \notag
 \\
 &=
 H(W_0|Y_j)
 -	H(W_0|X)
 +	H(W_j|W_0,Y_j)
 -	H(W_j|W_0,X)
 +	H(W_{\jc}|W_0,Y_{\jc})
 -	H(W_{\jc}|W_0,X)
 \notag
 \\
 &=
 H(W_0,W_j|Y_j)
 +	H(W_{\jc}|W_0,Y_{\jc})
 -	H(W_0|X)
 -	H(W_j|W_0,X)
 -	H(W_{\jc}|W_0,W_j,X)
 \notag
 \\
 &=
 H(W_0,W_j|Y_j)
 +	H(W_{\jc}|W_0,Y_{\jc})
 -	H(W_0,W_1,W_2|X),
\end{align}
where the first equality comes from
the fact that
the joint distribution of $(W_0,W_j,X,Y_j)$ is equal to
that of $(W'_0,W'_j,X,Y_j)$
for all $j\in\J$,
the third equality comes from the fact that
(\ref{eq:RCRNG-DSI3-markov-W}) implies
(\ref{eq:H(W0|X)})--(\ref{eq:H(Wic|W0,Wi,X)}),
and the fourth equality comes from
the fact that (\ref{eq:RCRNG-DSI3-ci})
implies
$H(W_{\jc}|W_0,X)=H(W_{\jc}|W_0,W_j,X)$.
From the above observations, we can conclude that
$\RCRNGDSIp$ derived as above,
which is the case of two decoders
introduced in Example \ref{example:dsi2},
is equal to the Heegard-Berger single-letter inner region $\RHBDSI$,
where the region is achievable with the code
using constrained-random number generators.
\end{example}

\begin{rem}
Similarly to the proof of $\RCRNGDSIp=\RHBDSI$,
we have the fact that the region $\RCRNGDSIp$ does not
change with the additional condition
(\ref{eq:RCRNG-DSI3-ci}) for $(W_0,W_{\J},Z_{\J})$.
Similarly, we have the fact that the region $\RHBDSI$ does not
change with the additional condition
$W'_1\markov (W'_0,X)\markov W'_2$
for $(W'_0,W'_{\J},Z'_{\J})$.
It should be noted that the discussion is valid 
when the distortion $d_j(X,Y_j,Z_j)$ (resp. $d_j(X,Y_j,Z_j)$)
depends only on the joint distribution of $(W_0,W_j,X,Y_j,Z_j)$
(resp. $(W_0,W_j,X,Y_j,Z'_j)$)
for all $j\in\J$.
It is a future challenge to investigate the case where
the distortion $d_j$ 
depends on the joint distribution of $(W_0,W_{\J},X,Y_{\J},Z_{\J})$.
\end{rem}

\section{Formulation of Distributed Source Coding by Jana and Blahut}
\label{sec:jb}

In the previous sections, we assumed that
all reproductions were allowed to be lossy.
In this section, we consider the formulation
of distributed source coding introduced by Jana and Blahut \cite{JB08}
(Fig. \ref{fig:jana-blahut}).
Although the formulation is included 
as a special case of that considered in the previous sections
by assuming $\fS=\{\{i\}:i\in\I\}$ and $|\J|=1$,
and letting $d_i^{(n)}(\xx_i,\yy_{\J},\zz_i)=\chi(\xx_i=\zz_i)$
and $D_i=0$ for some $i\in\I$,
it is worthwhile to present a concise representation of the region.

Let $\I$ be the index set of sources and encoders.
Let $\I_0$ be the index set of sources which are reproduced losslessly,
where $\I_0\subset\I$.
Let $\K$ be the index set of other (possibly lossy) reproductions,
where we assume that $\I_0\cap\K=\emptyset$.
In the following, the dependence of $\YY$ and $Y^n$ on $j$ is omitted
because $|\J|=1$ is assumed.

We assume that the $i$-th encoder observes source $X^n_i$
and transmits codeword $M^{(n)}_i$ to the decoder.
Let $\ZZ_{\K}\equiv\{\ZZ_k\}_{k\in\K}$ be a set of reproductions
other than $\XX_{\I_0}$,
where $\ZZ_k\equiv\{Z^n_k\}_{n=1}^{\infty}$.
We assume that the decoder reproduces $(X^n_{\I_0},Z^n_{\K})$
after observing the set of codewords
$M^{(n)}_{\I}\equiv\{M^{(n)}_i\}_{i\in\I}$
and the uncoded side information $Y^n$.
Let us introduce
the operational definition of the rate-distortion region,
which is analogous to the definition in \cite{JB08}.
\begin{df}
Rate-distortion pair $(R_{\I},D_{\K})$ is {\em achievable}
for a given set of distortion measures
$\{d^{(n)}_k\}_{k\in\K,n\in\NN}$
iff there is a sequence of codes
$\{
 (\{\vphi^{(n)}_i\}_{i\in\I},\{\psi^{(n)}_k\}_{k\in\I_0\cup\K})
 \}_{n=1}^{\infty}$
consisting of
encoding functions $\vphi^{(n)}_i:\X_i^n\to\M_i^{(n)}$
and reproducing functions
$\psi^{(n)}_k:\M^{(n)}_{\I}\times\Y^n\to\Z^{(n)}_k$
that satisfy
\begin{gather}
 \limsupn \frac {\log|\M^{(n)}_i|}n \leq R_i
 \ \text{for all $i\in\I$}
 \label{eq:rate-JB}
 \\
 \limn\Prob\lrsb{
	X^n_i\neq Z_i^n
 }=0
 \ \text{for all $i\in\I_0$}
 \label{eq:lossless-JB}
 \\
 \limn
 \Prob\lrsb{
	d^{(n)}_k(X_{\I}^n,Y^n,Z_k^n)
	> D_k+\delta
 }=0
 \ \text{for all $k\in\K$ and $\delta>0$},
 \label{eq:lossy-JB}
\end{gather}
where $\M^{(n)}_i$ is a finite set for all $i\in\I$
and $Z^n_k\equiv\psi_k^{(n)}(\{\vphi^{(n)}_i(X_i^n)\}_{i\in\I},Y^n)$
is the $k$-th reproduction for each $k\in\I_0\cup\K$.
The {\em rate-distortion region} $\ROPJB$
under the maximum-distortion criterion is defined as
the closure of the set of all achievable rate-distortion pairs.
\end{df}

Next, let $(\WW_{\I\setminus\I_0},\ZZ_{\K})$ be a set of general sources,
where $\WW_i\equiv \{W^n_i\}_{n=1}^{\infty}$ for each $i\in\I\setminus\I_0$
and $\ZZ_k\equiv \{Z^n_k\}_{n=1}^{\infty}$ for each $k\in\K$.
Let us define region $\RITJB$ as follows.
\begin{df}
\label{df:RIT}
Let $\RITJB(\WW_{\I\setminus\I_0},\ZZ_{\K})$
be defined as the set of all $(R_{\I},D_{\K})$ satisfying
\begin{align}
 \sum_{i\in\I'\cap\I_0}R_i
 &\geq
 \oH(\XX_{\I'\cap\I_0}|\WW_{\I\setminus\I_0},\XX_{\Ipc\cap\I_0},\YY)
 \label{eq:IT-RI'capI0}
 \\
 \sum_{i\in\I'\setminus\I_0}R_i
 &\geq
 \oH(\WW_{\I'\setminus\I_0}|\WW_{\Ipc\setminus\I_0},\YY)
 -
 \sum_{i\in\I'\setminus\I_0}
 \uH(\WW_i|\XX_i)
 \label{eq:IT-RI'setminusI0}
 \\
 D_k
 &\geq
 \od_k(\XX_{\I},\YY,\ZZ_k)
 \label{eq:IT-D}
\end{align}
for all $\I'\in2^{\I}\setminus\{\emptyset\}$ and $k\in\K$. 
Region $\RITJB$ is defined by the union of
$\RITJB(\WW_{\I\setminus\I_0},\ZZ_{\K})$
over all general sources $(\WW_{\I\setminus\I_0},\ZZ_{\K})$ 
satisfying the following conditions:
\begin{gather}
 (W^n_{[\I\setminus\I_0]\setminus\{i\}},X^n_{\I\setminus\{i\}},Y^n)
 \markov
 X^n_i
 \markov
 W^n_i
 \label{eq:IT-markov-encoder}
 \\
 X^n_{\I\setminus\I_0}
 \markov
 (W^n_{\I\setminus\I_0},X^n_{\I_0},Y^n)
 \markov
 Z^n_{\K}
 \label{eq:IT-markov-decoder}
\end{gather}
for all $i\in\I\setminus\I_0$ and $n\in\NN$.
Optionally, $Z^n_{\K}$ is allowed to be restricted to
being the deterministic function
of $(W^n_{\I\setminus\I_0},X^n_{\I_0},Y^n)$.
\end{df}

\begin{rem}
We can introduce auxiliary real-valued variables 
$\{r_i\}_{i\in\I\setminus\I_0}$
to obtain the bounds
\begin{align}
 0
 \leq
 r_i
 &\leq
 \uH(\WW_i|\XX_i)
 \label{eq:RIT-ri}
 \\
 \sum_{i\in\I'\cap\I_0}R_i
 &\geq
 \oH(\XX_{\I'\cap\I_0}|\WW_{\I\setminus\I_0},\XX_{\Ipc\cap\I_0},\YY)
 \label{eq:RIT-RI'capI_0}
 \\
 \sum_{i\in\I'\setminus\I_0}[r_i+R_i]
 &\geq
 \oH(\WW_{\I'\setminus\I_0}|\WW_{\Ipc\setminus\I_0},\YY)
 \label{eq:RIT-[r+R]I'setminusI_0}
\end{align}
for all $i\in\I\setminus\I_0$ and $\I'\in2^{\I}\setminus\{\emptyset\}$.
By using the Fourier-Motzkin method \cite[Appendix E]{EK11}
to eliminate $\{r_i\}_{i\in\I\setminus\I_0}$,
we have the fact that they are equivalent to
(\ref{eq:IT-RI'capI0}) and (\ref{eq:IT-RI'setminusI0})
for all $\I'\in2^{\I}\setminus\{\emptyset\}$.
\end{rem}

Furthermore, let us define region $\RCRNGJB$ as follows.
\begin{df}
\label{df:RCRNG}
Let $(\WW_{\I\setminus\I_0},\ZZ_{\K})$ be a set of general sources,
where $\WW_i\equiv \{W^n_i\}_{n=1}^{\infty}$ for each $i\in\I\setminus\I_0$
and $\ZZ_k\equiv \{Z^n_k\}_{n=1}^{\infty}$ for each $k\in\K$.
Let $\RCRNGJB(\WW_{\I\setminus\I_0},\ZZ_{\K})$
be defined as the set of all $(R_{\I},D_{\K})$ where 
there are real-valued variables
$\{r_i\}_{i\in\I\setminus\I_0}$ satisfying
\begin{align}
 0
 \leq
 r_i
 &\leq
 \uH(\WW_i|\XX_i)
 \label{eq:RCRNG-r}
 \\
 \sum_{i\in\I'\setminus\I_0}[r_i+R_i]
 +
 \sum_{i\in\I'\cap\I_0}R_i
 &\geq
 \oH(\WW_{\I'\setminus\I_0},\XX_{\I'\cap\I_0}
	|\WW_{\Ipc\setminus\I_0},\XX_{\Ipc\cap\I_0},\YY)
 \label{eq:RCRNG-r+R}
\end{align}
for all $i\in\I\setminus\I_0$
and $\I'\in2^{\I}\setminus\{\emptyset\}$,
and (\ref{eq:IT-D}) for all $k\in\K$.
Then region $\RCRNGJB$ is defined by the union of
$\RCRNGJB(\WW_{\I\setminus\I_0},\ZZ_{\K})$
over all general sources $(\WW_{\I\setminus\I_0},\ZZ_{\K})$
satisfying (\ref{eq:IT-markov-encoder}) and (\ref{eq:IT-markov-decoder})
for all $i\in\I$ and $n\in\NN$.
Optionally, $Z^n_{\K}$ is allowed to
be restricted to being the deterministic function
of $(W^n_{\I\setminus\I_0},X^n_{\I_0},Y^n)$.
\end{df}

\begin{rem}
By using the Fourier-Motzkin method \cite[Appendix E]{EK11},
we can eliminate auxiliary variables $\{r_i\}_{i\in\I\setminus\I_0}$
and obtain the bound
\begin{equation}
 \label{eq:RCRNG-JB-fme}
 \sum_{i\in\I'}R_i
 \geq
 \oH(\WW_{\I'\setminus\I_0},\XX_{\I'\cap\I_0}
	|\WW_{\Ipc\setminus\I_0},\XX_{\Ipc\cap\I_0})
 -
 \sum_{i\in\I'\setminus\I_0}
 \uH(\WW_i|\XX_i)
\end{equation}
for all $\I'\in2^{\I}\setminus\{\emptyset\}$,
which is equivalent to (\ref{eq:RCRNG-r}) and (\ref{eq:RCRNG-r+R})
for all $i\in\I\setminus\I_0$ and $\I'\in2^{\I}\setminus\{\emptyset\}$.
\end{rem}

We have the following theorem.
\begin{thm}
\label{thm:ROP=RIT=RCRNG-JB}
For a set of general correlated sources $(\XX_{\I},\YY)$,
we have
\begin{equation*}
 \ROPJB=\RITJB=\RCRNGJB.
\end{equation*}
\end{thm}
\begin{IEEEproof}
The proof of the theorem consists of the following three facts:
\begin{itemize}
 \item
 the converse $\ROPJB\subset\RITJB$,
 which is shown in Section \ref{sec:converse-jb},
 where (\ref{eq:IT-RI'capI0}) and (\ref{eq:IT-RI'setminusI0})
 are replaced by equivalent conditions 
 (\ref{eq:RIT-ri})--(\ref{eq:RIT-[r+R]I'setminusI_0});
 \item
 the relation $\RITJB\subset\RCRNGJB$
 derived immediately from the fact that
 (\ref{eq:RIT-ri})--(\ref{eq:RIT-[r+R]I'setminusI_0}) implies
 \begin{align}
	\sum_{i\in\I'\setminus\I_0}
	[r_i+R_i]
	+
	\sum_{i\in\I'\cap\I_0}
	R_i
	&\geq
	\oH(\WW_{\I'\setminus\I_0}|\WW_{\Ipc\setminus\I_0})
	+
	\oH(\XX_{\I'\cap\I_0}|
	 \WW_{\I\setminus\I_0},\XX_{\Ipc\cap\I_0},\YY)
	\notag
	\\
	&\geq
	\oH(\WW_{\I'\setminus\I_0}|
	 \WW_{\Ipc\setminus\I_0},\XX_{\Ipc\cap\I_0},\YY)
	+
	\oH(\XX_{\I'\cap\I_0}|
	 \WW_{\I\setminus\I_0},\XX_{\Ipc\cap\I_0},\YY)
	\notag
	\\
	&\geq
	\oH(\WW_{\I'\setminus\I_0},\XX_{\I'\cap\I_0}|
	 \WW_{\Ipc\setminus\I_0},\XX_{\Ipc\cap\I_0},\YY)
 \end{align}
 for all 
 $\I'\in 2^{\I}\setminus\{\emptyset\}$,
 where the second inequality comes from
 Lemma \ref{lem:oH(U|V)>oH(U|VV')}	in Appendix~\ref{sec:ispec},
 and the third inequality comes from
 Lemma \ref{lem:oH(UU'|V)<oH(U'|UV)+oH(U|V)} in Appendix~\ref{sec:ispec};
 \item
 the achievability $\RCRNGJB\subset\ROPJB$,
 which is shown in
 Section \ref{sec:construction}.
\end{itemize}
\end{IEEEproof}

The following examples are particular cases of the distributed
source coding problem.
In the examples, we discuss only the achievable regions,
which are actually optimal regions,
by specifying auxiliary random variables.
It should be noted that
the converse part can be shown from past studies.

\begin{example}
When $\I=\I_0$ and $\K=\emptyset$,
the rate-distortion region represents
the distributed lossless source coding region.
Since $\I'\setminus\I_0=\emptyset$
for all $\I'\in2^{\I}\setminus\{\emptyset\}$,
we have
\begin{equation*}
 \sum_{i\in\I'}
 R_i
 \geq
 \oH(\XX_{\I'}|\XX_{\Ipc},\YY)
\end{equation*}
for all $\I'\in 2^{\I}\setminus\{\emptyset\}$ from (\ref{eq:IT-RI'capI0}).
This region is given in \cite{MK95}
for the case where $\I=\I_0=\{1,2\}$ and $Y^n$ is a constant.
In particular,
when $\I=\I_0=\{1\}$,
the rate-distortion region represents
the point-to-point lossless source coding region introduced in \cite{HV93}
as
\begin{equation*}
 R_1
 \geq
 \oH(\XX_1).
\end{equation*}
It should be noted that the general region
for the case of two or more decoders is given in \cite{CRNG-MULTI}.
\end{example}

\begin{example}
When $\I=\K=\{1\}$, $\I_0=\emptyset$, and $Y^n$ is a constant,
the rate-distortion region
with given distortion measure
$d^{(n)}_1:\X^n_1\times\Z^n_1\to[0,\infty)$
and distortion level $D_1\in[0,\infty)$
represents the point-to-point rate-distortion region
under the maximum-distortion criterion introduced in \cite{CRNG,SV93}.
By letting $\ZZ_1=\WW_1$,
we have
\begin{align*}
 R_1
 &\geq
 \oH(\WW_1)
 -\uH(\WW_1|\XX_1)
 \\
 D_1
 &\geq
 \od_1(\XX_1,\WW_1)
\end{align*}
from (\ref{eq:IT-RI'setminusI0}) and (\ref{eq:IT-D}).
It is shown in \cite{CRNG} that
this region is equal to the region specified by
\begin{align*}
 R_1
 &\geq
 \oI(\XX_1;\WW_1)
 \\
 D_1
 &\geq
 \od_1(\XX_1,\WW_1)
\end{align*}
given in \cite{SV93}.
\end{example}

\begin{example}
When $\I=\K=\{1\}$ and $\I_0=\emptyset$,
the rate-distortion region
with given distortion measure $d^{(n)}_1:\X_1^n\times\Z_1^n\to[0,\infty)$
and distortion level $D_1\in[0,\infty)$
represents the region of lossy source coding
with non-causal side information at the decoder
introduced in \cite{IM02}.
We have
\begin{align*}
 R_1
 &\geq
 \oH(\WW_1|\YY)
 -\uH(\WW_1|\XX_1)
 \\
 D_1
 &\geq
 \od_1(\XX_1,\ZZ_1)
\end{align*}
from (\ref{eq:IT-RI'setminusI0}) and (\ref{eq:IT-D}).
The equality to the region specified by
\begin{align*}
 R_1
 &\geq
 \oI(\WW_1;\XX_1)
 -\uI(\WW_1;\YY)
 \\
 D_1
 &\geq
 \od_1(\XX_1,\ZZ_1)
\end{align*}
introduced in \cite{IM02} 
can be shown by using the achievability and converse
of the two regions.
\end{example}

\begin{example}
When $\I\setminus\I_0=\{\helper\}$, $\K=\emptyset$,
and $Y^n$ is a constant,
the rate-distortion region represents the region
of lossless source coding with a helper
that provides the coded side information.
We have the achievable region specified by
\begin{align*}
 \sum_{i\in\I'}
 R_i
 &\geq
 \oH(\XX_{\I'}|\WW_{\helper},\XX_{\I_0\setminus\I'})
 \\
 R_{\helper}
 &\geq
 \oH(\WW_{\helper})
 -
 \uH(\WW_{\helper}|\XX_{\helper})
\end{align*}
for all $\I'\subset2^{\I_0}\setminus\{\emptyset\}$
from (\ref{eq:IT-RI'capI0}) and (\ref{eq:IT-RI'setminusI0}).
This region is equal to the region specified by
\begin{align*}
 \sum_{i\in\I'}
 R_i
 &\geq
 \oH(\XX_{\I'}|\WW_{\helper},\XX_{\I_0\setminus\I'})
 \\
 R_{\helper}
 &\geq
 \oH(\WW_{\helper}|\XX_{\I_0})
 -
 \uH(\WW_{\helper}|\XX_{\helper})
 \\
 \sum_{i\in\I'}
 R_i
 +R_{\helper}
 &\geq
 \oH(\WW_{\helper},\XX_{\I'}|\XX_{\I_0\setminus\I'})
 - \uH(\WW_{\helper}|\XX_{\helper})
\end{align*}
for all $\I'\subset2^{\I_0}\setminus\{\emptyset\}$
from (\ref{eq:RCRNG-JB-fme}).
When $\I_0=\{1\}$,
these regions are also equal to the region specified by
\begin{align*}
 R_1
 &\geq
 \oH(\XX_1|\WW_{\helper})
 \\
 R_{\helper}
 &\geq
 \oI(\XX_{\helper};\WW_{\helper}),
\end{align*}
which is the region derived in \cite{MK95},
where the equality can be shown by using the achievability
and converse of these regions.
\end{example}

\begin{example}
When $\I=\K$ and $\I_0=\emptyset$,
the rate-distortion region
with given distortion measure $d^{(n)}_i:\X_i^n\times\Z_i^n\to[0,\infty)$
and distortion levels $D_i\in[0,\infty)$
represents the region of distributed lossy source coding,
where the conditions (\ref{eq:IT-RI'capI0})--(\ref{eq:IT-D}) 
are reduced to
\begin{align*}
 \sum_{i\in\I'}
 R_i
 &\geq
 \oH(\WW_{\I'}|\WW_{\Ipc},\YY)
 -
 \sum_{i\in\I'}
 \uH(\WW_i|\XX_i)
 \\
 D_i
 &\geq
 \od_i(\XX_i,\ZZ_i)
\end{align*}
for all $\I'\in2^{\I}\setminus\{\emptyset\}$ and $i\in\I$.
This is the rate-distortion region
defined by (\ref{eq:RIT-DSC-RI'}) and (\ref{eq:RIT-DSC-D}),
and alternative to that derived in \cite{YQ06a,YQ06b}.
It should be noted that our characterization
is simpler and more interpretable than that derived in \cite{YQ06a,YQ06b}.
\end{example}

\section{Proof of Converse}
\label{sec:converse}

This section shows the converse part of Theorems
\ref{thm:ROP=RIT=RCRNG=RCRNG'} and \ref{thm:ROP=RIT=RCRNG-JB}
based on the technique introduced in \cite{ISPEC-CONVERSE}.
It should be noted here that the introduction of
auxiliary variables $\{r_i\}_{i\in\I}$ simplifies the proof.

\subsection{Proof of $\ROP\subset\RCRNGDSC$}
\label{sec:converse-dsc}
This subsection argues the converse part $\ROP\subset\RCRNGDSC$.

Let us assume that $(R_{\I},D_{\K})\in\ROP$.
Then we have the fact that 
there is a sequence  $\{(\vphi^{(n)}_{\I},\psi^{(n)}_{\K})\}_{n=1}^{\infty}$
satisfying (\ref{eq:rate}) and (\ref{eq:lossy}).
From the definition of $\plimsupn$ in Appendix \ref{sec:ispec},
we have the fact that (\ref{eq:lossy}) 
implies (\ref{eq:RIT-DSC-D}) for all $k\in\K$.
Let
\begin{align*}
 W^n_i
 &\equiv
 \vphi^{(n)}_i(X^n_i)
 \\
 Z^n_k
 &\equiv
 \psi^{(n)}_k(W^n_{\I},Y^n_k)
\end{align*}
for $i\in\I$ and $k\in\K$.
Then we have the Markov chains
(\ref{eq:RIT-DSC-markov-encoder})
and (\ref{eq:RIT-DSC-markov-decoder}).

Let $r_i\equiv 0$ for each $i\in\I$.
Then, from Lemma \ref{lem:oH>uH>0} in Appendix \ref{sec:ispec}, we have
\begin{equation*}
 0
 \leq
 r_i
 \leq
 \uH(\WW_i|\XX_i)
\end{equation*}
for all $i\in\I$.
We have 
\begin{align}
 \sum_{i\in\I'_j}
 [r_i+R_i]
 &=
 \sum_{i\in\I'_j}
 R_i
 \notag
 \\
 &\geq
 \sum_{i\in\I'_j}
 \limsupn \frac{\log|\M^{(n)}_i|}n
 \notag
 \\
 &\geq
 \sum_{i\in\I'_j}
 \oH(\WW_i)
 \notag
 \\
 &\geq
 \oH(\WW_{\I'_j})
 \notag
 \\
 &\geq
 \oH(\WW_{\I'_j}|\WW_{\Ipcj},\YY_j)
 \label{eq:converse-[r+R]I'}
\end{align}
for all $j\in\J$ and $\I'_j\in2^{\I_j}\setminus\{\emptyset\}$,
where
the first inequality comes from (\ref{eq:rate}),
the second inequality comes from
Lemma \ref{lem:bound-by-cardinality} in Appendix \ref{sec:ispec},
the third inequality comes from
Lemmas \ref{lem:oH(UU'|V)<oH(U'|UV)+oH(U|V)}
and \ref{lem:oH(U|V)>oH(U|VV')} in Appendix \ref{sec:ispec},
and the last inequality comes from
Lemma \ref{lem:oH(U|V)>oH(U|VV')} in Appendix \ref{sec:ispec}.
\hfill\IEEEQED

\subsection{Proof of $\ROPJB\subset\RITJB$}
\label{sec:converse-jb}
This subsection argues the converse part $\ROPJB\subset\RITJB$.

Let us assume that $(R_{\I},D_{\K})\in\ROPJB$.
Then we have the fact that 
there is a sequence  $\{(\vphi^{(n)}_{\I},\psi^{(n)}_{\K})\}_{n=1}^{\infty}$
satisfying (\ref{eq:rate-JB})--(\ref{eq:lossy-JB}).
From the definition of $\plimsupn$ in Appendix \ref{sec:ispec},
we have the fact that (\ref{eq:lossy-JB}) 
implies (\ref{eq:IT-D}) for all $k\in\K$.
Let
\begin{align}
 W^n_i
 &\equiv
 \vphi^{(n)}_i(X^n_i)
 \label{eq:converse-W}
 \\
 Z^n_k
 &\equiv
 \psi^{(n)}_k(W^n_{\I},Y^n).
 \label{eq:converse-Z}
\end{align}
Then we have the Markov chains
(\ref{eq:IT-markov-encoder})
and (\ref{eq:IT-markov-decoder}).

Let $r_i\equiv 0$ for each $i\in\I_0$.
Then, from Lemma \ref{lem:oH>uH>0} in Appendix \ref{sec:ispec}, we have
\begin{equation*}
 0
 \leq
 r_i
 \leq
 \uH(\WW_i|\XX_i)
\end{equation*}
for all $i\in\I\setminus\I_0$.
We have 
\begin{align}
 \sum_{i\in\I'\setminus\I_0}
 [r_i+R_i]
 &=
 \sum_{i\in\I'\setminus\I_0}
 R_i
 \notag
 \\
 &\geq
 \sum_{i\in\I'\setminus\I_0}
 \limsupn \frac{\log|\M^{(n)}_i|}n
 \notag
 \\
 &\geq
 \sum_{i\in\I'\setminus\I_0}
 \oH(\WW_i)
 \notag
 \\
 &\geq
 \oH(\WW_{\I'\setminus\I_0})
 \notag
 \\
 &\geq
 \oH(\WW_{\I'\setminus\I_0}|\WW_{\Ipc\setminus\I_0},\YY)
 \label{eq:converse-[r+R]I'setminusI0}
\end{align}
for all $\I'\in2^{\I}\setminus\{\emptyset\}$,
where the first inequality comes from (\ref{eq:rate-JB}),
the second inequality comes from 
Lemma \ref{lem:bound-by-cardinality} in Appendix \ref{sec:ispec},
the third inequality comes from
Lemmas \ref{lem:oH(UU'|V)<oH(U'|UV)+oH(U|V)}
and \ref{lem:oH(U|V)>oH(U|VV')}
in Appendix \ref{sec:ispec},
and the last inequality comes from
Lemma \ref{lem:oH(U|V)>oH(U|VV')} in Appendix~\ref{sec:ispec}.

Furthermore, we have
\begin{align}
 \sum_{i\in\I'\cap\I_0}
 R_i
 &\geq
 \oH(\WW_{\I'\cap\I_0}
	|\WW_{[\I'\cap\I_0]^{\complement}},\XX_{\Ipc\cap\I_0},\YY)
 \notag
 \\
 &\geq
 \oH(\XX_{\I'\cap\I_0}
	|\WW_{[\I'\cap\I_0]^{\complement}},\XX_{\Ipc\cap\I_0},\YY)
 \notag
 \\
 &\geq
 \oH(\WW_{\Ipc\cap\I_0},\XX_{\I'\cap\I_0}
	|\WW_{\I\setminus\I_0},\XX_{\Ipc\cap\I_0},\YY)
 -
 \oH(\WW_{\Ipc\cap\I_0}
	|\WW_{\I\setminus\I_0},\XX_{\Ipc\cap\I_0},\YY)
 \notag
 \\
 &=
 \oH(\WW_{\Ipc\cap\I_0},\XX_{\I'\cap\I_0}
	|\WW_{\I\setminus\I_0},\XX_{\Ipc\cap\I_0},\YY)
 \notag
 \\
 &\geq
 \oH(\XX_{\I'\cap\I_0}
	|\WW_{\I\setminus\I_0},\XX_{\Ipc\cap\I_0},\YY)
 \label{eq:converse-RI'capI0}
\end{align}
for all $\I'\in2^{\I}\setminus\{\emptyset\}$,
where the first inequality
is derived similarly to the proof of (\ref{eq:converse-[r+R]I'setminusI0}),
the second inequality is shown by applying
Lemma \ref{lem:sw-bound} in Appendix \ref{sec:ispec}
together with the fact that
(\ref{eq:lossless-JB}), (\ref{eq:converse-Z}),
and Lemma~\ref{lem:fano} in Appendix \ref{sec:ispec} imply
\begin{align}
 0
 \leq
 \oH(\XX_{\I'\cap\I_0}
	|\WW_{\I'\cap\I_0},
	(\WW_{[\I'\cap\I_0]^{\complement}},\XX_{\Ipc\cap\I_0},\YY))
 &\leq
 \oH(\XX_{\I'\cap\I_0}|\WW_{\I},\YY)
 \notag
 \\
 &=0,
\end{align}
the third inequality comes from 
Lemma \ref{lem:oH(UU'|V)<oH(U'|UV)+oH(U|V)} in Appendix \ref{sec:ispec}
and the fact that
$[\I'\cap\I_0]^{\complement}$ is the disjoint union of
$\Ipc\cap\I_0$ and $\I\setminus\I_0$,
the equality comes from (\ref{eq:converse-W})
and Lemma \ref{lem:fano} in Appendix~\ref{sec:ispec},
and the last inequality comes from
Lemma \ref{lem:oH(UU'|V)>oH(U|V)} in Appendix~\ref{sec:ispec}.
Thus, we have the fact that $(R_{\I},D_{\K})\in\RITJB$.
\hfill\IEEEQED

\begin{rem}
We now show another converse, $\ROPJB\subset\RCRNGJB$,
directly without using
(\ref{eq:converse-[r+R]I'setminusI0})
and (\ref{eq:converse-RI'capI0}).
We have (\ref{eq:RCRNG-r+R}) as
\begin{align}
 \sum_{i\in\I'\setminus\I_0}
 [r_i+R_i]
 +
 \sum_{i\in\I'\cap\I_0}
 R_i
 &=
 \sum_{i\in\I'}
 R_i
 \notag
 \\
 &\geq
 \sum_{i\in\I'}
 \limsupn \frac{\log|\M^{(n)}_i|}n
 \notag
 \\
 &\geq
 \sum_{i\in\I'}
 \oH(\WW_i)
 \notag
 \\
 &\geq
 \oH(\WW_{\I'})
 \notag
 \\
 &\geq
 \oH(\WW_{\I'}|\WW_{\Ipc},\XX_{\Ipc\cap\I_0},\YY)
 \notag
 \\
 &\geq
 \oH(\WW_{\I'\setminus\I_0},\XX_{\I'\cap\I_0}
	|\WW_{\Ipc},\XX_{\Ipc\cap\I_0},\YY)
 \notag
 \\
 &\geq
 \oH(\WW_{\I'\setminus\I_0},\WW_{\Ipc\cap\I_0},\XX_{\I'\cap\I_0}
	|\WW_{\Ipc\setminus\I_0},\XX_{\Ipc\cap\I_0},\YY)
 -
 \oH(\WW_{\Ipc\cap\I_0}
	|\WW_{\Ipc\setminus\I_0},\XX_{\Ipc\cap\I_0},\YY)
 \notag
 \\
 &=
 \oH(\WW_{\I'\setminus\I_0},\WW_{\Ipc\cap\I_0},\XX_{\I'\cap\I_0}
	|\WW_{\Ipc\setminus\I_0},\XX_{\Ipc\cap\I_0},\YY)
 \notag
 \\
 &\geq
 \oH(\WW_{\I'\setminus\I_0},\XX_{\I'\cap\I_0}
	|\WW_{\Ipc\setminus\I_0},\XX_{\Ipc\cap\I_0},\YY)
 \label{eq:converse-RI'+rI'}
\end{align}
for all $\I'\in2^{\I}\setminus\{\emptyset\}$,
where the first equality comes from $r_i=0$ for all $i\in\I'\cap\I_0$.
the first inequality comes from (\ref{eq:rate-JB}),
the second inequality comes from
Lemma \ref{lem:bound-by-cardinality} in Appendix \ref{sec:ispec},
the third inequality comes from
Lemmas \ref{lem:oH(UU'|V)<oH(U'|UV)+oH(U|V)} and \ref{lem:oH(U|V)>oH(U|VV')}
in Appendix \ref{sec:ispec},
the fourth inequality comes from
Lemma \ref{lem:oH(U|V)>oH(U|VV')} in Appendix \ref{sec:ispec},
the fifth inequality is derived by applying
Lemma \ref{lem:sw-bound} in Appendix~\ref{sec:ispec}
together with the fact that
(\ref{eq:lossless-JB}), (\ref{eq:converse-Z}),
and Lemma \ref{lem:fano} in Appendix \ref{sec:ispec} imply 
\begin{align}
 0
 \leq
 \oH(\WW_{\I'\setminus\I_0},\XX_{\I'\cap\I_0}
	|\WW_{\I'},(\WW_{\Ipc},\XX_{\Ipc\cap\I_0},\YY))
 &\leq
 \oH(\WW_{\I'\setminus\I_0},\XX_{\I'\cap\I_0}|\WW_{\I},\YY)
 \notag
 \\
 &=0,
\end{align}
the sixth inequality comes from
Lemma \ref{lem:oH(UU'|V)<oH(U'|UV)+oH(U|V)} in Appendix \ref{sec:ispec},
the next equality comes from (\ref{eq:converse-W})
and Lemma \ref{lem:fano} in Appendix~\ref{sec:ispec},
and the last inequality comes from
Lemma \ref{lem:oH(UU'|V)>oH(U|V)} in Appendix \ref{sec:ispec}.
\end{rem}

\section{Proof of $\RCRNGDSC\subset\RCRNGMDC$}
\label{sec:RCRNGDSCsubsetRCRNGMDC}

Here, we show $\RCRNGDSC\subset\RCRNGMDC$
by showing that
for $(R_{\I},D_{\K})$ and $(\WW_{\I},\ZZ_{\K})$ satisfying
(\ref{eq:RIT-DSC-markov-encoder})--(\ref{eq:RCRNG-DSC-sum[ri+Ri]})
imply
(\ref{eq:RCRNG-MDC-sum[ri]})--(\ref{eq:RCRNG-MDC-markov-decoder}).
Let $\cS'\equiv\{i_1,\ldots,i_{|\cS'|}\}$,
where the order is arbitrary.
The relation (\ref{eq:RCRNG-MDC-sum[ri]}) is shown as
\begin{align}
 0\leq
 \sum_{i'\in\cS'}
 r_{i'}
 &\leq
 \sum_{i'\in\cS'}
 \uH(\WW_{i'}|\XX_{i'})
 \notag
 \\
 &=
 \sum_{i'\in\cS'}
 \uH(\WW_{i'}|\XX_{\cS})
 \notag
 \\
 &=
 \sum_{l=1}^{|\cS'|}
 \uH(\WW_{i_l}|\WW_{\{i_1,\ldots,i_{l-1}\}},\XX_{\cS})
 \notag
 \\
 &\leq
 \uH(\WW_{\cS'}|\XX_{\cS})
\end{align}
for all $\cS\in\fS$ and $\cS'\in2^{\cS}\setminus\{\emptyset\}$,
where the second inequality comes from (\ref{eq:RCRNG-DSC-ri}),
the first equality comes from the fact that 
$i'\in\cS'\subset\cS\in\fS$ implies $\XX_{i'}=\XX_{\cS}$,
the second equality comes from (\ref{eq:RIT-DSC-markov-encoder}),
and the last inequality comes from
Lemma \ref{lem:oH(UU'|V)<oH(U'|UV)+oH(U|V)} in Appendix \ref{sec:ispec}.
The relation (\ref{eq:RCRNG-MDC-sum[ri+Ri]}) is shown 
immediately from (\ref{eq:RCRNG-DSC-sum[ri+Ri]}).

Let $\cS\equiv\{i_1,\ldots,i_{|\cS|}\}$,
where the order is arbitrary.
Then the Markov chain (\ref{eq:RCRNG-MDC-markov-encoder})
is shown from the fact that
\begin{align}
 0
 &\leq
 I(W^n_{\cS};W^n_{\I\setminus\cS},X^n_{\I\setminus\cS},Y^n_{\J}
	|X^n_{\cS})
 \notag
 \\
 &=
 \sum_{l=1}^{|\cS|}
 I(W^n_{i_l};W^n_{\I\setminus\cS},X^n_{\I\setminus\cS},Y^n_{\J}
	|W^n_{\{i_1,\ldots,i_{l-1}\}},X^n_{\cS})
 \notag
 \\
 &\leq
 \sum_{l=1}^{|\cS|}
 I(W^n_{i_l};W^n_{\I\setminus\{i_l\}},X^n_{\I\setminus\{i_l\}},Y^n_{\J}
	|X^n_{i_l})
 \notag
 \\
 &=
 0,
\end{align}
where
the first equality comes from the chain rule
of conditional mutual information,
the next inequality comes from the relation
$I(U;U'|V,V')\leq I(U;U',V'|V)$,
and the last equality comes from (\ref{eq:RIT-DSC-markov-encoder}).
The Markov chain (\ref{eq:RCRNG-MDC-markov-decoder}) is shown
immediately from (\ref{eq:RIT-DSC-markov-decoder}).
From the above observations, we have $\RCRNGDSC\subset\RCRNGMDC$.
\hfill\IEEEQED

\section{Code Construction}
\label{sec:construction}

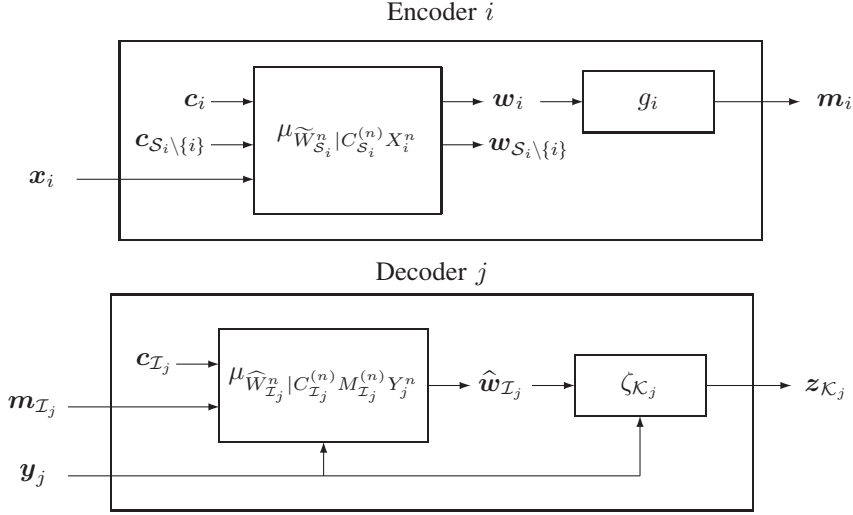
\begin{figure}[t]
\begin{center}
 \unitlength 0.57mm
 \begin{picture}(186,60)(0,0)
	\put(94,53){\makebox(0,0){Encoder $i$}}
	\put(20,0){\framebox(148,46){}}
	\put(40,32){\makebox(0,0)[r]{$\cc_i$}}
	\put(41,32){\vector(1,0){10}}
	\put(40,22){\makebox(0,0)[r]{$\cc_{\cS_i\setminus\{i\}}$}}
	\put(41,22){\vector(1,0){10}}
	\put(2,14){\makebox(0,0){$\xx_i$}}
	\put(10,14){\vector(1,0){41}}
	\put(51,6){\framebox(43,34){$\mu_{\tW^n_{\cS_i}|C^{(n)}_{\cS_i}X^n_i}$}}
	\put(106,32){\makebox(0,0)[l]{$\ww_i$}}
	\put(94,32){\vector(1,0){10}}
	\put(105,21){\makebox(0,0)[l]{$\ww_{\cS_i\setminus\{i\}}$}}
	\put(94,22){\vector(1,0){10}}
	\put(117,32){\vector(1,0){10}}
	\put(127,25){\framebox(30,14){$g_i$}}
	\put(157,32){\vector(1,0){20}}
	\put(185,32){\makebox(0,0){$\mm_i$}}
 \end{picture}
 \\
 \begin{picture}(186,65)(2,0)
	\put(94,57){\makebox(0,0){Decoder $j$}}
	\put(20,2){\framebox(148,50){}}
	\put(30,36){\makebox(0,0){$\cc_{\I_j}$}}
	\put(35,36){\vector(1,0){10}}
	\put(2,26){\makebox(0,0){$\mm_{\I_j}$}}
	\put(10,26){\vector(1,0){35}}
	\put(2,10){\makebox(0,0){$\yy_j$}}
	\put(10,10){\line(1,0){132}}
	\put(69,10){\vector(0,1){8}}
	\put(142,10){\vector(0,1){14}}
	\put(45,18){\framebox(48,26){
		$\mu_{\hW^n_{\I_j}|C^{(n)}_{\I_j}M^{(n)}_{\I_j}Y^n_j}$
	}}
	\put(93,31){\vector(1,0){10}}
	\put(110,31){\makebox(0,0){$\hww_{\I_j}$}}
	\put(117,31){\vector(1,0){10}}
	\put(127,24){\framebox(30,14){$\zeta_{\K_j}$}}
	\put(157,31){\vector(1,0){20}}
	\put(185,30){\makebox(0,0){$\zz_{\K_j}$}}
 \end{picture}
\end{center}
\caption{Construction of Source Code}
\label{fig:crng-code}
\end{figure}

This section introduces a source code
based on an idea drawn from \cite{CRNG,CRNGVLOSSY,CoCoNuTS-LOSSY}.
The code construction is illustrated in Fig.~\ref{fig:crng-code}.

For each $i\in\I$, let us introduce set $\C^{(n)}_i$
and function $f_i:\W^n_i\to\C^{(n)}_i$,
where the dependence of $f_i$ on $n$ is omitted.
For each $i\in\I$, let us introduce set
$\M^{(n)}_i$ and function $g_i:\W^n_i\to\M^{(n)}_i$,
where the dependence of $g_i$ on $n$ is omitted.
We can use sparse matrices as functions $f_i$ and $g_i$
by assuming that $\W_i^n$, $\C^{(n)}_i$, and $\M^{(n)}_i$
are linear spaces on the same finite field.

We define here a constrained-random number generator
used by the encoders, whose indexes belong to $\cS$,
to generate $\ww_{\cS}\in\W_{\cS}^n$.
For given $\xx_{\cS}$ and $\cc_{\cS}$,
let $\tW^n_{\cS}$ be a random variable corresponding to the distribution
\begin{align}
 \mu_{\tW^n_{\cS}|C^{(n)}_{\cS}X^n_{\cS}}(\ww_{\cS}|\cc_{\cS},\xx_{\cS})
 &\equiv
 \frac{
	\mu_{W^n_{\cS}|X^n_{\cS}}(\ww_{\cS}|\xx_{\cS})
	\chi(
	 f_{\cS}(\ww_{\cS})=\cc_{\cS}
	)
 }{
	\sum_{\ww_{\cS}}
	\mu_{W^n_{\cS}|X^n_{\cS}}(\ww_{\cS}|\xx_{\cS})
	\chi(
	 f_{\cS}(\ww_{\cS})=\cc_{\cS}
	)
 }.
 \label{eq:tWi}
\end{align}
The $i$-th encoder generates $\ww_{\cS_i}$
by use of this constrained-random number generator
and obtains vector $\ww_i$, which is a member of $\ww_{\cS_i}$.
We define stochastic encoding function
$\Phi^{(n)}_i:\X_i^n\to\M^{(n)}_i$ as
\begin{equation*}
 \Phi^{(n)}_i(\xx_i)
 \equiv
 g_i(\ww_i),
\end{equation*}
where the encoder claims an error when
the numerator of the righthand side of (\ref{eq:tWi}) is zero.
By using the technique described in \cite[Section VI]{CRNG},
we can represent
\begin{equation*}
 \Phi^{(n)}_i(\xx_i)
 =
 \vphi^{(n)}_i(\xx_i,B_i)
\end{equation*}
by using the deterministic function $\vphi^{(n)}_i$ and randomness $B_i$,
which is independent of $(X^n_{\I},Y^n_{\J},B_{\I\setminus\{i\}})$.
In the following, let $\mm_i$
be the codeword generated by the $i$-th encoder.

We define the constrained-random number generator used by the $j$-th decoder.
The $j$-th decoder generates $\hww_{\I_j}$
by using a constrained-random number generator with a distribution
given as
\begin{align*}
 \mu_{\hW^n_{\I_j}|C^{(n)}_{\I_j}M^{(n)}_{\I_j}Y^n_j}
 (\hww_{\I_j}|\cc_{\I_j},\mm_{\I_j},\yy_j)
 &\equiv
 \frac{
	\mu_{W_{\I_j}^n|Y^n_j}(\hww_{\I_j}|\yy_j)
	\chi((f,g)_{\I_j}(\hww_{\I_j})=(\cc_{\I_j},\mm_{\I_j}))
 }{
	\sum_{\hww_{\I_j}}
	\mu_{W_{\I_j}^n|Y^n_j}(\hww_{\I_j}|\yy_j)
	\chi((f,g)_{\I_j}(\hww_{\I_j})=(\cc_{\I_j},\mm_{\I_j}))
 }
\end{align*}
for given $\cc_{\I_j}$, $\mm_{\I_j}$, and $\yy_j$,
where
\begin{align*}
 \mu_{W_{\I_j}^n|Y^n_j}(\ww_{\I_j}|\yy_j)
 &\equiv
 \frac{
	\sum_{
	 \xx_{\I},
	 \ww_{\I\setminus\I_j}
	}
	\mu_{X^n_{\I}Y^n_j}(\xx_{\I},\yy_j)
	\prod_{\cS\in\fS}
	\mu_{W^n_{\cS}|X^n_{\cS}}(\ww_{\cS}|\xx_{\cS})}
 {
	\sum_{\xx_{\I}}
	\mu_{X^n_{\I}Y^n_j}(\xx_{\I},\yy_j)
 }.
\end{align*}

It should be noted that
we can also use either the maximum a posteriori probability decoder
or the typical-set decoder
instead of the constrained-random number generator.
We define the decoding function
$\Psi^{(n)}_j:\Prod_{i\in\I_j}\M^{(n)}_i\times\Y^n_j\to\Z^n_{\K_j}$
as
\begin{equation*}
 \Psi^{(n)}_j(\mm_{\I_j},\yy_j)
 \equiv
 \{\zeta^{(n)}_k(\hww_{\I_j},\yy_j)\}_{k\in\K_j}
\end{equation*}
by using functions
$\{\zeta^{(n)}_k:\W^n_{\I_j}\times\Y^n_j\to\Z^n_k\}_{k\in\K_j}$.
By using the technique described in \cite[Section VI]{CRNG},
we can represent
\begin{equation*}
 \Psi^{(n)}_j(\mm_{\I_j},\yy_j)
 =
 \psi^{(n)}_j(\mm_{\I_j},\yy_j,B_j)
\end{equation*}
for given deterministic function $\psi^{(n)}_j$ and randomness $B_j$,
which is independent of $(X^n_{\I},Y^n_{\J},B_{\I},B_{\J\setminus\{j\}})$.

Let
\begin{align}
 r_i
 &\equiv
 \frac{\log|\C^{(n)}_i|}n
 \label{eq:ri}
 \\
 R_i
 &\equiv
 \frac{\log|\M^{(n)}_i|}n
 \label{eq:Ri}
\end{align}
for each $i\in\I$,
where $R_i$ represents the transmission rate of the $i$-th encoder.
Let $\bb\equiv(\bb_{\I},\bb_{\J})$
be the output of the source $(B_{\I},B_{\J})$
used by the constrained-random number generators
and
$\hZ^n_k
\equiv
\psi_k^{(n)}(\{\vphi_i^{(n)}(X^n_i,\bb_i)\}_{i\in\I_j},Y^n_j,\bb_j)$.
For a given $\delta>0$, let $\Error(f_{\I},g_{\I},\cc_{\I},\bb)$ 
be the error probability defined as
\begin{align*}
 \Error(f_{\I},g_{\I},\cc_{\I},\bb)
 &\equiv
 \Prob\lrsb{
	d^{(n)}_j(X^n_{\I},Y^n_{\J},\hZ^n_k)
	> D_k+\delta
	\ \text{for some}\ k\in\K
 }.
\end{align*}
We introduce the following theorem,
which implies the achievability part $\RCRNGMDC\subset\ROP$
by taking closure.

\begin{thm}
\label{thm:crng}
For a given set of general correlated sources $(\XX_{\I},\YY_{\J})$
and rate-distortion pair $(R_{\I},D_{\K})$,
let us assume that general sources $\WW_{\I}$,
functions
$\{\{\zeta^{(n)}_k
	:\W^n_{\I_j}\times\Y^n_j\to\Z^n_k\}_{j\in\J,k\in\K_j}
 \}_{n=1}^{\infty}$,
and numbers $\{(r_i,R_i,)\}_{i\in\I}$ and $\{D_k\}_{k\in\K}$
satisfy
\begin{align}
 0
 \leq
 \sum_{i\in\cS'} r_i
 &<
 \uH(\WW_{\cS'}|\XX_{\cS})
 \label{eq:crng-sum[ri]}
 \\
 \sum_{i\in\I'_j}
 [R_i+r_i]
 &>
 \oH(\WW_{\I'_j}|\WW_{\Ipcj},\YY_j)
 \label{eq:crng-sum[Ri+ri]}
\end{align}
for all $\cS\in\fS$, $\cS'\in2^{\cS}\setminus\{\emptyset\}$,
$j\in\J$,
$\I'_j\in2^{\I_j}\setminus\{\emptyset\}$,
and (\ref{eq:RIT-DSC-D})
for all $k\in\K$ and $\delta>0$,
where the joint distribution of $(W^n_{\I},X^n_{\I},Y^n_{\J},Z^n_{\K})$
is given by
\begin{align}
 \mu_{W^n_{\I}X^n_{\I}Y^n_{\J}Z^n_{\K}}
 (\ww_{\I},\xx_{\I},\yy_{\J},\zz_{\K})
 &\equiv\!\!
 \lrB{
	\prod_{j\in\J}
	\prod_{k\in\K_j}
	\chi(\zz_k=\zeta^{(n)}_k(\ww_{\I_j},\yy))
 }\!\!
 \lrB{
	\prod_{\cS\in\fS}
	\mu_{W^n_{\cS}|X^n_{\cS}}(\ww_{\cS}|\xx_{\cS})
 }
 \mu_{X^n_{\I}Y^n_{\J}}(\xx_{\I},\yy_{\J}),
 \label{eq:crng-markov}
\end{align}
which is equivalent to the Markov conditions
(\ref{eq:RCRNG-MDC-markov-encoder}) and (\ref{eq:RCRNG-MDC-markov-decoder}).
Then for all $\delta>0$ and all sufficiently large $n$
there are functions (sparse matrices)
$f_{\I}\equiv\{f_i\}_{i\in\I}$ and $g_{\I}\equiv\{g_i\}_{i\in\I}$,
and vectors $cc_{\I}\equiv\{\cc_i\}_{i\in\I}$
and $\bb\equiv(\bb_{\I},\bb_{\J})$
such that
$\Error(f_{\I},g_{\I},\cc_{\I},\bb)<\delta$.
\end{thm}

\begin{rem}
\label{rem:interpretation}
Here, let us explain the interpretation of conditions 
(\ref{eq:crng-sum[ri]}) and (\ref{eq:crng-sum[Ri+ri]}). 
From (\ref{eq:crng-markov}),
the righthand side of (\ref{eq:crng-sum[ri]})
can be replaced by $\uH(\WW_{\cS'}|\XX_{\I},\YY_{\J})$.
Then the condition (\ref{eq:crng-sum[ri]})
represents the limit of the randomness
of source $\WW_{\I}$, where the randomness
is independent of the given source $(\XX_{\I},\YY_{\J})$.
Since the rate of $\cc_i\equiv f_i(\ww_i)$ is $r_i$,
which satisfies (\ref{eq:crng-sum[ri]}),
encoders and the decoder can share
the constant vectors $\cc_{\I}$,
which are generated independent of $(\xx_{\I},\yy_{\J})$.
Condition (\ref{eq:crng-sum[Ri+ri]})
represents the Slepian-Wolf region \cite{C75,SW73}
of the $j$-th decoder reproducing
the correlated sources $\WW_{\I_j}$
with decoder side information $\YY_j$,
where the encoding rate of source $\WW_i$
is reduced by $r_i$.
It should be noted that we consider Slepian-Wolf source codes
(hash property and constrained-random number generators)
as building blocks for code construction,
while codes for symmetric channel are considered
as building blocks in \cite{GGWK24}.
\end{rem}

Next, we consider 
the formulation by Jana and Blahut
discussed in Section \ref{sec:jb},
where some sources are reproduced without distortion.
To apply Theorem~\ref{thm:crng},
we assume that $\fS\equiv\{\{i\}: i\in\I\}$, $|\J|=1$,
and the set $\K$ in Theorem \ref{thm:crng} is divided into two disjoint
sets, $\I_0$ and $\K\setminus\I_0$,
where $\K\setminus\I_0$ corresponds to index set of the lossy reproductions.
We use the following definitions:
\begin{align*}
 W^n_i
 &\equiv
 X^n_i
 \\
 |\C_i^{(n)}|
 &\equiv
 1
 \\
 d_i^{(n)}(\xx_{\I},\yy,\zz_i)
 &\equiv
 \chi(\xx_i\neq \zz_i)
 \\
 \zeta_i^{(n)}(\ww_{\I},\yy)
 &\equiv
 \ww_i
\end{align*}
for each $i\in\I_0$ and $n\in\NN$.
Then we have
\begin{align*}
 \uH(\WW_i|\XX_i)
 &=0
 \\
 r_i
 &=0
 \\
 \tW^n_i
 &=
 X^n_i
 \\
 W^n_{\I}
 &=
 (W^n_{\I\setminus\I_0},X^n_{\I_0})
\end{align*}
for all $i\in\I_0$ and $n\in\NN$.
We also have the fact that
$C^{(n)}_i=f_i(W^n_i)$ is a constant
and (\ref{eq:RIT-DSC-D}) is satisfied for all $i\in\I_0$.
From Theorem \ref{thm:crng}, we have the fact that
there a valid code exists
when $\{r_i\}_{i\in\I\setminus\I_0}$ and $\{R_i\}_{i\in\I}$ satisfy
(\ref{eq:crng-sum[ri]}) for all $i\in\I$ and $\cS'\equiv\{i\}$,
and (\ref{eq:crng-sum[Ri+ri]})
for all $\I'\in 2^{\I}\setminus\{\emptyset\}$.
Since $r_i=0$ and $\uH(\WW_i|\XX_i)=0$ for all $i\in\I_0$,
conditions (\ref{eq:crng-sum[ri]}) and (\ref{eq:crng-sum[Ri+ri]})
are reduced to
\begin{align*}
 0
 \leq
 r_i
 &<
 \uH(\WW_i|\XX_i)
 \\
 \sum_{i\in\I'\cap\I_0}
 R_i
 +
 \sum_{i\in\I'\setminus\I_0}
 [R_i+r_i]
 &>
 \oH(\WW_{\I'\setminus\I_0},\XX_{\I'\cap\I_0}
	|\WW_{\Ipc\setminus\I_0},\XX_{\Ipc\cap\I_0},\YY)
\end{align*}
for all $i\in\I\setminus\I_0$ and $\I'\in2^{\I}\setminus\{\emptyset\}$,
which are equivalent to
\begin{equation*}
 \sum_{i\in\I'}
 R_i
 >
 \oH(\WW_{\I'\setminus\I_0},\XX_{\I'\cap\I_0}
	|\WW_{\Ipc\setminus\I_0},\XX_{\Ipc\cap\I_0},\YY)
 -
 \sum_{i\in\I'\setminus\I_0}
 \uH(\WW_i|\XX_i)
\end{equation*}
by eliminating $\{r_i\}_{i\in\I\setminus\I_0}$
using the Fourier-Motzkin method \cite[Appendix E]{EK11}.
Thus, by taking closure,
we have achievability $\RCRNGJB\subset\ROPJB$.

\section{Proof of Theorem~\ref{thm:crng}}
\label{sec:proof-crng}

In this section,
we use lemmas on the hash property introduced in 
\cite{CRNG,CRNGVLOSSY,HASH,HASH-BC,HASH-WTC,CRNG-MULTI,CRNG-CHANNEL};
a similar idea (for the special case of the random binning)
is found in \cite{YAG12}.
The definitions and lemmas used in the proof are introduced
in Appendices \ref{sec:crng-expectation}--\ref{sec:lossy-crp}.

We define
\begin{align*}
 \fC_{f_i}(\cc_i)
 &\equiv
 \{\ww_i: f_i(\ww_i)=\cc_i\}
 \\
 \fC_{f_{\I}}(\cc_{\I})
 &\equiv
 \{\ww_{\I}: f_i(\ww_i)=\cc_i\ \text{for all}\ i\in\I\}
 \\
 \fC_{(f,g)_{\I}}(\cc_{\I},\mm_{\I})
 &\equiv
 \{\ww_{\I}: f_i(\ww_i)=\cc_i, g_i(\ww_i)=\mm_i\ \text{for all}\ i\in\I\}
\end{align*}
for
$\cc_{\I}\equiv\{\cc_i\}_{i\in\I}$
and $\mm_{\I}\equiv\{\mm_i\}_{i\in\I}$,
where $(f,g)_{\I}(\ww_{\I})\equiv\{(f_i(\ww_i),g_i(\ww_i))\}_{i\in\I}$.
In the following, we use the following definitions:
\begin{align*}
 \E(f_{\cS},\cc_{\cS})
 &\equiv
 \lrb{
	\xx_{\cS}:
	\mu_{W_{\cS}|X_{\cS}}(\fC_{f_{\cS}}(\cc_{\cS})|\xx_{\cS})=0
 }
 \\
 \E(f_{\I},\cc_{\I})
 &\equiv
 \lrb{
	\xx_{\I}:
	\xx_{\cS}\in\E(f_{\cS},\cc_{\cS})\ \text{for some}\ \cS\in\fS
 }
 \\
 \E(D_{\K})
 &\equiv
 \lrb{
	(\xx_{\I},\yy_{\J},\zz_{\K}):
	d_k^{(n)}(\xx_{\I},\yy_{\J},\zz_k)>D_k+\delta
	\ \text{for some}\ k\in\K
 }
 \\
 \mu_{X_{\I}}
 (\xx_{\I})
 &\equiv
 \sum_{\yy_{\J}}
 \mu_{X_{\I}Y_{\J}}
 (\xx_{\I},\yy_{\J})
 \\
 \mu_{\tW_{\I}|X_{\I}C_{\I}}
 (\ww_{\I}|\xx_{\I},\cc_{\I})
 &\equiv
 \prod_{\cS\in\fS}
 \frac{
	\mu_{W_{\cS}|X_{\cS}}(\ww_{\cS}|\xx_{\cS})
	\chi(\cc_{\cS}=f_{\cS}(\ww_{\cS}))
 }
 {\mu_{W_{\cS}|X_{\cS}}(\fC_{f_{\cS}}(\cc_{\cS})|\xx_{\cS})}
 \\
 \mu_{\hW_{\I}|C_{\I}M_{\I}Y_{\J}}
 (\hww_{\I}|\cc_{\I},\mm_{\I},\yy_{\J})
 &\equiv
 \prod_{j\in\J}
 \frac{
	\mu_{W_{\I_j}|Y_j}(\hww_{\I_j}|\yy_j)
	\chi(\cc_{\I_j}=f_{\I_j}(\hww_{\I_j}),\mm_{\I_j}=g_{\I_j}(\hww_{\I_j}))
 }
 {\mu_{W_{\I_j}|Y_j}(\fC_{(f,g)_{\I_j}}(\cc_{\I_j},\mm_{\I_j})|\yy_j)}
 \\
 \mu_{Z_{\K_j}|W_{\I_j}Y_j}(\zz_{\K_j}|\ww_{\I_j},\yy_j)
 &\equiv
 \chi(\zz_k=\eta^{(n)}_k(\ww_{\I_j},\yy_j))
 \\
 \mu_{Z_{\K}|W_{\I}Y_{\J}}(\zz_{\K}|\ww_{\I},\yy_{\J})
 &\equiv
 \prod_{j\in\J}
 \prod_{k\in\K_j}
 \mu_{Z_{\K_j}|W_{\I_j}Y_j}(\zz_{\K_j}|\ww_{\I_j},\yy_j)
 \\
 \mu_{W_{\I}|X_{\I}}
 (\ww_{\I}|\xx_{\I})
 &\equiv
 \prod_{\cS\in\fS}
 \mu_{W_{\cS}|X_{\cS}}(\ww_{\cS}|\xx_{\cS})
 \\
 \mu_{W_{\I}X_{\I}Y_{\J}}
 (\ww_{\I},\xx_{\I},\yy_{\J})
 &\equiv
 \mu_{W_{\I}|X_{\I}}(\ww_{\I}|\xx_{\I})
 \mu_{X_{\I}Y_{\J}}(\xx_{\I},\yy_{\J})
 \\
 \mu_{X_{\I}Y_{\J}Z_{\K}}(\xx_{\I},\yy_{\J},\zz_{\K})
 &\equiv
 \mu_{X_{\I}Y_{\J}}(\xx_{\I},\yy_{\J})
 \sum_{\ww_{\I}}
 \mu_{Z_{\K}|W_{\I}Y_{\J}}(\zz_{\K}|\ww_{\I},\yy_{\J})
 \mu_{W_{\I}|X_{\I}}(\ww_{\I}|\xx_{\I}).
\end{align*}
It should be noted that we can let $\mu_{Z_{\K_j}|W_{\I_j}Y_j}$ 
be an arbitrary probability distribution in the following proof.

We let $B\equiv(B_{\I},B_{\J})$
and assume that $B$ is independent of $(\XX_{\I},\YY_{\J})$.
Since the expectation over random variable $B$
is the expectation of the random variable corresponding to
the constrained-random number generators,
shown in Appendix~\ref{sec:crng-expectation},
we have
\begin{align}
 &
 E_B\lrB{
	\Error(f_{\I},g_{\I},\cc_{\I},B)
 }
 \notag
 \\*
 &\leq
 \sum_{\xx_{\I}\in\E(f_{\I},\cc_{\I})}
 \mu_{X_{\I}}(\xx_{\I})
 \notag
 \\*
 &\quad
 +
 \sum_{\substack{
	 \xx_{\I},\yy_{\J},\ww_{\I},\hww_{\I},\zz_{\K}:\\
	 \xx_{\I}\notin\E(f_{\I},\cc_{\I})\\
	 \ww_{\I}\in\fC_{f_{\I}}(\cc_{\I})\\
	 (\xx_{\I},\yy_{\J},\zz_{\K})\in\E(D_{\K})
 }}
 \mu_{Z_{\K}|W_{\I}Y_{\J}}
 (\zz_{\K}|\hww_{\I},\yy_{\J})
 \mu_{\hW_{\I}|C_{\I}M_{\I}Y_{\J}}
 (\hww_{\I}|\cc_{\I},g_{\I}(\ww_{\I}),\yy_{\J})
 \mu_{\tW_{\I}|C_{\I}X_{\I}}
 (\ww_{\I}|\cc_{\I},\xx_{\I})
 \mu_{X_{\I}Y_{\J}}(\xx_{\I},\yy_{\J})
 \notag
 \\
 &\leq
 \sum_{\cS\in\fS}
 \sum_{\xx_{\cS}\in\E(f_{\cS},\cc_{\cS})}
 \mu_{X_{\cS}}(\xx_{\cS})
 \sum_{\xx_{\I\setminus\cS}}
 \mu_{X_{\I\setminus\cS|\X_{\cS}}}(\xx_{\I\setminus\cS}|\xx_{\cS})
 \notag
 \\
 &\quad
 +
 \sum_{\substack{
	 \xx_{\I},\yy_{\J},\ww_{\I},\hww_{\I},\zz_{\K}:\\
	 \xx_{\I}\notin\E(f_{\I},\cc_{\I})\\
	 \ww_{\I}\in\fC_{f_{\I}}(\cc_{\I})\\
	 (\xx_{\I},\yy_{\J},\zz_{\K})\in\E(D_{\K})
 }}
 \mu_{Z_{\K}|W_{\I}Y_{\J}}
 (\zz_{\K}|\hww_{\I},\yy_{\J})
 \mu_{\hW_{\I}|C_{\I}M_{\I}Y_{\J}}
 (\hww_{\I}|\cc_{\I},g_{\I}(\ww_{\I}),\yy_{\J})
 \notag
 \\*
 &\qquad\qquad\qquad\qquad\qquad\qquad\qquad\qquad\qquad\qquad
 \cdot
 \lrbar{
	\mu_{\tW_{\I}|C_{\I}X_{\I}}
	(\ww_{\I}|\cc_{\I},\xx_{\I})
	-
	\mu_{W_{\I}|X_{\I}}
	(\ww_{\I}|\xx_{\I})
	\prod_{i\in\I}|\im\F_i|
 }
 \mu_{X_{\I}Y_{\J}}(\xx_{\I},\yy_{\J})
 \notag
 \\
 &\quad
 +
 \sum_{\substack{
	 \xx_{\I},\yy_{\J},\ww_{\I},\hww_{\I},\zz_{\K}:\\
	 \xx_{\I}\notin\E(f_{\I},\cc_{\I})\\
	 \ww_{\I}\in\fC_{f_{\I}}(\cc_{\I})\\
	 \hww_{\I}\neq\ww_{\I}
 }}
 \!\!
 \mu_{Z_{\K}|W_{\I}Y_{\J}}
 (\zz_{\K}|\hww_{\I},\yy_{\J})
 \mu_{\hW_{\I}|C_{\I}M_{\I}Y_{\J}}
 (\hww_{\I}|\cc_{\I},g_{\I}(\zz_{\I}),\yy_{\J})
 \mu_{W_{\I}|X_{\I}}
 (\ww_{\I}|\xx_{\I})
 \mu_{X_{\I}Y_{\J}}(\xx_{\I},\yy_{\J})
 \prod_{i\in\I}|\im\F_i|
 \notag
 \\
 &\quad
 +
 \sum_{\substack{
	 \xx_{\I},\yy_{\J},\ww_{\I},\hww_{\I},\zz_{\K}:\\
	 \xx_{\I}\notin\E(f_{\I},\cc_{\I})\\
	 \ww_{\I}\in\fC_{f_{\I}}(\cc_{\I})\\
	 \hww_{\I}=\ww_{\I}\\
	 (\xx_{\I},\yy_{\J},\zz_{\K})\in\E(D_{\K})
 }}
 \!\!
 \mu_{Z_{\K}|W_{\I}Y_{\J}}
 (\zz_{\K}|\hww_{\I},\yy_{\J})
 \mu_{\hW_{\I}|C_{\I}M_{\I}Y_{\J}}
 (\hww_{\I}|\cc_{\I},g_{\I}(\ww_{\I}),\yy_{\J})
 \mu_{W_{\I}|X_{\I}}
 (\ww_{\I}|\xx_{\I})
 \mu_{X_{\I}Y}(\xx_{\I},\yy_{\J})
 \prod_{i\in\I}|\im\F_i|,
 \label{eq:proof-crng-dlc-error}
\end{align}
where,
in the first inequality,
we use the fact that
$\ww_{\I}\notin\fC_{f_{\I}}(\cc_{\I})$
implies $\mu_{\tW_{\I}|C_{\I}X_{\I}}(\ww_{\I}|\cc_{\I},\xx_{\I})=0$.

The first term on the righthand side of (\ref{eq:proof-crng-dlc-error})
is evaluated as
\begin{align*}
 E_{C_{\I}}\lrB{
	\text{the first term of (\ref{eq:proof-crng-dlc-error})}
 }
 &=
 \sum_{\cS\in\fS}
 \sum_{\substack{
	 \xx_{\cS},\cc_{\cS}:\\
	 \xx_{\cS}\in\E(f_{\cS},\cc_{\cS})
 }}
 \frac{\mu_{X_{\cS}}(\xx_{\cS})}{\prod_{s\in\cS}|\im\F_s|}.
\end{align*}

The second term on the righthand side of (\ref{eq:proof-crng-dlc-error})
is evaluated as
\begin{align}
 &
 E_{C_{\I}}\lrB{
	\text{the second term of (\ref{eq:proof-crng-dlc-error})}
 }
 \notag
 \\*
 &\leq
 \sum_{\substack{
	 \xx_{\I},\ww_{\I},\cc_{\I}:\\
	 \xx_{\I}\notin\E(f_{\I},\cc_{\I})\\
	 \ww_{\I}\in\fC_{f_{\I}}(\cc_{\I})
 }}
 \lrbar{
	\frac{
	 \mu_{\tW_{\I}|C_{\I}X_{\I}}
	 (\ww_{\I}|\cc_{\I},\xx_{\I})
	}{\prod_{\cS\in\fS}\prod_{i\in\cS}|\im\F_i|}
	-
	\mu_{W_{\I}|X_{\I}}
	(\ww_{\I}|\xx_{\I})
 }
 \mu_{X_{\I}}(\xx_{\I})
 \notag
 \\
 &=
 \sum_{\substack{
	 \xx_{\I},\ww_{\I},\cc_{\I}:\\
	 \xx_{\I}\notin\E(f_{\I},\cc_{\I})\\
	 \ww_{\I}\in\fC_{f_{\I}}(\cc_{\I})\\
 }}
 \lrbar{
	\prod_{\cS\in\fS}
	\frac{
	 1
	}{
	 \mu_{W_{\cS}|X_{\cS}}(\fC_{f_{\cS}}(\cc_{\cS})|\xx_{\cS})
	 \prod_{i\in\cS}|\im\F_i|
	}
	-
	1
 }
 \mu_{W_{\I}|X_{\I}}(\ww_{\I}|\xx_{\I})
 \mu_{X_{\I}}(\xx_{\I})
 \notag
 \\
 &\leq
 \sum_{\substack{
	 \xx_{\I},\ww_{\I},\cc_{\I}:\\
	 \xx_{\I}\notin\E(f_{\I},\cc_{\I})\\
	 \ww_{\I}\in\fC_{f_{\I}}(\cc_{\I})
 }}
 \mu_{W_{\I}|X_{\I}}
 (\ww_{\I}|\xx_{\I})
 \mu_{X_{\I}}(\xx_{\I})
 \sum_{l=1}^{|\fS|}
 \lrbar{
	\frac1{
	 \mu_{W_{\cS_l}|X_{\cS_l}}(\fC_{f_{\cS_l}}(\cc_{\cS_l})|\xx_{\cS_l})
	 \prod_{i\in\cS_l}|\im\F_i|
	}
	-
	1
 }
 \notag
 \\*
 &\qquad\qquad\qquad\qquad\qquad\qquad\qquad\qquad\qquad\quad
 \cdot
 \prod_{l'=l+1}^{|\fS|}
 \frac1{
	\mu_{W_{\cS_{l'}}|X_{\cS_{l'}}}
	(\fC_{f_{\cS_{l'}}}(\cc_{\cS_{l'}})|\xx_{\cS_{l'}})
	\prod_{i\in\cS_{l'}}|\im\F_i|
 }
 \notag
 \\*
 &\leq
 \sum_{l=1}^{|\fS|}
 \sum_{\substack{
	 \xx_{\cS_l},\ww_{\cS_l},\cc_{\cS_l}:\\
	 \xx_{\cS_l}\notin\E(f_{\cS_l},\cc_{\cS_l})\\
	 \ww_{\cS_l}\in\fC_{f_{\cS_l}}(\cc_{\cS_l})
 }}
 \mu_{W_{\cS_l}|X_{\cS_l}}
 (\ww_{\cS_l}|\xx_{\cS_l})
 \mu_{X_{\cS_l}}(\xx_{\cS_l})
 \lrbar{
	\frac1{
	 \mu_{W_{\cS_l}|X_{\cS_l}}(\fC_{f_{\cS_l}}(\cc_{\cS_l})|\xx_{\cS_l})
	 \prod_{i\in\cS_l}|\im\F_i|
	}
	-
	1
 }
 \notag
 \\
 &=
 \sum_{\cS\in\fS}
 \sum_{\substack{
	 \xx_{\cS},\cc_{\cS}:\\
	 \xx_{\cS}\notin\E(f_{\cS},\cc_{\cS})
 }}
 \mu_{X_{\cS}}(\xx_{\cS})
 \lrbar{
	\frac1
	{\prod_{i\in\cS}|\im\F_i|}
	-
	\mu_{W_{\cS}|X_{\cS}}(\fC_{f_{\cS}}(\cc_{\cS})|\xx_{\cS})
 }
 \notag
 \\
 &=
 \sum_{\cS\in\fS}
 \sum_{\substack{
	 \xx_{\cS},\cc_{\cS}
 }}
 \mu_{X_{\cS}}(\xx_{\cS})
 \lrbar{
	\frac1
	{\prod_{i\in\cS}|\im\F_i|}
	-
	\mu_{W_{\cS}|X_{\cS}}(\fC_{f_{\cS}}(\cc_{\cS})|\xx_{\cS})
 }
 -
 \sum_{\cS\in\fS}
 \sum_{\substack{
	 \xx_{\cS},\cc_{\cS}:\\
	 \xx_{\cS}\in\E(f_{\cS},\cc_{\cS})
 }}
 \frac{\mu_{X_{\cS}}(\xx_{\cS})}
 {\prod_{i\in\cS}|\im\F_i|}
\end{align}
by letting $\fS\equiv\{\cS_1,\cS_2,\ldots,\cS_{|\fS|}\}$,
where the second inequality comes from Lemma \ref{lem:diff-prod}
in Appendix \ref{sec:proof-lemma},
the third inequality comes from the fact that
\begin{align}
 &
 \sum_{\substack{
	 \ww_{\I\setminus\cS_l},\cc_{\I\setminus\cS_l}:\\
	 \xx_{\cS'}\notin\E(f_{\cS'},\cc_{\cS'})\ \text{for all}\ \cS'\in\fS\setminus\{\cS_l\}\\
	 \ww_{\I\setminus\cS_l}
	 \in\fC_{f_{\I\setminus\cS_l}}(\cc_{\I\setminus\cS_l})
 }}
 \lrB{
	\prod_{\cS'\in\fS\setminus\{\cS_l\}}
	\mu_{W_{\cS'}|X_{\cS'}}(\ww_{\cS'}|\xx_{\cS'})
 }
 \prod_{l'=l+1}^{|\fS|}
 \frac1{
	\mu_{W_{\cS_{l'}}|X_{\cS_{l'}}}
	(\fC_{f_{\cS_{l'}}}(\cc_{\cS_{l'}})|\xx_{\cS_{l'}})
	\prod_{i\in\cS_{l'}}|\im\F_i|
 }
 \notag
 \\*
 &\leq
 \lrB{
	\prod_{l'=1}^{l-1}
	\sum_{\substack{
		\ww_{\cS_{l'}},\cc_{\cS_{l'}}:\\
		\ww_{\cS_{l'}}\in\fC_{f_{\cS_{l'}}}(\cc_{\cS_{l'}})
	}}
	\mu_{W_{\cS_{l'}}|X_{\cS_{l'}}}(\ww_{\cS_{l'}}|\xx_{\cS_{l'}})
 }
 \lrB{
	\prod_{l'=l+1}^{|\fS|}
	\sum_{\substack{
		\ww_{\cS_{l'}},\cc_{\cS_{l'}}:\\
		\xx_{\cS_{l'}}\notin\E(f_{\cS_{l'}},\cc_{\cS_{l'}})\\
		\ww_{\cS_{l'}}\in\fC_{f_{\cS_{l'}}}(\cc_{\cS_{l'}})
	}}
	\frac{\mu_{W_{\cS_{l'}}|X_{\cS_{l'}}}(\ww_{\cS_{l'}}|\xx_{\cS_{l'}})
	}{
	 \mu_{W_{\cS_{l'}}|X_{\cS_{l'}}}
	 (\fC_{f_{\cS_{l'}}}(\cc_{\cS_{l'}})|\xx_{\cS_{l'}})
	 \prod_{i\in\cS_{l'}}|\im\F_i|}
 }
 \notag
 \\
 &\leq
 \prod_{l'=l+1}^{|\fS|}
 \sum_{\cc_{\cS_{l'}}}
 \frac1
 {\prod_{i\in\cS_{l'}}|\im\F_i|}
 \notag
 \\
 &=
 1,
\end{align}
and the last inequality comes from the fact that
$\mu_{W_{\cS}|X_{\cS}}(\fC_{f_{\cS}}(\cc_{\cS})|\xx_{\cS})=0$
when $\xx_{\cS}\in\E(f_{\cS},\cc_{\cS})$.

The third term on the righthand side of
(\ref{eq:proof-crng-dlc-error})
is evaluated as
\begin{align}
 &
 E_{C_{\I}}
 \lrB{
	\text{the third term of (\ref{eq:proof-crng-dlc-error})}
 }
 \notag
 \\*
 &\leq
 \sum_{\substack{
	 \xx_{\I},\yy_{\J},\ww_{\I},\hww_{\I},\zz_{\K},\cc_{\I}:\\
	 \ww_{\I}\in\fC_{f_{\I}}(\cc_{\I})\\
	 \hww_{\I}\neq\ww_{\I}
 }}
 \mu_{Z_{\K}|W_{\I}Y_{\J}}(\zz_{\K}|\hww_{\I},\yy_{\J})
 \mu_{\hW_{\I}|C_{\I}M_{\I}Y_{\J}}
 (\hww_{\I}|f_{\I}(\ww_{\I}),g_{\I}(\ww_{\I}),\yy_{\J})
 \mu_{W_{\I}X_{\I}Y_{\J}}(\ww_{\I},\xx_{\I},\yy_{\J})
 \notag
 \\
 &=
 \sum_{\substack{
	 \ww_{\I},\hww_{\I},\yy_{\J}:\\
	 \hww_{\I}\neq\ww_{\I}
 }}
 \mu_{\hW_{\I}|C_{\I}M_{\I}Y_{\J}}
 (\hww_{\I}|f_{\I}(\ww_{\I}),g_{\I}(\ww_{\I}),\yy_{\J})
 \mu_{W_{\I}Y_{\J}}(\ww_{\I},\yy_{\J})
 \notag
 \\
 &\leq
 \sum_{j\in\J}
 \sum_{\substack{
	 \ww_{\I_j},\hww_{\I_j},\yy_j:\\
	 \hww_{\I_j}\neq\ww_{\I_j}
 }}
 \mu_{\hW_{\I_j}|C_{\I_j}M_{\I_j}Y_j}
 (\hww_{\I_j}|f_{\I_j}(\ww_{\I_j}),g_{\I_j}(\ww_{\I_j}),\yy_j)
 \mu_{W_{\I_j}Y_j}(\ww_{\I_j},\yy_j),
 \label{eq:proof-crng-dlc-error-Z0}
\end{align}
where the last inequality comes from the union bound
with the fact that $\hww_{\I}\neq\ww_{\I}$ iff $\hww_{\I_j}\neq\ww_{\I_j}$ 
for some $j\in\J$.

The fourth term on the righthand side of (\ref{eq:proof-crng-dlc-error})
is evaluated as
\begin{align}
 &
 E_{C_{\I}}
 \lrB{
	\text{the fourth term of (\ref{eq:proof-crng-dlc-error})}
 }
 \notag
 \\*
 &\leq
 \sum_{\substack{
	 \xx_{\I},\yy_{\J},\ww_{\I},\zz_{\K},\cc_{\I}:\\
	 \ww_{\I}\in\fC_{f_{\I}}(\cc_{\I})\\
	 (\xx_{\I},\yy_{\J},\zz_{\K})\in\E(D_{\K})
 }}
 \mu_{Z_{\K}|W_{\I}Y_{\J}}
 (\zz_{\K}|\ww_{\I},\yy_{\J})
 \mu_{\hW_{\I}|C_{\I}M_{\I}Y_{\J}}
 (\ww_{\I}|\cc_{\I},g_{\I}(\ww_{\I}),\yy_{\J})
 \mu_{W_{\I}X_{\I}Y_{\J}}
 (\ww_{\I},\xx_{\I},\yy_{\J})
 \notag
 \\
 &\leq
 \sum_{\substack{
	 \xx_{\I},\yy_{\J},\ww_{\I},\zz_{\K},\cc_{\I}:\\
	 \ww_{\I}\in\fC_{f_{\I}}(\cc_{\I})\\
	 (\xx_{\I},\yy_{\J},\zz_{\K})\in\E(D_{\K})
 }}
 \mu_{Z_{\K}|W_{\I}Y_{\J}}
 (\zz_{\K}|\ww_{\I},\yy_{\J})
 \mu_{W_{\I}X_{\I}Y_{\J}}
 (\ww_{\I},\xx_{\I},\yy_{\J})
 \notag
 \\
 &=
 \sum_{\substack{
	 \xx_{\I},\yy_{\J},\zz_{\K}:\\
	 (\xx_{\I},\yy_{\J},\zz_{\K})\in\E(D_{\K})
 }}
 \mu_{X_{\I}Y_{\J}Z_{\K}}
 (\xx_{\I},\yy_{\J},\zz_{\K}),
 \label{eq:proof-crng-dlc-error-5}
\end{align}
where the second inequality comes from
$\mu_{\hW_{\I}|C_{\I}M_{\I}Y_{\J}}
(\ww_{\I}|\cc_{\I},g_{\I}(\ww_{\I}),\yy_{\J})
\leq 1$.

Finally, from Lemmas \ref{lem:channel-bcp} and \ref{lem:channel-crp}
in Appendices \ref{sec:lossy-bcp} and \ref{sec:lossy-crp}, respectively,
we have
\begin{align}
 &
 E_{(F,G)_{\I}C_{\I}B}\lrB{
	\Error(F_{\I},G_{\I},C_{\I},B)
 }
 \notag
 \\*
 &\leq
 \sum_{\cS\in\fS}
 E_{F_{\cS}}\lrB{
	\sum_{\substack{
		\xx_{\cS},\cc_{\cS}
	}}
	\mu_{X_{\cS}}(\xx_{\cS})
	\lrbar{
	 \frac1
	 {\prod_{i\in\cS}|\im\F_i|}
	 -
	 \mu_{W_{\cS}|X_{\cS}}(\fC_{F_{\cS}}(\cc_{\cS})|\xx_{\cS})
	}
 }
 \notag
 \\*
 &\quad
 +
 \sum_{j\in\J}
 E_{(F,G)_{\I}}\lrB{
	\sum_{\substack{
		\ww_{\I_j},\hww_{\I_j},\yy_j:\\
		\hww_{\I_j}\neq\ww_{\I_j}
	}}
	\mu_{\hW_{\I_j}|C_{\I_j}M_{\I_j}Y_j}
	(\hww_{\I_j}|(F,G)_{\I_j}(\ww_{\I_j}),\yy_j)
	\mu_{W_{\I_j}Y_j}(\ww_{\I_j},\yy_j)
 }
 \notag
 \\*
 &\quad
 +
 \sum_{\substack{
	 \xx_{\I},\yy_{\J},\zz_{\K}:\\
	 (\xx_{\I},\yy_{\J},\zz_{\K})\in\E(D_{\K})
 }}
 \mu_{X_{\I}Y_{\J}Z_{\K}}
 (\xx_{\I},\yy_{\J},\zz_{\K})
 \notag
 \\
 \begin{split}
	&\leq
	\sum_{\cS\in\fS}
	\sqrt{
	 \alpha_{F_{\cS}}-1
	 +
	 \sum_{\cS'\in2^{\cS}\setminus\{\emptyset\}}
	 \alpha_{F_{\cS\setminus\cS'}}
	 [\beta_{F_{\cS'}}+1]2^{-n\ugamma(\cS')}
	}
	+
	2
	\sum_{\cS\in\fS}
	\mu_{W_{\cS}X_{\cS}}(\uT_{W_{\cS}|X_{\cS}}^{\complement})
	\\*
	&\quad
	+2\sum_{j\in\J}
	\sum_{\I_j'\in2^{\I_j}\setminus\{\emptyset\}}
	\alpha_{(F,G)_{\I_j'}}\lrB{\beta_{(F,G)_{\Ipcj}}+1}
	2^{
	 -n\ogamma(\I'_j)
	}
	+2\sum_{j\in\J}
	\beta_{(F,G)_{\I_j}}
	+2\sum_{j\in\J}
	\mu_{W_{\I_j}Y_j}(\oT_{W_{\I_j}|Y_j}^{\complement})
	\\*
	&\quad
	+
	\sum_{k\in\K}
	\Prob\lrsb{
	 d^{(n)}_k(X^n_{\I},Y^n,Z^n_k)>D_k+\delta
	},
 \end{split}
 \label{eq:proof-lossy-error}
\end{align}
where $r_i$ and $R_i$ are defined by (\ref{eq:ri}) and (\ref{eq:Ri}),
respectively, and
\begin{align*}
 \uT_{W_{\cS}|X_{\cS}}
 &\equiv
 \lrb{
	(\ww_{\cS},\xx_{\cS}):
	\frac 1n
	\log_2\frac1{
	 \mu_{W_{\cS'}|X_{\cS}}(\ww_{\cS'}|\xx_{\cS})
	}
	\geq
	\uH(\WW_{\cS'}|\XX_{\cS})-\e
	\ \text{for all}\ \cS'\in2^{\cS}\setminus\{\emptyset\}
 }
 \\
 \oT_{W_{\I_j}|Y_j}
 &\equiv
 \lrb{
	(\ww_{\I_j},\yy_j):
	\begin{aligned}
	 \frac 1n
	 \log_2\frac1{
		\mu_{W_{\I'_j}|W_{\Ipcj}Y_j}(\ww_{\I'_j}|\ww_{\Ipcj},\yy_j)
	 }
	 \leq
	 \oH(\WW_{\I'_j}|\WW_{\Ipcj},\YY_j)+\e
	 \ \text{for all}\ \I'_j\in 2^{\I_j}\setminus\{\emptyset\}
	\end{aligned}
 }
 \\
 \ugamma(\cS')
 &\equiv
 \uH(\WW_{\cS'}|\XX_{\cS})-\sum_{i\in\cS'}r_i-\e
 \\
 \ogamma(\I'_j)
 &\equiv
 \sum_{i\in\I'_j}[r_i+R_i]-\oH(\WW_{\I'_j}|\WW_{\Ipcj},\YY_j)-\e.
\end{align*}
From (\ref{eq:RIT-DSC-D}),
the last term on the righthand side of (\ref{eq:proof-lossy-error})
goes to zero as $n\to\infty$.
From the definitions of limit inferior/superior in probability,
we have the fact that
$\mu_{X_{\cS}W_{\cS}}(\uT_{W_{\cS}|X_{\cS}}^{\complement})\to0$
for all $\cS\in\fS$
and $\mu_{W_{\I_j}Y_j}(\oT_{W_{\I_j}|Y_j}^{\complement})\to0$
for all $j\in\J$
as $n\to\infty$.
By assuming that
$\{(\F_{i,n},p_{F_{i},n})\}_{n=1}^{\infty}$ and
$\{(\G_{i,n},p_{G_{i,n}})\}_{n=1}^{\infty}$
have the hash property (Appendix \ref{sec:hash})
for all $i\in\I$,
we have
$\alpha_{F_{\cS\setminus\cS'}}\to1$,
$\beta_{F_{\cS'}}\to0$,
$\alpha_{(F,G)_{\I'_j}}\to1$,
$\beta_{(F,G)_{\I'_j}}\to0$
as $n\to\infty$
for all $\cS'\in\fS$, $j\in\J$,
and $\I'_j\in2^{\I_j}\setminus\{\emptyset\}$.
Since we can take $\e>0$ to be sufficiently small
so as to satisfy $\ugamma(\cS')>0$
for all $\cS\in\fS$ and $\cS'\subset\cS$,
and $\ogamma(\I'_j)>0$
for all $j\in\J$ and $\I'_j\in2^{\I_j}\setminus\{\emptyset\}$
under the conditions (\ref{eq:crng-sum[ri]}) and (\ref{eq:crng-sum[Ri+ri]}),
we have the fact that for 
all $\delta>0$ and sufficiently large $n$ 
there are functions $\{f_i\}_{i\in\I}$ and $\{g_i\}_{i\in\I}$,
and vectors $\{\cc_i\}_{i\in\I}$ and $\bb$
satisfying $\cc_i\in\im\F_i$
and $\Error(f_{\I},g_{\I},\cc_{\I},\bb)\leq\delta$.
\hfill\IEEEQED

\appendix
\subsection{Information-Spectrum Methods}
\label{sec:ispec}

This section introduces the information spectrum methods
introduced in \cite{HAN,HV93}.
Let $\{U_n\}_{n=1}^{\infty}$ be a general sequence of random variables,
where we do not assume conditions such as stationarity and ergodicity.
The generality of
the discussion does not change regardless of whether we assume or not
that the alphabet of $U_n$ is a Cartesian product,
that is, $\U_n\equiv\U^n$.
We do not assume the consistency of $\{U_n\}_{n=1}^{\infty}$
when $\U_n\equiv\U^n$.

First, we review the definition of the limit superior/inferior in
probability.
For $\{U_n\}_{n=1}^{\infty}$,
the {\em limit superior in probability} $\plimsupn U_n$
and the {\em limit inferior in probability} $\pliminfn U_n$ are defined
as
\begin{align*}
 \plimsupn U_n
 &\equiv 
 \inf\lrb{\theta: \limn \Prob\lrsb{U_n>\theta}=0}
 \\
 \pliminfn U_n
 &\equiv
 \sup\lrb{\theta: \limn \Prob\lrsb{U_n<\theta}=0}.
\end{align*}

In the following, we introduce the key inequalities used in the
proof of the converse part.
We have the following relations~\cite[Section 1.3]{HAN}:
\begin{align}
 \plimsupn U_n
 &\geq\pliminfn U_n
 \label{eq:plimsup>pliminf}
 \\
 \plimsupn\lrB{U_n+V_n}
 &\leq \plimsupn U_n + \plimsupn V_n
 \label{eq:plimsup-upper}
 \\
 \plimsupn\lrB{U_n+V_n}
 &\geq \plimsupn U_n + \pliminfn V_n
 \label{eq:plimsup-lower}
 \\
 \pliminfn\lrB{U_n+V_n}
 &\leq \plimsupn U_n + \pliminfn V_n
 \label{eq:pliminf-upper}
 \\
 \pliminfn\lrB{U_n+V_n}
 &\geq \pliminfn U_n + \pliminfn V_n
 \label{eq:pliminf-lower}
 \\
 \plimsupn U_n
 &=
 -\pliminfn[-U_n].
 \label{eq:plimsup-pliminf}
\end{align}
Here, we show the lemma for a sequence of constant random variables.
\begin{lem}[{\cite[Lemma 1]{ISPEC-CONVERSE}}]
\label{lem:constant}
For a sequence of constant random variables
$\{U_n\}_{n=1}^{\infty}=\{u_n\}_{n=1}^{\infty}$,
we have
\begin{align*}
 \plimsupn U_n
 &= 
 \limsupn u_n
 \\
 \pliminfn U_n
 &= 
 \liminfn u_n.
\end{align*}
\end{lem}
\begin{IEEEproof}
We show the lemma for the completeness of this paper.
It is sufficient to validate the first inequality because
the second equality can be shown by using the first inequality,
(\ref{eq:plimsup-pliminf}), and the relation
\begin{equation}
 \limsupn u_n = - \liminfn[-u_n].
\end{equation}

We have
\begin{align}
 \plimsupn U_n
 &= 
 \inf\lrb{\theta: \limn \Prob\lrsb{U_n>\theta}=0}
 \notag
 \\
 &=
 \inf\lrb{\theta:
	\begin{aligned}
	 &\Prob\lrsb{U_n>\theta}<\e
	 \ \text{for all $\e>0$}\\
	 &\text{and all sufficiently large $n$}
	\end{aligned}
 }
 \notag
 \\
 &=
 \inf\lrb{\theta:
	\begin{aligned}
	 &\Prob\lrsb{U_n>\theta}=0\\
	 &\text{for all sufficiently large $n$}
	\end{aligned}
 }
 \notag
 \\
 &=
 \inf\lrb{\theta:
	u_n\leq \theta\ 
	\text{for all sufficiently large $n$}
 }
 \notag
 \\
 &=
 \limsupn u_n,
\end{align}
where the third and the fourth equalities come from the fact that
$\Prob(U_n>\theta)\in\{0,1\}$
because $U_n=u_n$ with probability $1$.
\end{IEEEproof}

Next, let $\UU\equiv\{U_n\}_{n=1}^{\infty}$
be a general sequence of random variables,
which is called a {\em general source}.
For sequence $\{\mu_{U_n}\}_{n=1}^{\infty}$
of probability distributions corresponding to $\UU$,
we define the {\em spectral sup-entropy rate} $\oH(\UU)$ and
the {\em spectral inf-entropy rate} $\uH(\UU)$ as
\begin{align*}
 \oH(\UU)
 &\equiv
 \plimsupn\frac 1n\log_2\frac1{\mu_{U_n}(U_n)}
 \\
 \uH(\UU)
 &\equiv
 \pliminfn\frac 1n\log_2\frac1{\mu_{U_n}(U_n)}.
\end{align*}
For general sequence $\{\mu_{U_nV_n}\}_{n=1}^{\infty}$ of
the joint probability distributions
corresponding to $(\UU,\VV)=\{(U_n,V_n)\}_{n=1}^{\infty}$,
we define the {\em spectral conditional sup-entropy rate} $\oH(\UU|\VV)$,
the {\em spectral conditional inf-entropy rate} $\uH(\UU|\VV)$,
the {\em spectral sup-information rate} $\oI(\UU;\VV)$,
and the {\em spectral inf-information rate} $\uI(\UU;\VV)$
as
\begin{align*}
 \oH(\UU|\VV)
 &\equiv
 \plimsupn\frac 1n\log_2\frac1{\mu_{U_n|V_n}(U_n|V_n)}
 \\
 \uH(\UU|\VV)
 &\equiv
 \pliminfn\frac 1n\log_2\frac1{\mu_{U_n|V_n}(U_n|V_n)}
 \\
 \oI(\UU;\VV)
 &\equiv
 \plimsupn\frac 1n\log_2\frac{\mu_{U_n|V_n}(U_n|V_n)}{\mu_{U_n}(U_n)}
 \\
 \uI(\UU;\VV)
 &\equiv
 \pliminfn\frac 1n\log_2\frac{\mu_{U_n|V_n}(U_n|V_n)}{\mu_{U_n}(U_n)}.
\end{align*}
It should be noted here that
\begin{gather*}
 \oH(\UU)=\uH(\UU)=H(U)
 \\
 \oH(\UU|\VV)=\uH(\UU|\VV)=H(U|V)
 \\
 \oI(\UU;\VV)=\uI(\UU;\VV)=I(U;V)
\end{gather*}
if $(\UU,\VV)$ is a pair of stationary memoryless sources
with a pair of generic random variables $(U,V)$.

The following lemma is related to the non-negativity of the divergence between two distributions.
\begin{lem}[{\cite[Lemma 3.2.1, Definition 4.1.3]{HAN}}]
\label{lem:pliminf-div}
Let $\{\mu_{U_n}\}_{n=1}^{\infty}$
be a general sequence of probability distributions
corresponding to $\UU\equiv\{U_n\}_{n=1}^{\infty}$.
For each $n$,
let $\nu_n$ be an arbitrary probability distribution on $\U_n$.
Then we have
\begin{equation*}
 \pliminfn\frac 1n\log_2\frac{\mu_{U_n}(U_n)}{\nu_n(U_n)}\geq 0.
\end{equation*}
\end{lem}
\begin{IEEEproof}
For completeness, this paper proves the lemma
by following the proof given by \cite[Lemma 3.2.1, Definition 4.1.3]{HAN}.

For a given $\gamma>0$, we define $\E$ as
\begin{equation*}
 \E
 \equiv
 \lrb{
	\uu:
	\frac1n\log_2\frac{\mu_{U_n}(\uu)}{\nu_n(\uu)}
	\leq -\gamma
 }.
\end{equation*}
Then we have
\begin{align}
 \Prob\lrsb{
	\frac1n\log_2\frac{\mu_{U_n}(U_n)}{\nu_n(U_n)}
	\leq -\gamma
 }
 &=
 \sum_{\uu\in\E}
 \mu_{U_n}(\uu)
 \notag
 \\
 &\leq
 \sum_{\uu\in\E}
 \nu_n(\uu)2^{-n\gamma}
 \notag
 \\
 &\leq
 2^{-n\gamma},
\end{align}
which implies
\begin{equation*}
 \limn\Prob\lrsb{
	\frac1n\log_2\frac{\mu_{U_n}(U_n)}{\nu_n(U_n)}
	\leq -\gamma
 }
 =0.
\end{equation*}
From this inequality and the definition of $\pliminfn$, we have
\begin{equation*}
 \pliminfn\frac 1n\log_2\frac{\mu_{U_n}(U_n)}{\nu_n(U_n)}
 \geq -\gamma.
\end{equation*}
The lemma is verified by letting $\gamma\to0$.
\end{IEEEproof}

We show the following lemmas, which are used in the proof of
the converse theorem.
\begin{lem}
\label{lem:oH>uH>0}
\begin{equation*}
 \oH(\UU|\VV)\geq\uH(\UU|\VV)\geq 0.
 \label{eq:oH>uH>0}
\end{equation*}
\end{lem}
\begin{IEEEproof}
The first inequality comes from (\ref{eq:plimsup>pliminf}).
The second inequality comes from the fact that
$\mu_{U_n|V_n}(U_n|V_n)\leq 1$.
\end{IEEEproof}

\begin{lem}[{\cite[Lemma1]{CRNG-MULTI}}]
\label{lem:bound-by-cardinality}
If $\U_n$ is the alphabet of $U^n$, 
\begin{equation*}
 \oH(\UU)
 \leq
 \limsupn\frac{\log_2|\U_n|}n.
\end{equation*}
\end{lem}
\begin{IEEEproof}
Let $\nu_n$ be a uniform distribution on $\U_n$.
Then we have
\begin{align}
 \limsupn\frac{\log_2|\U_n|}n-\oH(\UU)
 &=
 \plimsupn\frac1n\log_2\frac1{\nu_n(U_n)}
 -\plimsupn\frac1n\log_2\frac1{\mu_{U_n}(U_n)}
 \notag
 \\
 &=
 \plimsupn\frac1n\log_2\frac1{\nu_n(U_n)}
 +\pliminfn\frac1n\log_2\mu_{U_n}(U_n)
 \notag
 \\
 &\geq
 \pliminfn\frac 1n\log_2\frac{\mu_{U_n}(U_n)}{\nu_n(U_n)}
 \notag
 \\
 &\geq
 0,
\end{align}
where the first equality comes from Lemma \ref{lem:constant}
and the fact that 
$\frac1n\log_2(1/\nu_n(U_n))$ is a constant random variable
satisfying $\frac1n\log_2(1/\nu_n(U_n))=\frac1n\log_2|\U_n|$,
the second equality comes from (\ref{eq:plimsup-pliminf}),
the first inequality comes from (\ref{eq:pliminf-upper}),
and the second inequality comes from  Lemma~\ref{lem:pliminf-div}.
\end{IEEEproof}

\begin{lem}
\label{lem:oH(UU'|V)<oH(U'|UV)+oH(U|V)}
For a triplet of general sources,
$(\UU,\UU',\VV)=\{(U_n,U'_n,V_n)\}_{n=1}^{\infty}$,
we have
\begin{align*}
 \oH(\UU,\UU'|\VV)
 &\leq
 \oH(\UU'|\UU,\VV) + \oH(\UU|\VV)
 \\
 \uH(\UU,\UU'|\VV)
 &\geq \uH(\UU'|\UU,\VV) + \uH(\UU|\VV).
\end{align*}
\end{lem}
\begin{IEEEproof}
We have
\begin{align}
 \oH(\UU,\UU'|\VV)
 &=
 \plimsupn\frac 1n\log_2
 \frac1{\mu_{U_nU'_n|V_n}(U_n,U'_n|V_n)}
 \notag
 \\
 &=
 \plimsupn\frac 1n\log_2
 \frac
 1
 {\mu_{U'_n|U_nV_n}(U'_n|U_n,V_n)\mu_{U_n|V_n}(U_n|V_n)}
 \notag
 \\
 &\leq
 \plimsupn\frac 1n\log_2
 \frac
 1
 {\mu_{U'_n|U_nV_n}(U'_n|U_n,V_n)}
 +
 \plimsupn\frac 1n\log_2
 \frac
 1
 {\mu_{U_n|V_n}(U_n|V_n)}
 \notag
 \\
 &=
 \oH(\UU'|\UU,\VV)
 +
 \oH(\UU|\VV),
\end{align} 
where the inequality comes from (\ref{eq:plimsup-upper}).
Similarly, we have
\begin{align}
 \uH(\UU,\UU'|\VV)
 &=
 \pliminfn\frac 1n\log_2
 \frac1{\mu_{U_nU'_n|V_n}(U_n,U'_n|V_n)}
 \notag
 \\
 &=
 \pliminfn\frac 1n\log_2
 \frac
 1
 {\mu_{U'_n|U_nV_n}(U'_n|U_n,V_n)\mu_{U_n|V_n}(U_n|V_n)}
 \notag
 \\
 &\geq
 \pliminfn\frac 1n\log_2
 \frac
 1
 {\mu_{U'_n|U_nV_n}(U'_n|U_n,V_n)}
 +
 \pliminfn\frac 1n\log_2
 \frac
 1
 {\mu_{U_n|V_n}(U_n|V_n)}
 \notag
 \\
 &=
 \uH(\UU'|\UU,\VV)
 +
 \uH(\UU|\VV),
\end{align}
where the inequality comes from (\ref{eq:pliminf-lower}).
\end{IEEEproof}

\begin{lem}
\label{lem:oH(UU'|V)>oH(U|V)}
For a triplet of general sources,
$(\UU,\UU',\VV)=\{(U_n,U'_n,V_n)\}_{n=1}^{\infty}$,
we have
\begin{equation*}
 \oH(\UU,\UU'|\VV)\geq \oH(\UU|\VV).
\end{equation*}
\end{lem}
\begin{IEEEproof}
We have
\begin{align}
 \oH(\UU,\UU'|\VV)
 &=
 \plimsupn\frac 1n\log_2
 \frac1{\mu_{U_nU'_n|V_n}(U_n,U'_n|V_n)}
 \notag
 \\
 &=
 \plimsupn\frac 1n\log_2
 \frac
 1
 {\mu_{U'_n|U_nV_n}(U'_n|U_n,V_n)\mu_{U_n|V_n}(U_n|V_n)}
 \notag
 \\
 &\geq
 \pliminfn\frac 1n\log_2
 \frac
 1
 {\mu_{U'_n|U_nV_n}(U'_n|U_n,V_n)}
 +
 \plimsupn\frac 1n\log_2
 \frac
 1
 {\mu_{U_n|V_n}(U_n|V_n)}
 \notag
 \\
 &=
 \uH(\UU'|\UU,\VV)
 +
 \oH(\UU|\VV)
 \notag
 \\
 &\geq
 \oH(\UU|\VV),
\end{align}
where the first inequality comes from (\ref{eq:plimsup-lower})
and the second inequality comes from
Lemma \ref{lem:oH>uH>0}.
\end{IEEEproof}

\begin{lem}[{\cite[Lemma2]{CRNG-MULTI}}]
\label{lem:oH(U|V)>oH(U|VV')}
For a triplet of general sources,
$(\UU,\VV,\VV')=\{(U_n,V_n,V'_n)\}_{n=1}^{\infty}$,
we have
\begin{equation*}
 \oH(\UU|\VV)\geq \oH(\UU|\VV,\VV').
\end{equation*}
\end{lem}
\begin{IEEEproof}
We have
\begin{align}
 \oH(\UU|\VV)-\oH(\UU|\VV,\VV')
 &=
 \plimsupn\frac 1n\log_2
 \frac1{\mu_{U_n|V_n}(U_n|V_n)}
 -
 \plimsupn\frac 1n\log_2\frac1{\mu_{U_n|V_nV'_n}(U_n|V_n,V'_n)}
 \notag
 \\
 &=
 \plimsupn\frac 1n\log_2
 \frac1{\mu_{U_n|V_n}(U_n|V_n)}
 +
 \pliminfn\frac 1n\log_2\mu_{U_n|V_nV'_n}(U_n|V_n,V'_n)
 \notag
 \\
 &\geq
 \pliminfn\frac 1n\log_2
 \frac{\mu_{U_n|V_nV'_n}(U_n|V_n,V'_n)}{\mu_{U_n|V_n}(U_n|V_n)}
 \notag
 \\
 &=
 \pliminfn\frac 1n\log_2
 \frac{\mu_{U_nV_nV'_n}(U_n,V_n,V'_n)}
 {\mu_{U_n|V_n}(U_n|V_n)\mu_{V_nV'_n}(V_n,V'_n)}
 \notag
 \\
 &\geq
 0,
\end{align}
where the second equality comes from (\ref{eq:plimsup-pliminf}),
the first inequality comes from (\ref{eq:pliminf-upper}),
and the second inequality comes from Lemma~\ref{lem:pliminf-div}.
\end{IEEEproof}

The following lemma means that
when complimentary information $\VV$ eliminates
the uncertainty of $\UU$ for given $\VV'$,
the uncertainty of $\VV$ for given $\VV'$
is greater than the uncertainty of $\UU$ for given $\VV'$
before the observation of $\VV$.
\begin{lem}
\label{lem:sw-bound}
For the triplet of general sources
$(\UU,\VV,\VV')=\{(U_n,V_n,V'_n)\}_{n=1}^{\infty}$
satisfying
$\oH(\UU|\VV,\VV')=0$,
we have
\begin{equation*}
 \oH(\VV|\VV')\geq\oH(\UU|\VV').
\end{equation*}
\end{lem}
\begin{IEEEproof}
We have
\begin{align}
 \oH(\VV|\VV')
 &=
 \plimsupn\frac1n\log\frac1{\mu_{V_nV'_n}(V_n|V'_n)}
 \notag
 \\
 &=
 \plimsupn
 \frac1n\log
 \frac{
	\mu_{V'_n}(V'_n)
	\mu_{U_nV_nV'_n}(U_n,V_n,V'_n)
 }{
	\mu_{V_nV'_n}(V_n,V'_n)\mu_{U_nV_nV'_n}(U_n,V_n,V'_n)
 }
 \notag
 \\
 &=
 \plimsupn\frac1n
 \log\frac{
	\mu_{U_n|V_nV'_n}(U_n|V_n,V'_n)
 }{
	\mu_{V_n|U_nV'_n}(V_n|U_n,V'_n)
	\mu_{U_n|V'_n}(U_n|V'_n)
 }
 \notag
 \\
 &\geq
 \plimsupn\frac1n
 \log\frac{
	1
 }{
	\mu_{U_n|V'_n}(U_n|V'_n)
 }
 +
 \pliminfn\frac1n
 \log\frac{
	\mu_{U_n|V_nV'_n}(U_n|V_n,V'_n)
 }{
	\mu_{V_n|U_nV'_n}(V_n|U_n,V'_n)
 }
 \notag
 \\
 &\geq
 \plimsupn\frac1n
 \log\frac{
	1
 }{
	\mu_{U_n|V'_n}(U_n|V'_n)
 }
 +
 \pliminfn\frac1n
 \log\frac{
	1
 }{
	\mu_{V_n|U_nV'_n}(V_n|U_n,V'_n)
 }
 +
 \pliminfn\frac1n
 \log \mu_{U_n|V_nV'_n}(U_n|V_n,V'_n)
 \notag
 \\
 &=
 \plimsupn\frac1n
 \log\frac{
	1
 }{
	\mu_{U_n|V'_n}(U_n|V'_n)
 }
 +
 \pliminfn\frac1n
 \log\frac{
	1
 }{
	\mu_{V_n|U_nV'_n}(V_n|U_n,V'_n)
 }
 -
 \plimsupn
 \frac1n
 \log\frac{
	1
 }{
	\mu_{U_n|V_nV'_n}(U_n|V_n,V'_n)
 }
 \notag
 \\
 &=
 \oH(\UU|\VV')
 +\uH(\VV|\UU,\VV')
 -\oH(\UU|\VV,\VV')
 \notag
 \\
 &\geq
 \oH(\UU|\VV'),
\end{align}
where the first inequality comes from (\ref{eq:plimsup-lower}),
the second inequality comes from (\ref{eq:pliminf-lower}),
the next equality comes from (\ref{eq:plimsup-pliminf}),
and the last inequality comes from Lemma \ref{lem:oH>uH>0}
and assumption $\oH(\UU|\VV,\VV')=0$.
\end{IEEEproof}

We introduce the following lemma, which is analogous to Fano inequality.
\begin{lem}[{\cite[Lemma 4]{K08}\cite[Lemma 7]{CRNG}}]
\label{lem:fano}
Let $(\UU,\VV)\equiv\{(U_n,V_n)\}_{n=1}^{\infty}$
be a sequence of two random variables.
If there is a sequence $\{\psi_n\}_{n=1}^{\infty}$ of functions
satisfying the condition
\begin{equation}
 \limn \Prob(\psi_n(V_n)\neq U_n)=0,
 \label{eq:fano-error}
\end{equation}
then
\begin{equation*}
 \oH(\UU|\VV)=0.
\end{equation*}
\end{lem}
\begin{IEEEproof}
We introduce this lemma following the proof of \cite[Lemma 1.3.2]{HAN} for the completeness of this paper.

Let $\{\psi_n\}_{n=1}^{\infty}$ be a sequence of deterministic
functions satisfying (\ref{eq:fano-error}).
For $\gamma>0$, let
\begin{align*}
 \E
 &\equiv\lrb{
	(\uu,\vv): \psi_n(\vv)\neq\uu
 }
 \\
 \E'
 &\equiv\lrb{
	(\uu,\vv): \frac 1n\log_2\frac1{\mu_{U_n|V_n}(\uu|\vv)}\geq\gamma
 }.
\end{align*}
Then we have
\begin{align}
 \Prob\lrsb{
	\frac 1n\log_2\frac1{\mu_{U_n|V_n}(U_n|V_n)} > \gamma
 }
 &\leq
 \mu_{U_nV_n}(\E')
 \notag
 \\
 &=
 \mu_{U_nV_n}(\E\cap\E')
 +\mu_{U_nV_n}(\E^{\complement}\cap\E')
 \notag
 \\
 &=
 \mu_{U_nV_n}(\E\cap\E')
 +\sum_{(\uu,\vv)\in\E^{\complement}\cap\E'}\mu_{U_nV_n}(\uu,\vv)
 \notag
 \\
 &=
 \mu_{U_nV_n}(\E\cap\E')
 +
 \sum_{\vv\in\V_n}\mu_{V_n}(\vv)
 \sum_{\substack{
	 \uu\in\U_n:\\
	 \psi_n(\vv)=\uu\\
	 (\uu,\vv)\in\E'
 }}
 \mu_{U_n|V_n}(\uu|\vv)
 \notag
 \\
 &\leq
 \mu_{U_nV_n}(\E)
 +
 \sum_{\vv\in\V_n}\mu_{V_n}(\vv)
 \sum_{\uu\in\U_n: \psi_n(\vv)=\uu}
 2^{-n\gamma}
 \notag
 \\
 &=
 \Prob(\psi_n(V_n)\neq U_n)+2^{-n\gamma},
\end{align}
where the second inequality comes from the definition of $\E'$
and the last equality comes from the fact that
for all $\vv$ there is a unique $\uu$ satisfying $\psi_n(\vv)=\uu$.
From this inequality and (\ref{eq:fano-error}), we have
\begin{equation*}
 \limn\Prob\lrsb{
	\frac 1n\log_2\frac1{\mu_{U_n|V_n}(U_n|V_n)}>\gamma
 }=0.
\end{equation*}
Then we have
\begin{equation*}
 0\leq\oH(\UU|\VV)\leq \gamma
\end{equation*}
from the definition of $\oH(\UU|\VV)$.
We have $\oH(\UU|\VV)=0$ by letting $\gamma\to0$.
\end{IEEEproof}

\subsection{Common Randomness}
\label{sec:common}

The following lemma shows an equivalent condition under which
two encoders can generate common randomness $X_0$
from shared randomness $B_0$
and the respective observations $X_1$ and $X_2$.
\begin{lem}
\label{lem:common}
For a given triplet of random variables $(X_0,X_1,X_2)$,
the following two conditions are equivalent:
\begin{enumerate}
 \item Given triplet of random variables $(X_0,X_1,X_2)$ satisfies
 \begin{gather*}
	X_2\markov X_1\markov X_0
	\\
	X_1\markov X_2\markov X_0.
 \end{gather*}
 \item
 There are functions $\xi_1$, $\xi_2$, and source $B_0$
 such that $B_0$ is independent of $(X_1,X_2)$,
 $\xi_1(X_1,B_0)=\xi_2(X_2,B_0)$ with probability $1$,
 and the joint distribution of $(X_0,X_1,X_2)$ is the same
 as that of $(\hX_0,X_1,X_2)$ by letting $\hX_0\equiv \xi_i(X_i,B_0)$,
 which does not depend on the choice of $i\in\{1,2\}$.
 That is, both generators can simulate identical (synchronized) source $X_0$
 without communication.
\end{enumerate}
\end{lem}
\begin{rem}
When $H(B_0)=0$ and $H(\hX_0)>0$,
$\hX_0$ is called the common part of $X_1$ and $X_2$
\cite[Sec. 14.1.3]{EK11}.
The maximum value of $H(\hX_0)$ over functions $\xi_1$ and $\xi_2$
satisfying $\xi_1(X_1)=\xi_2(X_2)$ with probability $1$
is known as the G\'acs-K\"orner common information \cite{GK73}.
It should be noted that this lemma does not focus on
the maximum value of $H(\hX_0)$
because it can be infinite by forwarding the unlimited shared randomness.
\end{rem}
\begin{IEEEproof}
First, we show the fact that 
Condition 2) implies Condition 1).
Let us assume Condition 2).
Then we have
\begin{align}
0
\leq
I(X_0;X_{\ic}|X_i)
&=
I(\hX_0;X_{\ic}|X_i)
\notag
\\
&=
I(\xi_i(X_i,B_0);X_{\ic}|X_i)
\notag
\\
&\leq
I(X_i,B_0;X_{\ic}|X_i)
\notag
\\
&=
I(B_0;X_{\ic}|X_i)
\notag
\\
&\leq
I(B_0;X_1,X_2)
\notag
\\
&=
0
\end{align}
for any $(i,\ic)\in\{(1,2),(2,1)\}$,
where the last equality comes from the fact that
$B_0$ is independent of $(X_1,X_2)$.
This yields Condition 1).

Next, we show the fact that Condition 1) implies Condition 2).
From Condition 1), we have the fact that
\begin{equation}
p_{X_0|X_1}(x_0|x_1)=p_{X_0|X_1,X_2}(x_0|x_1,x_2)=p_{X_0|X_2}(x_0|x_2)
\label{eq:common-markov}
\end{equation}
for all $(x_0,x_1,x_2)\in\X_0\times\X_1\times\X_2$
satisfying $p_{X_1X_2}(x_1,x_2)>0$.
For each $i\in\{1,2\}$ and $x_i\in\X_i$,
let $\{\Q(x_0|x_i)\}_{x_0\in\X_0}$ be a partition
of the interval $[0,1]$,
where the width of partition $\Q(x_0|x_i)$ is $p_{X_0|X_i}(x_0|x_i)$.
Let $B_0$ be a random variable subject to
the uniform distribution on the interval $[0,1]$
independent of $(X_1,X_2)$.
For each $i\in\{1,2\}$,
let $\xi_i$ be defined so as to satisfy
\[
\xi_i(x_i,u)
=x_0\ \text{iff}\ u\in\Q(x_0|x_i),
\]
where it is well-defined because $\{\Q(x_0|x_i)\}_{x_0\in\X_0}$
forms a partition of $[0,1]$ for every $i\in\{1,2\}$ and $x_i\in\X_i$.
From (\ref{eq:common-markov}),
we have the fact that
two partitions $\{\Q(x_0|x_1)\}_{x_0\in\X_0}$
and $\{\Q(x_0|x_2)\}_{x_0\in\X_0}$ are identical
when $p_{X_1X_2}(x_1,x_2)>0$.
This implies that
$\xi_1(X_1,B_0)=\xi_2(X_2,B_0)$ with probability $1$.
Let $\hX_0\equiv g_i(X_i,B_0)$,
which does not depend on the choice of $i\in\{1,2\}$.
Then the joint distribution $p_{\hX_0X_1X_2}$ of $(\hX_0,X_1,X_2)$
is given as
\begin{align}
p_{\hX_0X_1X_2}(x_0,x_1,x_2)
&=
\int p_{B_0}(db_0)
p_{X_1X_2}(x_1,x_2)\chi(x_0=\xi_1(x_1,b_0)=\xi_2(x_2,b_0))
\notag
\\
&=
p_{B_0}(\Q(x_0|x_i))p_{X_1X_2}(x_1,x_2)
\notag
\\
&=
p_{X_0|X_i}(x_0|x_i)p_{X_1X_2}(x_1,x_2)
\notag
\\
&=
p_{X_0X_1X_2}(x_0,x_1,x_2)
\end{align}
for all $i\in\{1,2\}$ and
$(x_0,x_1,x_2)\in\X_0\times\X_1\times\X_2$,
where the last equality comes from (\ref{eq:common-markov}).
\end{IEEEproof}

\subsection{Expectation of Constrained-Random Number Generator}
\label{sec:crng-expectation}

The following fact is used in the proof of Theorem \ref{thm:crng}.
\begin{lem}
Let us assume that 
the constrained-random number generator
(deterministic function) $\mathrm{crng}:\B\to\U$
generates random number $U=\mathrm{crng}(B)$
by using random source $B$.
Then we have
\begin{equation*}
E_B\lrB{
\lambda(\mathrm{crng}(B))
}
=
E_U\lrB{\lambda(U)}
\end{equation*}
for any (integrable) function $\lambda$ on $\U$.
\end{lem}
\begin{IEEEproof}
We have
\begin{align}
E_B\lrB{
\lambda(\mathrm{crng}(B))
}
&=
\sum_{b\in\B}
\mu_B(b)
\lambda(\mathrm{crng}(b))
\notag
\\
&=
\sum_{b\in\B}
\mu_B(b)
\sum_{u\in\U}
\lambda(u)
\chi(\mathrm{crng}(b)=u)
\notag
\\
&=
\sum_{u\in\U}
\lambda(u)
\sum_{b\in\B}
\mu_B(b)
\chi(\mathrm{crng}(b)=u)
\notag
\\
&=
\sum_{u\in\U}
\lambda(u)
\mu_U(u)
\notag
\\
&=
E_U\lrB{\lambda(U)}
\end{align}
for any (integrable) function $\lambda$ on $\U$.
\end{IEEEproof}

\subsection{$(\aalpha,\bbeta)$-hash property}
\label{sec:hash}

In this section, we review the $(\aalpha,\bbeta)$-hash property
introduced in \cite{CRNG}\cite{HASH-BC} and show two basic lemmas.

\begin{df}[{\cite[Definition~3]{CRNG}}]
Let $\F_n$ be a set of functions on $\W^n$.
For probability distribution $p_{F_n}$ on $\F_n$,
we call the pair $(\F_n,p_{F_n})$ an {\em ensemble}.
Then, $(\F_n,p_{F_n})$ has $(\alpha_{F_n},\beta_{F_n})$-{\em hash property}
if there is a pair $(\alpha_{F_n},\beta_{F_n})$
depending on $p_{F_n}$ such that
\begin{equation}
\sum_{\substack{
\ww'\in\W^n\setminus\{\ww\}:
\\
p_{F_n}(\{f: f(\ww) = f(\ww')\})>\frac{\alpha_{F_n}}{|\im\F_n|}
}}
p_{F_n}\lrsb{\lrb{f: f(\ww) = f(\ww')}}
\leq
\beta_{F_n}
\label{eq:hash}
\end{equation}
for any $\ww\in\W^n$,
where $\im\F_n\equiv \bigcup_{f\in\F_n}\{f(\ww): \ww\in\W^n\}$.
Consider the following conditions for two sequences
$\aalpha_F\equiv\{\alpha_{F_n}\}_{n=1}^{\infty}$ and
$\bbeta_F\equiv\{\beta_{F_n}\}_{n=1}^{\infty}$,
\begin{align}
\limn \alpha_{F_n}
&=1
\label{eq:alpha}
\\
\limn \beta_{F_n}
&=0.
\label{eq:beta}
\end{align}
Then, we say that 
$(\bcF,\bp_F)$ has $(\aalpha_F,\bbeta_F)$-{\em hash property}
if $\aalpha_F$ and $\bbeta_F$ satisfy (\ref{eq:hash})--(\ref{eq:beta}).
Throughout this paper,
we omit the dependence of $\F$ and $F$ on $n$.
\end{df}

It should be noted that
when $\F$ is a $2$-universal class of hash functions \cite{CW79}
and  $p_F$ is the uniform distribution on $\F$,
then $(\bcF,\bp_F)$ has $(\one,\zero)$-hash property.
Random binning \cite{C75} and the set of all linear functions \cite{CSI82}
are $2$-universal classes of hash functions.
It is proved in \cite[Section III-B]{HASH-BC} that
an ensemble of sparse matrices (with logarithmic column weight)
has hash property.

First, we introduce the lemma for a joint ensemble.
\begin{lem}[
{\cite[Lemma 4 of the extended version]{HASH-BC}\cite[Lemma 3]{CRNG}}
]
\label{lem:hash-FG}
Let $(\F,p_F)$ and $(\G,p_G)$ be ensembles of functions
on the same set $\W^n$.
Assume that $(\F,p_F)$ (resp. $(\G,p_G)$) has an $(\alpha_F,\beta_F)$-hash
(resp. $(\alpha_G,\beta_G)$-hash) property.
Let $(f,g)\in\F\times\G$ be a function defined as
\begin{equation*}
(f,g)(\ww)\equiv(f(\ww),g(\ww))\quad\text{for each}\ \ww\in\W^n.
\end{equation*}
Let $p_{(F,G)}$  be a joint distribution on $\F\times\G$ defined as
\begin{equation*}
p_{(F,G)}(f,g)\equiv p_F(f)p_G(g)\quad\text{for each}\ (f,g)\in\F\times\G.
\end{equation*}
Then ensemble $(\F\times\G, p_{(F,G)})$ has 
$(\alpha_{(F,G)},\beta_{(F,G)})$-hash property,
where $\alpha_{(F,G)}$ and $\beta_{(F,G)}$ are defined as
\begin{align*}
\alpha_{(F,G)}
&\equiv
\alpha_F\alpha_G
\\
\beta_{(F,G)}
&\equiv
\beta_F+\beta_G.
\end{align*}
\end{lem}
\begin{IEEEproof}
We show this lemma for the completeness of this paper.
Let
\begin{align*}
p_{F,\ww,\ww'}&\equiv p_F(\{f: f(\ww)=f(\ww')\})
\\
p_{G,\ww,\ww'}&\equiv p_G(\{g: g(\ww)=g(\ww')\})
\\
p_{(F,G),\ww,\ww'}&\equiv p_{(F,G)}(
\{(f,g): (f,g)(\ww)=(f,g)(\ww')\}
).
\end{align*}
Then we have
\begin{align}
\sum_{\substack{
\ww'\in\W^n\setminus\{\ww\}:
\\
p_{(F,G),\ww,\ww'}>\frac{\alpha_{(F,G)}}{|\im\F\times\G|}
}}
p_{(F,G),\ww,\ww'}
&\leq
\sum_{\substack{
\ww'\in\W^n\setminus\{\ww\}:
\\
p_{F,\ww,\ww'}p_{G,\ww,\ww'}>\frac{\alpha_F\alpha_G}{|\im\F||\im\G|}
}}
p_{F,\ww,\ww'}p_{G,\ww,\ww'}
\notag
\\
&=
\sum_{\substack{
\ww'\in\W^n\setminus\{\ww\}:
\\
p_{F,\ww,\ww'}p_{G,\ww,\ww'}>\frac{\alpha_F\alpha_G}{|\im\F||\im\G|}
\\
p_{F,\ww,\ww'}>\frac{\alpha_F}{|\im\F|}
}}
p_{F,\ww,\ww'}p_{G,\ww,\ww'}
+
\sum_{\substack{
\ww'\in\W^n\setminus\{\ww\}:
\\
p_{F,\ww,\ww'}p_{G,\ww,\ww'}>\frac{\alpha_F\alpha_G}{|\im\F||\im\G|}
\\
p_{F,\ww,\ww'}\leq\frac{\alpha_F}{|\im\F|}
}}
p_{F,\ww,\ww'}p_{G,\ww,\ww'}
\notag
\\
&\leq
\sum_{\substack{
\ww'\in\W^n\setminus\{\ww\}:
\\
p_{F,\ww,\ww'}>\frac{\alpha_F}{|\im\F|}
}}
p_{F,\ww,\ww'}p_{G,\ww,\ww'}
+
\sum_{\substack{
\ww'\in\W^n\setminus\{\ww\}:
\\
p_{G,\ww,\ww'}>\frac{\alpha_G}{|\im\G|}
}}
p_{F,\ww,\ww'}p_{G,\ww,\ww'}
\notag
\\
&\leq
\sum_{\substack{
\ww'\in\W^n\setminus\{\ww\}:
\\
p_{F,\ww,\ww'}>\frac{\alpha_F}{|\im\F|}
}}
p_{F,\ww,\ww'}
+
\sum_{\substack{
\ww'\in\W^n\setminus\{\ww\}:
\\
p_{G,\ww,\ww'}>\frac{\alpha_G}{|\im\G|}
}}
p_{G,\ww,\ww}
\notag
\\
&=
\beta_F+\beta_G
\notag
\\
&=
\beta_{(F,G)},
\end{align}
where the first inequality comes from the fact that
$F$ and $G$ are mutually independent
and $\im\F\times\G\subset\im\F\times\im\G$,
and the last inequality comes from the fact that
$p_{F,\ww,\ww'}\leq 1$ and $p_{G,\ww,\ww'}\leq 1$.
Then we have the fact that
$(\F\times\G,p_{(F,G)})$ has 
$(\alpha_{(F,G)},\beta_{(F,G)})$-hash property.
\end{IEEEproof}

Next, we introduce lemmas that are multiple extensions of
the {\it balanced-coloring property}
and the {\it collision-resistant property}.
We use the following notations.
For each $i\in\I$,
let $\F_i$ be a set of functions on $\W_i^n$ and $\cc_i\in\im\F_i$.
Let $\W_{\I'}^n\equiv\Prod_{i\in\I'}\W_i^n$ and 
\begin{align*}
\alpha_{F_{\I'}}
&\equiv
\prod_{i\in\I'}\alpha_{F_i}
\\
\beta_{F_{\I'}}
&\equiv
\prod_{i\in\I'}\lrB{\beta_{F_i}+1}-1,
\end{align*}
where $\prod_{i\in\emptyset}\theta_i\equiv1$.
It should be noted that
\begin{align*}
\limn \alpha_{F_{\I'}}=1
\\
\limn \beta_{F_{\I'}}=0
\end{align*}
for every $\I'\subset\I$
when $(\aalpha_{F_i},\bbeta_{F_i})$ satisfies
(\ref{eq:alpha}) and (\ref{eq:beta}) for all $i\in\I$.
For $\T\subset\W_{\I}^n$ and $\ww_{\I'}\in\W^n_{\I'}$,
let $\T_{\I'}$ and $\T_{\I'^{\complement}|\I'}(\ww_{\I'})$ be defined as
\begin{align*}
&
\T_{\I'}
\equiv\{\ww_{\I'}:
(\ww_{\I'},\ww_{\I'^{\complement}})\in\T
\ \text{for some}\ \ww_{\I'^{\complement}}\in\W_{\I'^{\complement}}
\}
\\
&
\T_{\Ipc|\I'}(\ww_{\I'})
\equiv
\{\ww_{\Ipc}: (\ww_{\I'},\ww_{\Ipc})\in\T\},
\end{align*}
where $\Ipc\equiv\I\setminus\I'$.
Let
$p_{\ww_i,\ww'_i}
\equiv
p_{F_i}\lrsb{\lrb{
f_i:
f_i(\ww_i)=f_i(\ww'_i)
}}$.
In the following, we use the relation
\begin{align}
\sum_{\substack{
\ww_i\in\W^n_i:
\\
p_{\ww_i,\ww'_i}
>\frac{\alpha_{F_i}}{|\im\F_i|}
}}
p_{\ww_i,\ww'_i}
&=
\sum_{\substack{
\ww_i\in\W^n_i\setminus\{\ww'_i\}:
\\
p_{\ww_i,\ww'_i}
>\frac{\alpha_{F_i}}{|\im\F_i|}
}}
p_{\ww_i,\ww'_i}
+
p_{\ww'_i,\ww'_i}
\notag
\\
&\leq
\beta_{F_i}+1
\label{eq:proof-beta}
\end{align}
for all $\ww'_i\in\W_i^n$, which comes from (\ref{eq:hash})
and the fact that $p_{\ww'_i,\ww'_i}=1$.

The following lemma is related to the {\em balanced-coloring property},
which is an extension of \cite[Lemma 4]{HASH-WTC},
the leftover hash lemma~\cite{IZ89} and
the balanced-coloring lemma~\cite[Lemma 3.1]{AC98}\cite[Lemma 17.3]{CK11}.
This lemma implies that there is an assignment that divides a set equally.
\begin{lem}[{\cite[Lemma 4 in the extended version]{CRNG-MULTI}}]
\label{lem:mBCP}
For each $i\in\I$, let $\F_i$ be a set
of functions on $\W_i^n$
and $p_{F_i}$ be the probability distribution on $\F_i$,
where $(\F_i,p_{F_i})$ satisfies (\ref{eq:hash}).
We assume that random variables $\{F_i\}_{i\in\I}$ are mutually independent.
Then
\begin{align*}
E_{F_{\I}}\lrB{
\sum_{\cc_{\I}}
\lrbar{
\frac{Q(\T\cap\fC_{F_{\I}}(\cc_{\I}))}
{Q(\T)}
-
\frac1
{\prod_{i\in\I}|\im\F_i|}
}
}
&\leq
\sqrt{
\alpha_{F_{\I}}-1
+
\sum_{\I'\in2^{\I}\setminus\{\emptyset\}}
\alpha_{F_{\Ipc}}
\lrB{\beta_{F_{\I'}}+1}
\lrB{\prod_{i\in\I'}|\im\F_i|}
\cdot
\frac{
\oQ_{\Ipc}
}
{Q(\T)}
}
\end{align*}
for any function $Q:\W_{\I}\to[0,\infty)$ and $\T\subset\W_{\I}^n$,
where
\begin{equation}
\oQ_{\Ipc}
\equiv
\begin{cases}
\displaystyle
\max_{\ww_{\I}\in\T}Q(\ww_{\I})
&\!\!\text{if}\ \Ipc=\I
\\
\displaystyle
\max_{\ww_{\I'}\in\T_{\I'}}
\!\!\!
\sum_{\ww_{\Ipc}\in\T_{\Ipc|\I'}(\ww_{\I'})}
\!\!\!
Q(\ww_{\I'},\ww_{\Ipc})
&\!\!\text{if}\ \emptyset\neq\Ipc\subsetneq\I
\\
Q(\T)
&\!\!\text{if}\ \Ipc=\emptyset.
\end{cases}
\label{eq:maxQJ}
\end{equation}
\end{lem}
\begin{IEEEproof}
We show this lemma for the completeness of the paper.
First, we have
\begin{align}
\sum_{\substack{
\ww_{\I}\in\T:
\\
p_{\ww_i,\ww'_i}
>\frac{\alpha_{F_i}}{|\im\F_i|}
\ \text{for all}\ i\in\I'
\\
p_{\ww_i,\ww'_i}
\leq\frac{\alpha_{F_i}}{|\im\F_i|}
\ \text{for all}\  i\in\Ipc
}}
Q(\ww_{\I})
\prod_{i\in\I} p_{\ww_i,\ww'_i}
&=
\sum_{\substack{
\ww_{\I'}\in\T_{\I'}:
\\
p_{\ww_i,\ww'_i}
>\frac{\alpha_{F_i}}{|\im\F_i|}
\\
\text{for all}\  i\in\I'
}}
\lrB{\prod_{i\in\I'} p_{\ww_i,\ww'_i}}
\sum_{\substack{
\ww_{\Ipc}\in\T_{\Ipc|\I'}\lrsb{\ww_{\I'}}:
\\
p_{\ww_i,\ww'_i}
\leq\frac{\alpha_{F_i}}{|\im\F_i|}
\\
\text{for all}\  i\in\Ipc
}}
Q(\ww_{\I'},\ww_{\Ipc})
\prod_{i\in\Ipc}p_{\ww_i,\ww'_i}
\notag
\\
&\leq
\lrB{\prod_{i\in\Ipc}\frac{\alpha_{F_i}}{|\im\F_i|}}
\sum_{\substack{
\ww_{\I'}\in\T_{\I'}:
\\
p_{\ww_i,\ww'_i}
>\frac{\alpha_{F_i}}{|\im\F_i|}
\\
\text{for all}\  i\in\I'
}}
\lrB{
\prod_{i\in\I'} p_{\ww_i,\ww'_i}
}
\sum_{
\ww_{\Ipc}
\in\T_{\Ipc|\I'}\lrsb{\ww_{\I'}}
}
Q(\ww_{\I'},\ww_{\Ipc})
\notag
\\
&\leq
\oQ_{\Ipc}
\lrB{\prod_{i\in\Ipc}\frac{\alpha_{F_i}}{|\im\F_i|}}
\prod_{i\in\I'}
\lrB{
\sum_{\substack{
\ww_i\in\W_i^n:
\\
p_{\ww_i,\ww'_i}
>\frac{\alpha_{F_i}}{|\im\F_i|}
}}
p_{\ww_i,\ww'_i}
}
\notag
\\
&\leq
\oQ_{\Ipc}
\lrB{\prod_{i\in\Ipc}\frac{\alpha_{F_i}}{|\im\F_i|}}
\prod_{i\in\I'}
\lrB{\beta_{F_i}+1}
\notag
\\
&=
\frac{
\alpha_{F_{\Ipc}}\lrB{\beta_{F_{\I'}}+1}
\oQ_{\Ipc}
}
{\prod_{i\in\Ipc}\lrbar{\im\F_i}}
\label{eq:lemma-mbcp}
\end{align}
for all $(\ww'_{\I},\I')$ satisfying
$\ww'_{\I}\in\T$ and $\emptyset\neq\I'\subsetneq\I$,
where the second inequality comes from (\ref{eq:maxQJ})
and the third inequality comes from (\ref{eq:proof-beta}).
It should be noted that (\ref{eq:lemma-mbcp})
is valid for the cases of $\Ipc=\emptyset$ and $\Ipc=\I$
by letting $\oQ_{\emptyset}\equiv Q(\T)$
and $\oQ_{\I}\equiv\max_{\ww_{\I}\in\T}Q(\ww_{\I})$,
respectively,
because
\begin{align}
\sum_{\substack{
\ww_{\I}\in\T:
\\
p_{\ww_i,\ww'_i}
\leq\frac{\alpha_{F_i}}{|\im\F_i|}
\ \text{for all}\  i\in\I
}}
Q(\ww_{\I})
\prod_{i\in\I} p_{\ww_i,\ww'_i}
&\leq
\frac{\alpha_{F_{\I}}Q(\T)}
{\prod_{i\in\I}\lrbar{\im\F_i}}
\notag
\\
&=
\frac{\alpha_{F_{\I}}\lrB{\beta_{F_{\emptyset}}+1}\oQ_{\emptyset}}
{\prod_{i\in\I}\lrbar{\im\F_i}}
\end{align}
and
\begin{align}
\sum_{\substack{
\ww_{\I}\in\T:
\\
p_{\ww_i,\ww'_i}
>\frac{\alpha_{F_i}}{|\im\F_i|}
\ \text{for all}\  i\in\I
}}
Q(\ww_{\I})
\prod_{i\in\I} p_{\ww_i,\ww'_i}
&\leq
\lrB{\max_{\ww_{\I}\in\T}Q(\ww_{\I})}
\sum_{\substack{
\ww_{\I}\in\T:
\\
p_{\ww_i,\ww'_i}
>\frac{\alpha_{F_i}}{|\im\F_i|}
\ \text{for all}\  i\in\I
}}
\prod_{i\in\I} p_{\ww_i,\ww'_i}
\notag
\\
&\leq
\lrB{\max_{\ww_{\I}\in\T}Q(\ww_{\I})}
\prod_{i\in\I}
\lrB{
\sum_{\substack{
\ww_i\in\W_i^n:
\\
p_{\ww_i,\ww'_i}
>\frac{\alpha_{F_i}}{|\im\F_i|}
}}
p_{\ww_i,\ww'_i}
}
\notag
\\
&\leq
\lrB{\max_{\ww_{\I}\in\T}Q(\ww_{\I})}
\prod_{i\in\I}
\lrB{\beta_{F_i}+1}
\notag
\\
&=
\frac{\alpha_{F_{\emptyset}}\lrB{\beta_{F_{\I}}+1}\oQ_{\I}}
{\prod_{i\in\emptyset}\lrbar{\im\F_i}}.
\end{align}
Then we have
\begin{align}
\sum_{\ww_{\I}\in\T}Q(\ww_{\I})
\prod_{i\in\I}p_{\ww_i,\ww'_i}
&\leq
\sum_{\I'\subset\I}
\sum_{\substack{
\ww_{\I}\in\T:
\\
p_{\ww_i,\ww'_i}
>\frac{\alpha_{F_i}}{|\im\F_i|}
\ \text{for all}\  i\in\I'
\\
p_{\ww_i,\ww'_i}
\leq\frac{\alpha_{F_i}}{|\im\F_i|}
\ \text{for all}\  i\in\Ipc
}}
Q(\ww_{\I})
\prod_{i\in\I} p_{\ww_i,\ww'_i}
\notag
\\
&\leq
\sum_{\I'\subset\I}
\frac{
\alpha_{F_{\Ipc}}\lrB{\beta_{F_{\I'}}+1}
\oQ_{\Ipc}
}{\prod_{i\in\Ipc}\lrbar{\im\F_i}}
\notag
\\
&
=
\frac{\alpha_{F_{\I}}Q(\T)}
{\prod_{i\in\I}\lrbar{\im\F_i}}
+
\sum_{\I'\in2^{\I}\setminus\{\emptyset\}}
\frac{
\alpha_{F_{\Ipc}}\lrB{\beta_{F_{\I'}}+1}
\oQ_{\I'}
}
{\prod_{i\in\Ipc}\lrbar{\im\F_i}}
\label{eq:proof-BCP-multi2}
\end{align}
for all $\ww'_{\I}\in\T$,
where the equality comes from the fact that $\oQ_{\emptyset}=Q(\T)$ and
$\beta_{F_{\emptyset}}=0$.

Next, let $C_{\I}$ be a random variable
subject to the uniform distribution on $\Prod_{i\in\I}\im\F_i$.
From (\ref{eq:proof-BCP-multi2}), we have
\begin{align}
&
E_{F_{\I}C_{\I}}\lrB{
\lrB{
\sum_{\ww_{\I}\in\T}Q(\ww_{\I})\chi(F_{\I}(\ww_{\I})=C_{\I})
}^2
}
\notag
\\*
&=
\sum_{\ww'_{\I}\in\T}Q(\ww'_{\I})
\sum_{\ww_{\I}\in\T}Q(\ww_{\I})
E_{F_{\I}}\lrB{
\chi(F_{\I}(\ww_{\I})=F_{\I}(\ww'_{\I}))
E_{C_{\I}}\lrB{\chi(F_{\I}(\ww_{\I})=C_{\I})}
}
\notag
\\
&=
\frac1{\prod_{i\in\I}\lrbar{\im\F_i}}
\sum_{\ww'_{\I}\in\T}Q(\ww'_{\I})
\sum_{\ww_{\I}\in\T}Q(\ww_{\I})
\prod_{i\in\I}p_{\ww_i,\ww'_i}
\notag
\\
&\leq
\frac{\alpha_{F_{\I}}Q(\T)^2}
{\lrB{\prod_{i\in\I}\lrbar{\im\F_i}}^2}
+
\frac{Q(\T)}
{\prod_{i\in\I}\lrbar{\im\F_i}}
\sum_{\I'\in2^{\I}\setminus\{\emptyset\}}
\frac{
\alpha_{F_{\Ipc}}\lrB{\beta_{F_{\I'}}+1}
\oQ_{\Ipc}
}
{\prod_{i\in\Ipc}\lrbar{\im\F_i}}.
\label{eq:lemma-multi}
\end{align}
Then we have
\begin{align}
&
E_{F_{\I}C_{\I}}\lrB{
\lrB{
\frac{Q\lrsb{\T\cap\fC_{F_{\I}}(C_{\I})}
\prod_{i\in\I}|\im\F_i|}
{Q(\T)}
-1}^2
}
\notag
\\*
&=
E_{F_{\I}C_{\I}}\lrB{
\lrB{\sum_{\ww_{\I}\in\T}
\frac{Q(\ww)\chi(F_{\I}(\ww_{\I})=C_{\I})
\prod_{i\in\I}|\im\F_i|
}{Q(\T)}
}^2
}
-2
E_{F_{\I}C_{\I}}\lrB{
\sum_{\ww_{\I}\in\T}
\frac{Q(\ww)\chi(F_{\I}(\ww_{\I})=C_{\I})
\prod_{i\in\I}|\im\F_i|
}{Q(\T)}
}
+1
\notag
\\
&=
E_{F_{\I}C_{\I}}\lrB{
\lrB{\sum_{\ww_{\I}\in\T}
\frac{Q(\ww)\chi(F_{\I}(\ww_{\I})=C_{\I})
\prod_{i\in\I}|\im\F_i|
}{Q(\T)}
}^2
}
-2
\sum_{\ww_{\I}\in\T}
\frac{
Q(\ww)
E_{F_{\I}C_{\I}}\lrB{\chi(F_{\I}(\ww_{\I})=C_{\I})}
\prod_{i\in\I}|\im\F_i|
}{Q(\T)}
+1
\notag
\\
&=
\frac{\displaystyle
\lrB{\prod_{i\in\I}|\im\F_i|}^2
}{Q(\T)^2}
E_{F_{\I}C_{\I}}\lrB{
\lrB{\sum_{\ww_{\I}\in\T}
Q(\ww_{\I})\chi(F_{\I}(\ww_{\I})=C_{\I})
}^2
}
-1
\notag
\\
&\leq
\alpha_{F_{\I}}-1
+
\sum_{\I'\in2^{\I}\setminus\{\emptyset\}}
\alpha_{F_{\Ipc}}\lrB{\beta_{F_{\I'}}+1}
\lrB{\prod_{i\in\I'}\lrbar{\im\F_i}}
\cdot
\frac{\oQ_{\Ipc}}
{Q(\T)},
\end{align}
where the inequality comes from (\ref{eq:lemma-multi}).

Finally, the lemma is confirmed by
\begin{align}
E_{F_{\I}}\lrB{
\sum_{\cc_{\I}}
\left|
\frac{Q\lrsb{\T\cap\fC_{F_{\I}}(\cc_{\I})}}{Q(\T)}
-\frac 1{\prod_{i\in\I}|\im\F_i|}
\right|
}
&=
E_{F_{\I}C_{\I}}\lrB{
\left|
\frac{Q\lrsb{\T\cap\fC_{F_{\I}}(C_{\I})}
\prod_{i\in\I}|\im\F_i|
}{Q(\T)}
-1
\right|
}
\notag
\\
&=
E_{F_{\I}C_{\I}}\lrB{
\sqrt{
\lrB{
\frac{
Q\lrsb{\T\cap\fC_{F_{\I}}(C_{\I})}
\prod_{i\in\I}|\im\F_i|
}{Q(\T)}
-1}^2
}
}
\notag
\\
&\leq
\sqrt{
E_{F_{\I}C_{\I}}\lrB{
\lrB{\frac{
Q\lrsb{\T\cap\fC_{F_{\I}}(C_{\I})}
\prod_{i\in\I}|\im\F_i|
}{Q(\T)}
-1}^2
}
}
\notag
\\
&\leq
\sqrt{
\alpha_{F_{\I}}-1
+
\sum_{\I'\in2^{\I}\setminus\{\emptyset\}}
\alpha_{F_{\Ipc}}
\lrB{\beta_{F_{\I'}}+1}
\lrB{\prod_{i\in\I'}|\im\F_i|}
\cdot
\frac{\oQ_{\Ipc}}
{Q(\T)}
},
\label{eq:proof-BCP-multi}
\end{align}
where the first inequality comes from the Jensen inequality.
\end{IEEEproof}

The following lemma is a multiple extension of
the {\em collision-resistant property}.
This lemma implies that
there is an assignment such that every bin contains at most one item.
\begin{lem}[{\cite[Lemma 7 in the extended version]{HASH-BC}}]
\label{lem:mCRP}
For each $i\in\I$, let $\F_i$ be a set of functions on $\W_i^n$
and $p_{F_i}$ be the probability distribution on $\F_i$,
where $(\F_i,p_{F_i})$ satisfies (\ref{eq:hash}).
We assume that random variables $\{F_i\}_{i\in\I}$ are mutually independent.
Then
\begin{align*}
p_{F_{\I}}\lrsb{\lrb{
f_{\I}:
\lrB{\T\setminus\{\ww_{\I}\}}\cap\fC_{f_{\I}}(f_{\I}(\ww_{\I}))
\neq
\emptyset
}}
\leq
\sum_{\I'\in2^{\I}\setminus\{\emptyset\}}
\frac{
\alpha_{F_{\I'}}\lrB{\beta_{F_{\Ipc}}+1}
\oO_{\I'}
}
{
\prod_{i\in\I'}\lrbar{\im\F_i}
}
+\beta_{F_{\I}}
\end{align*}
for all $\T\subset\W_{\I}^n$ and $\ww_{\I}\in\W_{\I}^n$,
where
\begin{equation}
\oO_{\I'}
\equiv
\begin{cases}
1
&\text{if}\ \I'=\emptyset,
\\
\displaystyle\max_{\ww_{\Ipc}\in\T_{\Ipc}}
\lrbar{\T_{\I'|\Ipc}\lrsb{\ww_{\Ipc}}}
&\text{if}\ \emptyset\neq\I'\subsetneq\I,
\\
|\T|
&\text{if}\ \I'=\I.
\end{cases}
\label{eq:oO}
\end{equation}
\end{lem}
\begin{IEEEproof}
We show this lemma for the completeness of the paper.
First, we have
\begin{align}
\sum_{\substack{
\ww_{\I}\in\T:
\\
p_{\ww_i,\ww'_i}
\leq\frac{\alpha_{F_i}}{|\im\F_i|}
\ \text{for all}\  i\in\I'
\\
p_{\ww_i,\ww'_i}
>\frac{\alpha_{F_i}}{|\im\F_i|}
\ \text{for all}\ i\in\Ipc
}}
\prod_{i\in\I} p_{\ww_i,\ww'_i}
&=
\sum_{\substack{
\ww_{\Ipc}\in\T_{\Ipc}:
\\
p_{\ww_i,\ww'_i}
>\frac{\alpha_{F_i}}{|\im\F_i|}
\\
\text{for all}\  i\in\Ipc
}}
\lrB{\prod_{i\in\Ipc} p_{\ww_i,\ww'_i}}
\sum_{\substack{
\ww_{\I'}\in\T_{\I'|\Ipc}(\ww_{\Ipc}):
\\
p_{\ww_i,\ww'_i}
\leq\frac{\alpha_{F_i}}{|\im\F_i|}
\\
\text{for all}\ i\in\I'
}}
\prod_{i\in\I'}p_{\ww_i,\ww'_i}
\notag
\\
&\leq
\lrB{\prod_{i\in\I'}\frac{\alpha_{F_i}}{|\im\F_i|}}
\sum_{\substack{
\ww_{\Ipc}\in\T_{\Ipc}:
\\
p_{\ww_i,\ww'_i}
>\frac{\alpha_{F_i}}{|\im\F_i|}
\\
\text{for all}\ i\in\Ipc
}}
\lrB{
\prod_{i\in\Ipc} p_{\ww_i,\ww'_i}
}
\sum_{\substack{
\ww_{\I'}\in\T_{\I'|\Ipc}(\ww_{\Ipc}):
\\
p_{\ww_i,\ww'_i}
\leq\frac{\alpha_{F_i}}{|\im\F_i|}
\\
\text{for all}\ i\in\I'
}}
1
\notag
\\
&\leq
\oO_{\I'}
\lrB{\prod_{i\in\I'}\frac{\alpha_{F_i}}{|\im\F_i|}}
\prod_{i\in\Ipc}
\lrB{
\sum_{\substack{
\ww_i\in\W_i^n:
\\
p_{\ww_i,\ww'_i}
>\frac{\alpha_{F_i}}{|\im\F_i|}
}}
p_{\ww_i,\ww'_i}
}
\notag
\\
&\leq
\oO_{\I'}
\lrB{\prod_{i\in\I'}\frac{\alpha_{F_i}}{|\im\F_i|}}
\prod_{i\in\Ipc}
\lrB{\beta_{F_i}+1}
\notag
\\
&=
\frac{
\alpha_{F_{\I'}}\lrB{\beta_{F_{\Ipc}}+1}
\oO_{\I'}
}
{\prod_{i\in\I'}\lrbar{\im\F_i}}
\label{eq:lemma-mcrp}
\end{align}
for all $(\ww'_{\I},\I')$ satisfying
$\ww'_{\I}\in\T$ and $\emptyset\neq\I'\subsetneq\I$,
where the second inequality comes from (\ref{eq:oO})
and the third inequality comes from (\ref{eq:proof-beta}).
It should be noted that (\ref{eq:lemma-mcrp})
is valid for cases $\I'=\emptyset$ and $\I'=\I$
by letting $\oO_{\emptyset}\equiv 1$
and $\oO_{\I}\equiv |\T|$, respectively, 
because
\begin{align}
\sum_{\substack{
\ww_{\I}\in\T:
\\
p_{\ww_i,\ww'_i}
>\frac{\alpha_{F_i}}{|\im\F_i|}
\ \text{for all}\ i\in\I
}}
\prod_{i\in\I} p_{\ww_i,\ww'_i}
&\leq
\prod_{i\in\I}
\lrB{
\sum_{\substack{
\ww_i\in\W_i^n:
\\
p_{\ww_i,\ww'_i}
>\frac{\alpha_{F_i}}{|\im\F_i|}
}}
p_{\ww_i,\ww'_i}
}
\notag
\\
&\leq
\prod_{i\in\I}
\lrB{\beta_{F_i}+1}
\notag
\\
&=
\frac{\alpha_{F_{\emptyset}}\lrB{\beta_{F_{\I}}+1}\oO_{\emptyset}}
{\prod_{i\in\emptyset}\lrbar{\im\F_i}}
\end{align}
and
\begin{align}
\sum_{\substack{
\ww_{\I}\in\T:
\\
p_{\ww_i,\ww'_i}
\leq\frac{\alpha_{F_i}}{|\im\F_i|}
\ \text{for all}\ i\in\I
}}
\prod_{i\in\I} p_{\ww_i,\ww'_i}
&\leq
\frac{\alpha_{F_{\I}}|\T|}
{\prod_{i\in\I}\lrbar{\im\F_i}}
\notag
\\
&=
\frac{\alpha_{F_{\I}}\lrB{\beta_{F_{\emptyset}}+1}\oO_{\I}}
{\prod_{i\in\I}\lrbar{\im\F_i}}.
\end{align}
Then we have
\begin{align}
p_{F_{\I}}\lrsb{\lrb{
f_{\I}:
\lrB{\T\setminus\{\ww_{\I}\}}\cap\fC_{F_{\I}}(f_{\I}(\ww_{\I}))
\neq
\emptyset
}}
&\leq
\sum_{\ww'_{\I}\in\T\setminus\{\ww_{\I}\}}
p_{F_{\I}}\lrsb{\lrb{
f_{\I}:
f_{\I}(\ww_{\I})=f_{\I}(\ww'_{\I})
}}
\notag
\\
&=
\sum_{\ww'_{\I}\in\T\setminus\{\ww_{\I}\}}
p_{F_{\I}}\lrsb{\lrb{
f_{\I}:
f_i(\ww_i)=f_i(\ww'_i)
\ \text{for all}\ i\in\I
}}
\notag
\\
&=
\sum_{\ww'_{\I}\in\T\setminus\{\ww_{\I}\}}
\prod_{i\in\I} p_{\ww_i,\ww'_i}
\notag
\\
&=
\sum_{\ww'_{\I}\in\T}
\prod_{i\in\I} p_{\ww_i,\ww'_i}
-
\prod_{i\in\I} p_{\ww_i,\ww_i}
\notag
\\
&=
\sum_{\I'\subset\I}
\sum_{\substack{
\ww_{\I}\in\T:
\\
p_{\ww_i,\ww'_i}
\leq\frac{\alpha_{F_i}}{|\im\F_i|}
\ \text{for all}\ i\in\I'
\\
p_{\ww_i,\ww'_i}
>\frac{\alpha_{F_i}}{|\im\F_i|}
\ \text{for all}\ i\in\Ipc
}}
\prod_{i\in\I} p_{\ww_i,\ww'_i}
-1
\notag
\\
&\leq
\sum_{\I'\subset\I}
\frac{\alpha_{F_{\I'}}\lrB{\beta_{F_{\Ipc}}+1}\oO_{\I'}}
{\prod_{i\in\I}|\im\F_i|}
-1
\notag
\\
&=
\sum_{\I'\in2^{\I}\setminus\{\emptyset\}}
\frac{
\alpha_{F_{\I'}}\lrB{\beta_{F_{\Ipc}}+1}\oO_{\I'}
}
{\prod_{i\in\I'}|\im\F_i|}
+
\beta_{F_{\I}}
\end{align}
for all $\T\subset\W_{\I'}^n$ and $\ww_{\I'}\in\W_{\I'}^n$,
where the third equality comes from the fact that $p_{\ww_i,\ww_i}=1$,
the second inequality comes from (\ref{eq:lemma-mcrp}),
and the last equality comes from the fact that
$\alpha_{F_{\emptyset}}=1$,
$\beta_{F_{\emptyset^{\complement}}}=\beta_{F_{\I}}$,
$\prod_{i\in\emptyset}|\im\F_i|=1$,
and $\oO_{\emptyset}=1$.
\end{IEEEproof}

\subsection{The First and the Second Terms of (\ref{eq:proof-lossy-error})}
\label{sec:lossy-bcp}

Here, we elucidate the first and the second terms
on the righthand side of (\ref{eq:proof-lossy-error}).

Let us assume that
ensembles $(\F_i,p_{F_i})$ and $(\G_i,p_{G_i})$ have the hash property 
((\ref{eq:hash}) in Appendix \ref{sec:hash}) for every $i\in\cS$,
where their dependence on $n$ is omitted.
In the following, we omit the dependence of $W$ and $X$ on $n$
when it appears in the subscript of $\mu$.
Moreover, we omit the dependence of $\alpha$ and $\beta$ on $n$.

\begin{lem}[{\cite[Eq. (50)]{ICC}}]
\label{lem:channel-bcp}
For a given set $\cS$,
let $\{(W^n_{\cS},X^n_{\cS}\}_{n=1}^{\infty}$ be general correlated
sources, where $(W^n_{\cS},X^n_{\cS})\equiv\{(W^n_i,X^n_i)\}_{i\in\cS}$.
Let $\uT$ be defined as
\begin{equation*}
\uT
\equiv
\lrb{
(\ww_{\cS},\xx_{\cS}):
\begin{aligned}
&\frac 1n
\log_2\frac1{
\mu_{W_{\cS'}|X_{\cS}}(\ww_{\cS'}|\xx_{\cS})
}
\geq
\uH(\WW_{\cS'}|\XX_{\cS})-\e
\\
&\text{for all}\ \cS'\in2^{\cS}\setminus\{\emptyset\}
\end{aligned}
}.
\end{equation*} 
Then we have
\begin{align*}
&
E_{F_{\cS}}\lrB{
\sum_{
\xx_{\cS},\cc_{\cS}
}
\mu_{X_{\cS}}(\xx_{\cS})
\lrbar{
\mu_{W_{\cS}|X_{\cS}}(
\fC_{F_{\cS}}(\cc_{\cS})
|\xx_{\cS}
)
-
\frac{
1
}{
\prod_{i\in\cS}
|\im\F_i|
}
}
}
\\*
&\leq
\sqrt{
\alpha_{F_{\cS}}-1
+\sum_{\cS'\in2^{\cS}\setminus\{\emptyset\}}
\alpha_{F_{\cS\setminus\cS'}}
[\beta_{F_{\cS'}}+1]
\lrB{
\prod_{i\in\cS'}|\im\F_i|
}
2^{-n[\uH(\WW_{\cS'}|\XX_{\cS})-\e]}
}
+
2\mu_{W_{\cS}X_{\cS}}(\uT^{\complement}).
\end{align*}
\end{lem}
\begin{IEEEproof}
Let $\uT(\xx_{\cS})$ be defined as
\begin{equation*}
\uT
(\xx_{\cS})
\equiv
\lrb{
\ww_{\cS}:
(\ww_{\cS},\xx_{\cS})
\in\uT
}.
\end{equation*}
Then we have
\begin{align}
&
E_{F_{\cS}}\lrB{
\sum_{
\xx_{\cS},\cc_{\cS}
}
\mu_{X_{\cS}}(\xx_{\cS})
\lrbar{
\mu_{W_{\cS}|X_{\cS}}(
\fC_{F_{\cS}}(\cc_{\cS})
|\xx_{\cS}
)
-
\frac{
1
}{
\prod_{i\in\cS}
|\im\F_i|
}
}
}
\notag
\\*
&\leq
E_{F_{\cS}}
\lrB{
\sum_{
\xx_{\cS},\cc_{\cS}
}
\mu_{X_{\cS}}(\xx_{\cS})
\lrbar{
\mu_{W_{\cS}|X_{\cS}}(
\uT(\xx_{\cS})
\cap\fC_{F_{\cS}}(\cc_{\cS})
|\xx_{\cS}
)
-
\frac{
\mu_{W_{\cS}|X_{\cS}}(
\uT(\xx_{\cS})
|\xx_{\cS}
)
}{
\prod_{i\in\cS}
|\im\F_i|
}
}
}
\notag
\\*
&\quad
+
E_{F_{\cS}}
\lrB{
\sum_{
\xx_{\cS},\cc_{\cS}
}
\mu_{W_{\cS}|X_{\cS}}(
\uT(\xx_{\cS})^{\complement}
\cap\fC_{F_{\cS}}(\cc_{\cS})
|\xx_{\cS}
)
\mu_{X_{\cS}}(\xx_{\cS})
}
+
E_{F_{\cS}}
\lrB{
\sum_{
\xx_{\cS},\cc_{\cS}
}
\frac{
\mu_{W_{\cS}|X_{\cS}}(
\uT(\xx_{\cS})^{\complement}
|\xx_{\cS}
)
\mu_{X_{\cS}}(\xx_{\cS})
}{
\prod_{i\in\cS}
|\im\F_i|
}
}
\notag
\\
&=
\sum_{
\xx_{\cS}
}
\mu_{W_{\cS}|X_{\cS}}(
\uT(\xx_{\cS})
|\xx_{\cS}
)
\mu_{X_{\cS}}(\xx_{\cS})
E_{F_{\cS}}\lrB{
\sum_{
\cc_{\cS}
}
\lrbar{
\frac{
\mu_{W_{\cS}|X_{\cS}}(
\uT(\xx_{\cS})
\cap\fC_{F_{\cS}}(\cc_{\cS})
|\xx_{\cS}
)
}{
\mu_{W_{\cS}|X_{\cS}}(
\uT(\xx_{\cS})
|\xx_{\cS}
)
}
-
\frac1
{\prod_{i\in\cS}|\im\F_i|}
}
}
\notag
\\*
&\quad
+
2
\sum_{
\xx_{\cS}
}
\mu_{W_{\cS}|X_{\cS}}(
\uT(\xx_{\cS})^{\complement}
|\xx_{\cS}
)
\mu_{X_{\cS}}(\xx_{\cS})
\notag
\\
&\leq
\sum_{
\xx_{\cS}
}
\mu_{W_{\cS}|X_{\cS}}(
\uT(\xx_{\cS})
|\xx_{\cS}
)
\mu_{X_{\cS}}(\xx_{\cS})
\sqrt{
\alpha_{F_{\cS}}-1
+
\sum_{\cS'\in2^{\cS}\setminus\{\emptyset\}}
\alpha_{F_{\cS\setminus\cS'}}
[\beta_{F_{\cS'}}+1]
\lrB{
\prod_{i\in\cS'}|\im\F_i|
}
\frac{
2^{-n[\uH(\WW_{\cS'}|\XX_{\cS})-\e]}
}{
\mu_{W_{\cS}|X_{\cS}}(
\uT(\xx_{\cS})
|\xx_{\cS}
)
}
}
\notag
\\*
&\quad
+
2\mu_{W_{\cS}X_{\cS}}(\uT^{\complement})
\notag
\\
&\leq
\sum_{
\xx_{\cS}
}
\mu_{X_{\cS}}(\xx_{\cS})
\sqrt{
\alpha_{F_{\cS}}-1
+
\sum_{\cS'\in2^{\cS}\setminus\{\emptyset\}}
\alpha_{F_{\cS\setminus\cS'}}
[\beta_{F_{\cS'}}+1]
\lrB{
\prod_{i\in\cS'}|\im\F_i|
}
2^{-n[\uH(\WW_{\cS'}|\XX_{\cS})-\e]}
}
+
2\mu_{W_{\cS}X_{\cS}}(\uT^{\complement})
\notag
\\
&=
\sqrt{
\alpha_{F_{\cS}}-1
+\sum_{\cS'\in2^{\cS}\setminus\{\emptyset\}}
\alpha_{F_{\cS\setminus\cS'}}
[\beta_{F_{\cS'}}+1]
\lrB{
\prod_{i\in\cS'}|\im\F_i|
}
2^{-n[\uH(\WW_{\cS'}|\XX_{\cS})-\e]}
}
+
2\mu_{W_{\cS}X_{\cS}}(\uT^{\complement}),
\label{eq:proof-crng-bcp-5}
\end{align}
where the first inequality comes from the triangular inequality
and the second inequality comes from
Lemma~\ref{lem:mBCP} in Appendix~\ref{sec:hash}
by letting
\begin{align*}
\T
&\equiv
\uT(\xx_{\cS})
\\
Q(\cdot)
&\equiv
\mu_{W_{\cS}|X_{\cS}}(\cdot|\xx_{\cS})
\end{align*}
and using the relations
\begin{align}
\T_{\cS'}
&\subset
\lrb{
\ww_{\cS'}:
\frac 1n
\log_2\frac1{\mu_{W_{\cS'}|X_{\cS}}(\ww_{\cS'}|\xx_{\cS})}
\geq
\uH(\WW_{\cS'}|\XX_{\cS})-\e
}
\notag
\\
\oQ_{\cS'^{\complement}}
&=
\max_{\ww_{\cS'}\in\T_{\cS'}}
\sum_{\ww_{\cS'^{\complement}}
\in\T_{\cS'^{\complement}|\cS'}(\ww_{\cS'})}
\mu_{W_{\cS'}W_{\cS'^{\complement}}|X_{\cS}}
(\ww_{\cS'},\ww_{\cS'^{\complement}}|\xx_{\cS})
\notag
\\
&\leq
\max_{\ww_{\cS'}\in\T_{\cS'}}
\mu_{W_{\cS'}|X_{\cS}}(\ww_{\cS'}|\xx_{\cS})
\notag
\\
&\leq
2^{-n[\uH(\WW_{\cS'}|\XX_{\cS})-\e]}.
\end{align}
\end{IEEEproof}

\subsection{From the Third to the Fifth Terms of
(\ref{eq:proof-lossy-error})}
\label{sec:lossy-crp}

Here, we elucidate the third to the fifth terms
on the righthand side of (\ref{eq:proof-lossy-error}).
We omit the dependence on $j\in\J$.

Let us assume that
ensembles $(\F_i,p_{F_i})$ and $(\G_i,p_{G_i})$
have the hash property
((\ref{eq:hash}) given in Appendix \ref{sec:hash}) for every $i\in\I$,
where their dependence on $n$ is omitted.
For brevity, the above joint ensemble $(\F_i\times\G_i,p_{(F,G)_i})$
is renamed $(\F_i,p_{F_i})$ in the following.
Furthermore, we omit the dependence of $C$, $W$, $\hW$, and $Y$ on $n$,
when it appears in the subscript of $\mu$.
We omit the dependence of $\alpha$ and $\beta$ on $n$.

Let us assume that $(\WW_{\I},\YY,\CC_{\I},\hWW_{\I})$
satisfies the Markov chain
\begin{equation}
W^n_{\I}\markov (C^{(n)}_{\I},Y^n)\markov\hW^n_{\I}
\label{eq:markov-decoding}
\end{equation}
and
\begin{equation}
f_i(W^n_i)=C^{(n)}_i
\label{eq:source-encoding}
\end{equation}
for all $i\in\I$ and $n\in\NN$.
For a given $\e>0$, let $\oT$ and $\oT(\yy)$  be defined as
\begin{align*}
\oT
&\equiv
\lrb{
(\ww_{\I},\yy):
\begin{aligned}
&
\frac1n\log_2
\frac1{\mu_{W_{\I'}|W_{\Ipc}Y}
(\ww_{\I'}|\ww_{\Ipc},\yy)}
\leq
\oH(\WW_{\I'}|\WW_{\Ipc},\YY)+\e
\\
&\text{for all}\ \I'\in2^{\I}\setminus\{\emptyset\}
\end{aligned}
}
\\
\oT(\yy)
&\equiv
\{\ww_{\I}: (\ww_{\I},\yy)\in\oT\}.
\end{align*}

First, we show the following lemma.
\begin{lem}[{\cite[Eq. (58)]{CRNG-MULTI}}]
\label{lem:tsdecoding}
Let us assume that $\chww_{\I}(\cc_{\I},\yy)$
outputs one of the elements
in $\oT(\yy)\cap\fC_{f_{\I}}(\cc_{\I})$,
where it outputs an arbitrary element of $\W_{\I}^n$
when $\oT(\yy)\cap\fC_{f_{\I}}(\cc_{\I})=\emptyset$.
Then we have
\begin{align*}
E_{F_{\I}}\lrB{
\mu_{W_{\I}Y}\lrsb{
\lrb{
(\ww_{\I},\yy): \chww_{\I}(F_{\I}(\ww_{\I}),\yy)\neq\ww_{\I}
}
}
}
&\leq
\sum_{\I'\in2^{\I}\setminus\{\emptyset\}}
\alpha_{F_{\I'}}\lrB{\beta_{F_{\Ipc}}+1}
\frac{
2^{
n\lrB{\oH(\WW_{\I'}|\WW_{\Ipc},\YY)+\e}
}
}{
\prod_{i\in\I'}|\im\F_i|
}
+\beta_{F_{\I}}
+\mu_{W_{\I}Y}(\oT^{\complement}).
\end{align*}
\end{lem}
\begin{IEEEproof}
Assume that $\ww_{\I}\in\oT(\yy)$
and $\chww_{\I}(f_{\I}(\ww_{\I}),\yy)\neq\ww_{\I}$.
Since $\ww_{\I}\in\fC_{f_{\I}}(f_{\I}(\ww_{\I}))$,
we have $\oT(\yy)\cap\fC_{f_{\I}}(f_{\I}(\ww_{\I}))\neq\emptyset$
and
$\chww_{\I}(f_{\I}(\ww_{\I}),\yy)\in\lrB{\oT(\yy)\setminus\{\ww_{\I}\}}
\cap\fC_{f_{\I}}(f_{\I}(\ww_{\I}))$,
which implies
$\lrB{\oT(\yy)\setminus\{\ww_{\I}\}}
\cap\fC_{f_{\I}}(f_{\I}(\ww_{\I}))\neq\emptyset$.
Then we have
\begin{align}
E_{F_{\I}}\lrB{
\chi(\chww_{\I}(F_{\I}(\ww_{\I}),\yy)\neq\ww_{\I})
}
&\leq
p_{F_{\I}}\lrsb{\lrb{
f_{\I}:
\lrB{\oT(\yy)\setminus\{\ww_{\I}\}}
\cap\C_{f_{\I}}(f_{\I}(\ww_{\I}))\neq\emptyset
}}
\notag
\\
&\leq
\sum_{\I'\in2^{\I}\setminus\{\emptyset\}}
\frac{
\alpha_{F_{\I'}}\lrB{\beta_{F_{\Ipc}}+1}
\oO_{\I'}
}{
\prod_{i\in\I'}\lrbar{\im\F_i}
}
+\beta_{F_{\I}}
\notag
\\
&\leq
\sum_{\I'\in2^{\I}\setminus\{\emptyset\}}
\alpha_{F_{\I'}}\lrB{\beta_{F_{\Ipc}}+1}
\frac{
2^{
n\lrB{\oH(\WW_{\I'}|\WW_{\Ipc},\YY)+\e}
}
}{
\prod_{i\in\I'}\lrbar{\im\F_i}
}
+\beta_{F_{\I}},
\end{align}
where
the second inequality comes from Lemma~\ref{lem:mCRP} 
in Appendix \ref{sec:hash}
by letting $\T\equiv\oT(\yy)$
and the third inequality comes from the fact that
\begin{equation*}
\oO_{\I'}
\leq
2^{n\lrB{\oH(\WW_{\I'}|\WW_{\Ipc},\YY)+\e}}.
\end{equation*}
We have
\begin{align}
&
E_{F_{\I}}\lrB{
\mu_{W_{\I}Y}\lrsb{
\lrb{
(\ww_{\I},\yy): \chww_{\I}(F_{\I}(\ww_{\I}),\yy)\neq\ww_{\I}
}
}
}
\notag
\\*
&=
E_{F_{\I}}\lrB{
\sum_{
\ww_{\I},\yy
}
\mu_{W_{\I}Y}(\ww_{\I},\yy)
\chi(\chww_{\I}(F_{\I}(\ww_{\I}),\yy)\neq \ww_{\I})
}
\notag
\\*
&=
\sum_{(\ww_{\I},\yy)\in\oT}
\mu_{W_{\I}Y}(\ww_{\I},\yy)
E_{F_{\I}}\lrB{
\chi(\chww_{\I}(F_{\I}(\zz_{\I}),\yy)\neq\ww_{\I})
}
+
\sum_{(\ww_{\I},\yy)\in\oT^{\complement}}
\mu_{W_{\I}Y}(\ww_{\I},\yy)
E_{F_{\I}}\lrB{
\chi(\chww_{\I}(F_{\I}(\ww_{\I}),\yy)\neq \ww_{\I})
}
\notag
\\
&\leq
\sum_{\I'\in2^{\I}\setminus\{\emptyset\}}
\alpha_{F_{\I'}}\lrB{\beta_{F_{\Ipc}}+1}
\frac{
2^{
n\lrB{\oH(\WW_{\I'}|\WW_{\I\setminus\I'},\YY)+\e}
}
}{
\prod_{i\in\I'}\lrbar{\im\F_i}
}
+\beta_{F_{\I}}
+\mu_{Z_{\I}W}(\oT^{\complement}).
\end{align}
\end{IEEEproof}

Next, we introduce the following lemma on the stochastic decision.
\begin{lem}[{
\cite[Lemma 20 in the extended version]{CRNG-GAS}
\cite[Corollary 2]{SDECODING}\cite[Lemma 4]{CRNG-CHANNEL}
}]
\label{lem:sdecoding}
Let $(U,V)$ be a pair consisting of state $U\in\U$
and observation $V\in\V$.
Let $\mu_{UV}$ be the joint distribution of $(U,V)$
and $\mu_{U|V}$ be defined as
\begin{equation*}
\mu_{U|V}(u|v)
\equiv
\frac{\mu_{UV}(u,v)}
{\mu_V(v)}
\end{equation*}
when $\mu_V(v)\equiv\sum_u\mu_{UV}(u,v)\neq 0$.
We make a stochastic decision $\hU$ with $\mu_{U|V}$
guessing state $U$,
that is, the joint distribution of $(U,V,\hU)$ is given as
\begin{equation*}
\mu_{UV\hU}(u,v,\hu)\equiv\mu_{UV}(u,v)\mu_{U|V}(\hu|v).
\end{equation*}
The decision error probability yielded by this rule
is at most twice the decision error probability
of {\em any} (possibly stochastic) decision, 
that is,
\begin{align*}
\sum_{\substack{
u\in\U,v\in\V,\hu\in\U:
\\
\hu\neq u
}}
\mu_{UV}(u,v)
\mu_{U|V}(\hu|v)
&\leq
2
\sum_{\substack{
u\in\U,v\in\V,\chu\in\U:
\\
\chu\neq u
}}
\mu_{UV}(u,v)
\mu_{\chU|V}(\chu|v)
\end{align*}
for arbitrary probability distribution $\mu_{\chU|V}$,
where $\hU$ is the inferred value of $U$ by the decision rule.
\end{lem}
\begin{IEEEproof}
Here, we show this lemma directly for the completeness of this paper.
We have
\begin{align}
\sum_{\substack{
u\in\U,v\in\V,\hu\in\U:
\\
\hu\neq u
}}
\mu_{UV}(u,v)
\mu_{U|V}(\hu|v)
&=
\sum_{\substack{
u\in\U,v\in\V
}}
\mu_{UV}(u,v)
[1-\mu_{U|V}(u|v)]
\notag
\\
&=
\sum_{\substack{
u\in\U,v\in\V
}}
[\mu_{U|V}(u|v)-\mu_{U|V}(u|v)^2]
\mu_V(v)
\notag
\\
&\leq
\sum_{\substack{
u\in\U,v\in\V
}}
[\mu_{U|V}(u|v)-\mu_{U|V}(u|v)^2]
\mu_V(v)
+\sum_{\substack{
u\in\U,v\in\V
}}
[\mu_{U|V}(u|v)-\mu_{\chU|V}(u|v)]^2
\mu_V(v)
\notag
\\*
&\quad
+\sum_{\substack{
u\in\U,v\in\V
}}
[\mu_{U|V}(u|v)-\mu_{\chU|V}(u|v)]
\mu_V(v)
+\sum_{\substack{
u\in\U,v\in\V
}}
\mu_{\chU|V}(u|v)[1-\mu_{\chU|v}(u|v)]
\mu_V(v)
\notag
\\
&=
\sum_{\substack{
u\in\U,v\in\V
}}
2\mu_{U|V}(u|v)
[1-\mu_{\chU|V}(u|v)]
\mu_V(v)
\notag
\\
&=
2\sum_{\substack{
u\in\U,v\in\V
}}
\mu_{UV}(u,v)
[1-\mu_{\chU|V}(u|v)]
\notag
\\
&=
2\sum_{\substack{
u\in\U,v\in\V,\chu\in\U:
\\
\chu\neq u
}}
\mu_{UV}(u,v)
\mu_{\chU|V}(\chu|v),
\end{align}
where the inequality comes from the fact that
the third term on the right hand side is zero because
$\sum_{u\in\U}\mu_{U|V}(u|v)=\sum_{u\in\U}\mu_{\chU|V}(u|v)=1$
and the last term on the right hand side is not negative because
$\mu_{\chU|V}(u|v)\in[0,1]$ for all $u\in\U$ and $v\in\V$.
\end{IEEEproof}

Finally, we show the following lemma,
which we can derive from the third to the fifth terms
of (\ref{eq:proof-lossy-error}).
\begin{lem}[{\cite[Lemma 2 in the extended version]{ICC}}]
\label{lem:channel-crp}
Assume that $(W^n_{\I},Y^n,C^{(n)}_{\I},\hW^n_{\I})$
satisfy (\ref{eq:markov-decoding}) and (\ref{eq:source-encoding}).
Then the expectation of the probability of the event
$\hWW_{\I}\neq\WW_{\I}$ is evaluated as
\begin{align*}
&
E_{F_{\I}}\lrB{
\sum_{\substack{
\ww_{\I},\yy,\hww_{\I},\cc_{\I}:\\
\hww_{\I}\neq\ww_{\I}
}}
\mu_{\hW_{\I}|C_{\I}Y}
(\hww_{\I}|F_{\I}(\ww_{\I}),\yy)
\mu_{W_{\I}Y}(\ww_{\I},\yy)
}
\notag
\\*
&\leq
2
\sum_{\I'\in2^{\I}\setminus\{\emptyset\}}
\alpha_{F_{\I'}}\lrB{\beta_{F_{\Ipc}}+1}
2^{
-n\lrB{
\sum_{i\in\I'}r_i-\oH(\WW_{\I'}|\WW_{\Ipc},\YY)
-\e
}
}
+2\beta_{F_{\I}}
+2\mu_{W_{\I}Y}(\oT^{\complement}).
\end{align*}
\end{lem}
\begin{IEEEproof}
For given $f_{\I}$,
the joint distribution of $(W_{\I}^n,Y^n,C^{(n)}_{\I})$ is given as
\begin{equation}
\mu_{W_{\I}C_{\I}Y}(\ww_{\I},\cc_{\I},\yy)
=
\chi(f_{\I}(\ww_{\I})=\cc_{\I})
\mu_{W_{\I}Y}(\ww_{\I},\yy)
\label{eq:joint-WICIY}
\end{equation}
from (\ref{eq:source-encoding}).
Then we have
\begin{align}
\mu_{W_{\I}|C_{\I}Y}(\ww_{\I}|\cc_{\I},\yy)
&\equiv
\frac{
\mu_{W_{\I}C_{\I}Y}(\ww_{\I},\cc_{\I},\yy)
}{
\sum_{\ww_{\I}}
\mu_{W_{\I}C_{\I}Y}(\ww_{\I},\cc_{\I},\yy)
}
\notag
\\
&=
\frac{
\mu_{W_{\I}Y}(\ww_{\I},\yy)\chi(f_{\I}(\ww_{\I})=\cc_{\I})
}{
\sum_{\ww_{\I}}
\mu_{W_{\I}Y}(\ww_{\I},\yy)
\chi(f_{\I}(\ww_{\I})=\cc_{\I})
}
\notag
\\
&=
\mu_{\hW_{\I}|C_{\I}Y}(\ww_{\I}|\cc_{\I},\yy),
\end{align}
that is, the constrained-random number generator
is a stochastic decision with
$\mu_{W_{\I}|C_{\I}Y}$.
By letting
\begin{equation}
\mu_{\chW_{\I}|C_{\I}Y}(\chww_{\I}|\cc_{\I},\yy)
\equiv
\chi(\chww_{\I}(\cc_{\I},\yy)=\chww_{\I}),
\end{equation}
we have the fact that
\begin{align}
&
\sum_{\substack{
\ww_{\I},\yy,\hww_{\I}:\\
\hww_{\I}\neq\ww_{\I}
}}
\mu_{\hW_{\I}|C_{\I}Y}
(\hww_{\I}|f_{\I}(\ww_{\I}),\yy)
\mu_{W_{\I}Y}(\ww_{\I},\yy)
\notag
\\*
&=
\sum_{\substack{
\ww_{\I},\yy,\hww_{\I},\cc_{\I}:\\
\hww_{\I}\neq\ww_{\I}
}}
\mu_{\hW_{\I}|C_{\I}Y}(\hww_{\I}|\cc_{\I},\yy)
\chi(f_{\I}(\ww_{\I})=\cc_{\I})
\mu_{W_{\I}Y}(\ww_{\I},\yy)
\notag
\\
&=
\sum_{\substack{
\ww_{\I},(\cc_{\I},\yy),\hww_{\I}:\\
\hww_{\I}\neq\ww_{\I}
}}
\mu_{W_{\I}|C_{\I}Y}(\hww_{\I}|\cc_{\I},\yy)
\mu_{W_{\I}C_{\I}Y}(\ww_{\I},\cc_{\I},\yy)
\notag
\\
&\leq
2
\sum_{\substack{
\ww_{\I},(\cc_{\I},\yy),\chww_{\I}:\\
\chww_{\I}\neq\ww_{\I}
}}
\mu_{\chW_{\I}|C_{\I}Y}(\chww_{\I}|\cc_{\I},\yy)
\mu_{W_{\I}C_{\I}Y}(\ww_{\I},\cc_{\I},\yy)
\notag
\\
&=
2
\sum_{\substack{
\ww_{\I},\yy,\chww_{\I},\cc_{\I}:\\
\chww_{\I}\neq\ww_{\I}
}}
\mu_{W_{\I}Y}(\ww_{\I},\yy)
\chi(f_{\I}(\ww_{\I})=\cc_{\I})
\chi(\chww_{\I}(\cc_{\I},\yy)=\chww_{\I})
\notag
\\
&=
2
\sum_{\substack{
\ww_{\I},\yy:
\\
\chww_{\I}(f_{\I}(\ww_{\I}),\yy)\neq\ww_{\I}
}}
\mu_{W_{\I}Y}(\ww_{\I},\yy)
\notag
\\
&=
2
\mu_{W_{\I}Y}
\lrsb{
\lrb{
(\ww_{\I},\yy):
\chww_{\I}(f_{\I}(\ww_{\I}),\yy)\neq\ww_{\I}
}
},
\end{align}
where the second and the third equalities
come from (\ref{eq:joint-WICIY})
and the inequality comes from Lemma~\ref{lem:sdecoding}.
This yields the fact that
\begin{align}
&
E_{F_{\I}}\lrB{
\sum_{\substack{
\ww_{\I},\yy,\hww_{\I}:\\
\hww_{\I}\neq\ww_{\I}
}}
\mu_{\hW_{\I}|C_{\I}Y}
(\hww_{\I}|f_{\I}(\ww_{\I}),\yy)
\mu_{W_{\I}Y}(\ww_{\I},\yy)
}
\notag
\\*
&\leq
2
E_{F_{\I}}\lrB{
\mu_{W_{\I}Y}
\lrsb{
\lrb{
(\ww_{\I},\yy):
\chww_{\I}(F_{\I}(\ww_{\I}),\yy)\neq\ww_{\I}
}
}
}
\notag
\\
&\leq
2 \sum_{\I'\in2^{\I}\setminus\{\emptyset\}}
\alpha_{F_{\I'}}\lrB{\beta_{F_{\Ipc}}+1}
2^{
-n\lrB{
\sum_{i\in\I'}r_i-\oH(\WW_{\I'}|\WW_{\Ipc},\YY)-\e
}
}
+2 \beta_{F_{\I}}
+2 \mu_{W_{\I}Y}(\oT^{\complement}),
\end{align}
where the last inequality comes from Lemma \ref{lem:tsdecoding}
and the relation $r_i=\log_2(|\C_i|)/n=\log_2(|\im\F_i|)/n$.
\end{IEEEproof}

\subsection{Proof of Lemma}
\label{sec:proof-lemma}

\begin{lem}[{\cite[Lemma 19 in the extended version]{CRNG-GAS}}]
\label{lem:diff-prod}
For any sequence $\{\theta_l\}_{l=1}^L$,
we have
\begin{equation}
\prod_{l=1}^L \theta_l - 1
=
\sum_{l=1}^L
\lrB{\theta_l - 1}
\prod_{l'=l+1}^L\theta_{l'},
\label{eq:diff-prod}
\end{equation}
where $\prod_{l'=L+1}^L\theta_{l'}\equiv1$.
For any sequence $\{\theta_l\}_{l=1}^L$ of non-negative numbers,
we have
\begin{equation}
\lrbar{
\prod_{l=1}^L \theta_l - 1
}
\leq
\sum_{l=1}^L
\lrbar{\theta_l - 1}
\prod_{l'=l+1}^L\theta_{l'}.
\label{eq:diff-prod-ineq}
\end{equation}
\end{lem}
\begin{IEEEproof}
First, we show (\ref{eq:diff-prod}) by induction.
When $L=1$, (\ref{eq:diff-prod}) is trivial.
Assuming that (\ref{eq:diff-prod}) is satisfied,
we have
\begin{align}
\prod_{l=1}^{L+1} \theta_l - 1
&=
\prod_{l=1}^{L+1} \theta_l
-
\theta_{L+1}
+
\theta_{L+1}
- 1
\notag
\\
&=
\lrB{
\prod_{l=1}^L \theta_l
-
1
}
\theta_{L+1}
+
\lrB{
\theta_{L+1}
- 1
}
\notag
\\
&=
\lrB{
\sum_{l=1}^L
\lrB{\theta_l - 1}
\prod_{l'=l+1}^L\theta_{l'}
}
\theta_{L+1}
+
\lrB{
\theta_{L+1}
- 1
}
\notag
\\
&=
\sum_{l=1}^{L+1}
\lrB{\theta_l - 1}
\prod_{l'=l+1}^{L+1}\theta_{l'},
\end{align}
where the third equality comes from the assumption,
and the last equality comes from the fact that
\begin{equation*}
\lrB{\theta_{L+1}-1}\prod_{l'=L+1+1}^{L+1}\theta_{l'}
=\theta_{L+1}-1.
\end{equation*}

We have (\ref{eq:diff-prod-ineq}) from (\ref{eq:diff-prod}) as
\begin{align}
\lrbar{\prod_{l=1}^{L+1} \theta_l - 1}
&=
\lrbar{
\sum_{l=1}^{L+1}
\lrB{\theta_l - 1}
\prod_{l'=l+1}^{L+1}\theta_{l'}
}
\notag
\\
&\leq
\sum_{l=1}^{L+1}
\lrbar{
\lrB{\theta_l - 1}
\prod_{k'=l+1}^{L+1}\theta_{l'}
}
\notag
\\
&=
\sum_{l=1}^{L+1}
\lrbar{\theta_l - 1}
\prod_{l'=l+1}^{L+1}\theta_{l'},
\end{align}
where
the inequality comes from the triangle inequality
and the last equality comes from the fact that
$\{\theta_l\}_{l=1}^L$ is a sequence of non-negative numbers.
\end{IEEEproof}


\begin{thebibliography}{99}
\bibitem{AC98}
R.\ Ahlswede and I.\ Csisz\'{a}r,
``Common randomness in information theory and cryptography
--- Part II: CR capacity,''
{\it IEEE Trans.\ Inform.\ Theory},
vol.\ IT-44, no.1, pp.\ 225--240, Jan.\ 1998.
\bibitem{AK75}
R.\ Ahlswede and J.\ K\"orner,
``Source coding with side information and a converse for degraded broadcast channels,''
{\it IEEE Trans.\ Inform. Theory},
vol.\ IT-21, no.\ 6, pp.\ 629--637, Nov.\ 1975.
\bibitem{B78}
T.\ Berger,
``Multiterminal source coding,''
{\it The Information Theory Approach to Communications},
pp.\ 171--231, Springer-Verlag, New York, 1978.
\bibitem{CW79}
J.\ L.\ Carter and M.\ N.\ Wegman,
``Universal classes of hash functions,''
{\it J.\ Comput.\ Syst.\ Sci.},
vol.\ 18, pp.\ 143--154, 1979.
\bibitem{C75}
T.\ M.\ Cover,
``A proof of the data compression theorem of Slepian and Wolf for ergodic source,''
{\it IEEE Trans. Inform Theory},
vol.~IT-21, no.~2, pp.~226--228, Mar.\ 1975.
\bibitem{CSI82}
I.\ Csisz\'{a}r,
``Linear codes for sources and source networks:
Error exponents, universal coding,''
{\it IEEE Trans.\ Inform.\ Theory},
vol.\ IT-28, no.\ 4, pp.\ 585--592, Jul.\ 1982.
\bibitem{CK11}
I.\ Csisz\'{a}r and J.\ K\"{o}rner,
{\it Information Theory: Coding Theorems for Discrete Memoryless Systems
2nd Ed.},
Cambridge University Press, 2011.
\bibitem{EC91}
W.\ H.\ R.\ Equitz and T.\ M.\ Cover,
``Successive refinement of information,''
{\it IEEE Trans.\ Inform.\ Theory},
vol.\ IT-37, no.\ 2, pp.\ 269--275, Mar.\ 1991.
\bibitem{GK73}
P.\ G\'acs and J.\ K\"orner,
``Common information is far less than mutual information,''
{\it Probl. Control Inf. Theory},
vol.\ 2, no.\ 2, pp.\ 149--162, 1973.
\bibitem{EC82}
A.\ El Gamal and T.\ M.\ Cover,
``Achievable rates for multiple descriptions,''
{\it IEEE Trans.\ Inform.\ Theory},
vol.~IT-28, no.~6, pp.~851--857, Nov.\ 1982.
\bibitem{EK11}
A.\ El Gamal and Y.H. Kim,
{\it Network Information Theory},
Cambridge University Press, 2011.
\bibitem{GP79}
S.\ I.\ Gelfand and M.\ S.\ Pinsker, 
``Source coding with incomplete side information,''
(in Russian) {\it Probl.\ Pered.\ Inform.},
vol.\ 15, no.\ 2, pp.\ 45--57, 1979.
\bibitem{GGWK24}
N.\ Ghaddar, S.\ Ganguly, L.\ Wang, and Y.H.\ Kim,
``A Lego-brick approach to coding for network communication,''
{\it IEEE Trans.\ Inform.\ Theory},
vol.~IT-70, no.~2, pp.~865--903, Feb.\ 2024.
\bibitem{HAN}
T.S.\ Han,
{\it Information-Spectrum Methods in Information Theory},
Springer, 2003.
\bibitem{HK80}
T.S.\ Han and K.\ Kobayashi,
``A unified achievable rate region for a general class of
multiterminal source coding systems,''
{\it IEEE Trans.\ Inform.\ Theory},
vol.~IT-26, no.~3, pp.~277--288, May 1980.
\bibitem{HV93}
T.S.\ Han and S.\ Verd\'u,
``Approximation theory of output statistics,''
{\it IEEE Trans.\ Inform.\ Theory},
vol.~IT-39, no.~3, pp.~752--772, May 1993.
\bibitem{HB85}
C.\ Heegerd and T.\ Berger,
``Rate distortion when side information may be absent,''
{\it IEEE Trans.\ Inform.\ Theory},
vol.~IT-31, no.~6, pp.~727--734, Nov.\ 1985.
\bibitem{IZ89}
R.\ Impagliazzo and D.\ Zuckerman,
``How to recycle random bits,''
{\it 30th IEEE Symp. Fund. Computer Sci.},
Oct.~30--Nov.~1, 1989, pp.~248--253.
\bibitem{IM02}
K.\ Iwata and J.\ Muramatsu,
``An information-spectrum approach to rate-distortion function
with side information,''
{\it IEICE Trans.\ Fundamentals},
vol.\ E85-A, no.\ 6, pp.\ 1387--1395, Jun.\ 2002.
\bibitem{JB08}
S.\ Jana and R.\ Blahut,
``Canonical description for multiterminal source coding,''
{\it Proc.\ 2008 IEEE Int.\ Symp.\ Inform.\ Theory},
Toronto, Canada, July 6--11, pp.\ 697--701, 2008.
\bibitem{KU11}
W.\ Kang and S.\ Ulukus,
``A new data processing inequality and its applications in distributed
source and channel coding,''
{\it IEEE Trans.\ Inform. Theory},
vol.\ IT-57, no.\ 1, pp.\ 56--69, Jan.\ 2011.
\bibitem{K08}
H.\ Koga,
``Coding theorems on the threshold scheme for a general source,''
{\it IEEE Trans.\ Inform.\ Theory},
vol.~IT-54, no.~6, pp.~2658--2677, Jun.~2008.
\bibitem{KM79}
J.\ K\"orner and K.\ Marton,
``How to encode the modulo-two sum of binary sources,''
{\it IEEE Trans.\ Inform.\ Theory},
vol.\ IT-25, no.\ 2, pp. 219-–221, Mar 1979.
\bibitem{MU12}
T.\ Matsuta and T.\ Uyematsu,
``A general formula of rate-distortion functions for source coding
with side information at many decoders,''
{\it Proc.\ 2012 IEEE Int.\ Symp.\ Inform.\ Theory},
Cambridge, U.S.A., July 1--6, pp.\ 174--178, 2012.
\bibitem{MK95}
S.\ Miyake and F.\ Kanaya,
``Coding theorems on correlated general sources,''
{\it IEICE Trans.\ Fundamentals}, vol.\ E78-A, no.\ 9,
pp.\ 1063--1070, Sept.\ 1995.
\bibitem{CRNG}
J.\ Muramatsu,
``Channel coding and lossy source coding using a generator
of constrained random numbers,''
{\it IEEE Trans.\ Inform.\ Theory},
vol.~IT-60, no.~5, pp.~2667--2686, May 2014.
\bibitem{CRNGVLOSSY}
J.\ Muramatsu,
``Variable-length lossy source code using a constrained-random-number
generator,''
{\it IEEE Trans.\ Inform.\ Theory},
vol.~IT-61, no.~6, pp.~3574--3592, Jun.\ 2015.
\bibitem{ICC}
J.\ Muramatsu,
``On the achievability of interference channel coding,''
{\it Proc.\ 2022 Int.\ Symp.\ Inform.\ Theory and its Applicat.},
Tsukuba, Japan, Oct.\ 17--19, pp.\ 39--43, 2022.
Extended version is available at {\tt arXiv:2201.10756[cs.IT]}.
\bibitem{CRNG-GAS}
J.\ Muramatsu,
``Channel codes for relayless networks
with general message access structure,''
{\it Proc.\ 2023 IEEE Inform.\ Theory Workshop},
Saint-Malo, France, Apr.\ 23--28, 2023.
Extended version is available at {\tt arXiv:2206.00792[cs.IT]}.
\bibitem{CRNG-DSC}
J.\ Muramatsu,
``Distributed source coding using constrained-random-number generators,''
{\it Proc.\ 2024 IEEE Int.\ Symp.\ Inform.\ Theory},
Atens, Greece, Jul.\ 7--12, 2024.
Extended version is available at {\tt arXiv:2401.13232[cs.IT]}.
\bibitem{ISPEC-CONVERSE}
J.\ Muramatsu,
``A simple proof of multi-letter converse theorem
for distributed lossless source coding,''
to appear in
{\it Proc.\ 2024 Int.\ Symp.\ Inform.\ Theory and its Applicat.},
Taipei, Taiwan, Nov.\ 10--13, 2024.
\bibitem{HASH}
J.\ Muramatsu and S.\ Miyake,
``Hash property and coding theorems for sparse matrices and
maximal-likelihood coding,''
{\it IEEE Trans.\ Inform.\ Theory},
vol.\ IT-56, no. 5, pp.\ 2143--2167, May 2010.
Corrections: vol.\ IT-56, no.\ 9, p.\ 4762, Sep.\ 2010,
vol.\ IT-59, no.\ 10, pp.\ 6952--6953, Oct.\ 2013.
\bibitem{HASH-BC}
J.\ Muramatsu and S.\ Miyake,
``Construction of broadcast channel code
based on hash property,''
{\it Proc.\ 2010 IEEE Int.\ Symp.\ Inform.\ Theory},
Austin, U.S.A., June 13--18, pp.\ 575--579, 2010.
Extended version is available at {\tt arXiv:1006.5271[cs.IT]}.
\bibitem{HASH-WTC}
J.\ Muramatsu and S.\ Miyake,
``Construction of strongly secure wiretap channel code
based on hash property,''
{\it Proc.\ 2011 IEEE Int.\ Symp.\ Inform.\ Theory},
St. Petersburg, Russia, Jul.\ 31--Aug. 5, 2011, pp.612--616.
\bibitem{SDECODING}
J.\ Muramatsu and S.\ Miyake,
``On the error probability of stochastic decision and stochastic
decoding,''
{\it Proc.\ 2017 IEEE Int.\ Symp.\ Inform.\ Theory},
Aachen, Germany, Jun.\ 25--30, 2017, pp.\ 1643--1647.
Extended version is available at
{\tt arXiv:1701.04950[cs.IT]}.
\bibitem{CRNG-MULTI}
J.\ Muramatsu and S.\ Miyake,
``Multi-terminal codes using constrained-random-number generator,''
{\it Proc.\ 2018 Int.\ Symp.\ Inform.\ Theory and its Applicat.},
Singapore, Oct.~28--31, 2018, pp.\ 612--616.
Extended version is available at {\tt arXiv:1801.02875v2[cs.IT]}.
\bibitem{CRNG-CHANNEL}
J.\ Muramatsu and S.\ Miyake,
``Channel code using constrained-random number generator revisited,''
{\it IEEE Trans.\ Inform.\ Theory},
vol.\ IT-65, no. 1, pp.\ 500--510, Jan.\ 2019.
\bibitem{CoCoNuTS-LOSSY}
J.\ Muramatsu and S.\ Miyake,
``Lossy source code based on CoCoNuTS,''
{\it Proc.\ Workshop on Concepts in Inform. Theory},
Rotterdam, the Netherlands, July.\ 3--5, p.\ 36, 2019.
\bibitem{SP18}
F.\ Shirani and S.\ S.\ Pradhan,
``An achievable rate-distortion region for multiple descriptions
source coding based on coset codes,''
{\it IEEE Trans.\ Inform.\ Theory},
vol.\ IT-64, no.\ 5, pp.\ 3781--3809, May 2018.
\bibitem{SP21}
F.\ Shirani and S.\ S.\ Pradhan,
``A new achievable rate-distortion region for distributed source coding,''
{\it IEEE Trans.\ Inform.\ Theory},
vol.\ IT-67, no.\ 7, pp.\ 4485--4503, Jul.\ 2021.
\bibitem{SW73}
D.\ Slepian and J.\ K.\ Wolf,
``Noiseless coding of correlated information sources,''
{\it IEEE Trans.\ Inform.\ Theory},
vol.~IT-19, no.~4, pp.~471--480, Jul.~1973.
\bibitem{SV93}
Y.\ Steinberg and S.\ Verd\'u,
``Simulation of random process and rate-distortion theory,''
{\it IEEE Trans.\ Inform.\ Theory},
vol.\ IT-42, no.\ 1, pp.\ 63--86, Jan.\ 1996.
\bibitem{TCG11}
R.\ Timo, T.\ Chan, and A.\ Grant,
``Rate distortion with side information at many decoders,''
{\it IEEE Trans.\ Inform.\ Theory},
vol.\ IT-57, no.\ 8, pp.\ 5240--5257, Aug.\ 2011.
\bibitem{T78}
S.-Y.\ Tung,
{\it Multiterminal source coding},
Ph.D. thesis, Cornell University, Ithaca, NY., 1978.
\bibitem{VAR16}
K.\ B.\ Viswanatha, E.\ Akyol, and K.\ Rose,
``Conbinatorial message sharing and a new achievable region
for multiple descriptions,''
{\it IEEE Trans.\ Inform.\ Theory},
vol.\ IT-62, no.\ 2, pp.\ 769--791, Feb.\ 2016.
\bibitem{VKG03}
R.\ Venkataramani, G.\ Kramer, and V.\ K.\ Goyal,
``Multiple description coding with many channels,''
{\it IEEE Trans.\ Inform.\ Theory},
vol.\ IT-49, no.\ 9, pp.\ 2106--2114, Sep.\ 2003.
\bibitem{WA08}
A.\ B.\ Wagner and V.\ Anantharam,
``An improved outer bound for multiterminal source coding,''
{\it IEEE Trans.\ Inform. Theory},
vol.\ IT-54, no.\ 5, pp.\ 1919--1937, May.\ 2008.
\bibitem{WKA11}
A.\ B.\ Wagner, B.\ G.\ Kelly, and Y.\ Altu\u{g},
``Distributed rate-distortion with common components,''
{\it IEEE Trans.\ Inform. Theory},
vol.\ IT-57, no.\ 7, pp.\ 4035--4057, Jul.\ 2011.
\bibitem{WCZCP11}
J.\ Wang, J.\ Chen, L.\ Zhao, P.\ Cuff, and H.\ H.\ Permuter,
``On the role of the refinement layer in multiple description coding
and scalable coding,''
{\it IEEE Trans.\ Inform. Theory},
vol.\ IT-57, no.\ 3, pp.\ 1443--1456, May.\ 2011.
\bibitem{W75}
A.\ D.\ Wyner,
``On source coding with side information at the decoder,''
{\it IEEE Trans.\ Inform.\ Theory},
vol.\ IT-21, no.\ 3, pp.\ 294--300, May.\ 1975.
\bibitem{WZ76}
A.\ D.\ Wyner and J.\ Ziv,
``The rate-distortion function for source coding with side information at the decoder,''
{\it IEEE Trans.\ Inform.\ Theory},
vol.\ IT-22, no.\ 1, pp.\ 1--10, Jan.\ 1976.
\bibitem{YAG12}
M.\ H.\ Yassaee, M.\ R.\ Aref, and A.\ Gohari,
``Achievability proof via output statistics of random binning,''
{\it IEEE Trans.\ Inform.\ Theory},
vol.~IT-60, no.~11, pp.~6760--6786, Nov.~2014.
\bibitem{YQ06a}
S.\ Yang and P.\ Qin,
``An information-spectrum approach to
multiterminal rate-distortion theory,''
Available at {\tt arXiv:cs/0605006v1[cs.IT]}, 2006.
\bibitem{YQ06b}
S.\ Yang and P.\ Qin,
``An enhanced covering lemma for multiterminal source coding,''
{\it Proc.\ 2006 IEEE Inform.\ Theory Workshop},
Chengdu, China, Oct.\ 22--26, pp.\ 303--307, 2006.
\bibitem{ZB87}
Z.\ Zhang and T.\ Berger,
``New results in binary multiple descriptions,''
{\it IEEE Trans.\ Inform.\ Theory},
vol.~IT-33, no.~4, pp.~502--521, Nov.~1987.
\end{thebibliography}
\end{document}